\documentclass[times,review,preprint]{elsarticle}

\usepackage{jcomp}
\usepackage{framed,multirow}
\usepackage{lineno}
\usepackage{amssymb}
\usepackage{amsmath}
\usepackage{mathrsfs} 
\usepackage{latexsym}
\biboptions{sort&compress}
\usepackage{bm}
\usepackage{subfigure}
\usepackage{tikz}
\tikzset{>=stealth}
\usepackage{url}
\usepackage{xcolor}
\definecolor{newcolor}{rgb}{.8,.349,.1}

\journal{Journal of Computational Physics}

\begin{document}

\verso{Tu \textit{et al}}

\begin{frontmatter}

\title{A Novel Algorithm to Solve for an Underwater Line Source Sound Field Based on Coupled Modes and a Spectral Method}

\author[1]{Houwang \snm{Tu}}
\ead{tuhouwang@nudt.edu.cn}
\author[2]{Yongxian \snm{Wang}\corref{cor1}}
\cortext[cor1]{Corresponding author: \url{yxwang@nudt.edu.cn}}
\author[3]{Chunmei \snm{Yang}}
\ead{ycm@fio.org.cn}
\author[1]{Xiaodong \snm{Wang}}
\author[2]{Shuqing \snm{Ma}}
\author[2]{Wenbin \snm{Xiao}}
\author[2]{Wei \snm{Liu}}

\address[1]{College of Computer, National University of Defense Technology, Changsha, 410073, China}
\address[2]{College of Meteorology and Oceanography, National University of Defense Technology, Changsha, 410073, China}
\address[3]{Key Laboratory of Marine Science and Numerical Modeling, First Institute of Oceanography, Ministry of Natural Resources, Qingdao, 266061, China}


\begin{abstract}
A high-precision numerical sound field is the basis of underwater target detection, positioning and communication. A line source in a plane is a common type of sound source in computational ocean acoustics. The exciting waveguide in a range-dependent ocean environment is often structurally complicated; however, traditional algorithms often assume that the waveguide has a simple seabed boundary and that the line source is located at a horizontal range of 0 m, although this ideal situation is rarely encountered in the actual ocean. In this paper, a novel algorithm is designed that can solve for the sound field excited by a line source at any position in a range-dependent ocean environment. The proposed algorithm uses the classic stepwise approximation approach to address the range dependence of the environment and uses the Chebyshev--Tau spectral method to solve for the horizontal wavenumbers and modes of approximately range-independent segments. Once the modal information of these flat segments has been obtained, a global matrix is constructed to solve for the coupling coefficients of all segments, and finally, the complete sound field is synthesized. Numerical experiments using a robust numerical program developed based on this algorithm verify the correctness and usability of our novel algorithm and software. Furthermore, a detailed analysis and test of the computational cost of this algorithm show that it is efficient.
\end{abstract}

\begin{keyword}
\KWD Chebyshev--Tau spectral method \sep Coupled modes \sep Range-dependent
\sep Line source \sep Computational ocean acoustics
\end{keyword}

\end{frontmatter}


\section{Introduction}
Research on the theory of underwater sound propagation has led to a series of breakthroughs in the development and utilization of underwater acoustics. A particularly salient example is the development of computational ocean acoustics, which has long been a research field of great academic value and far-reaching significance in ocean acoustics and has achieved incredible advances in recent decades, motivated by the urgent need for numerical sound fields and rapid advances in computer technology. Various computational ocean acoustic models based on, for example, rays, normal modes, parabolic approximation and wavenumber integration have been applied to calculate sound fields \cite{Jensen2011}. The goal of computational ocean acoustics is to solve the Helmholtz equation derived from the wave equation; therefore, since different models rely on different assumptions and apply different transformations to the Helmholtz equation, the scopes of application and the advantages/disadvantages of these models vary \cite{Etter2013}. Nevertheless, based on a comprehensive summary of the various available ocean acoustic propagation models, Buckingham asserted that the normal mode solution to the Helmholtz equation, including coupled modes, can generally serve as a benchmark solution for the acoustic propagation problem \cite{Buckingham1992}.

The normal mode model was proposed by Pekeris to calculate a homogeneous waveguide within an acoustic half-space \cite{Pekeris1948}. In Pekeris's branch cut approach, the sound field is expressed as a finite number of discrete normal modes and a continuous spectrum corresponding to the contour integral. However, due to the complexity of integrating this continuous spectrum and its minimal impact on the far field, it is typically ignored in traditional normal modes. As an alternative approach, because the classical normal modes can only be used to solve the sound field problem for range-independent waveguides, Pierce and Miller introduced coupled modes into ocean acoustics to extend the solution scope of the normal modes to range-dependent waveguides \cite{Pierce1954,Miller1954}. There are two main strategies for considering coupled modes. One is the consistent coupled modes, which was initially developed by Athanassoulis et al. and applied to 3D problems in layered ocean environments excited by point sources \cite{Athanassoulis2008}; later, Belibassakis et al. formulated a corresponding 2D propagation problem in a vertical plane \cite{Belibassakis2014b}. Using another classical approach proposed earlier, Evans used stepwise approximation to establish the elliptical bidirectional model of a line source that is commonly used in planar geometry waveguides and, on this basis, developed the widely used numerical program COUPLE \cite{Evans1983,Evans1986,Couple}.

Considering the range dependence of stepwise approximation, Porter et al. studied the issue of energy conservation and proposed energy conserving correction schemes such as sound pressure matching, radial particle velocity matching and impedance matching \cite{Porter1991}. The most recent version of the COUPLE, COUPLE07, is often used as a benchmark for testing other models and methods. However, due to unreasonable range normalization, the results of COUPLE may diverge in actual calculations. To overcome this problem, Luo and Yang proposed a global matrix coupled mode solution and established a two-way model to solve for the coupling coefficients \cite{Luowy2012a,Luowy2012c,Luowy2012d,Yangcm2012,Yangcm2015}. This technique is highly computationally efficient and numerically stable and thus is expected to become a representative method of coupled modes. Consequently, this article partially adopts this method and further optimizes it. 

The finite difference method is classically employed to solve for local normal modes. Representative programs implenting this method are KRAKEN and KRAKENC \cite{Kraken2001}. However, KRAKEN cannot accurately calculate the sound field in the case of strong absorption, and KRAKENC finds roots directly on the complex plane, resulting in poor stability and the possibility that some roots may be missed due to changes in medium parameters. Notably, Evans also incorporated the Galerkin method to solve for local normal modes in COUPLE \cite{Couple}. This method starts from the weak form of the modal equation, applies essential boundary conditions to eliminate the derivative term in the partial integral, and finally discretizes the problem into a generalized matrix eigenvalue problem. However, this approach encounters the following problems: first, the lower boundary of the waveguide must be a pressure release boundary, which is not flexible enough to deal with complex ocean environments; and second, the basis/weight function of the Galerkin method must be constructed by solving a nonsingular Sturm--Liouville problem. 

As a class of high-precision methods for solving differential equations, spectral methods were introduced into computational ocean acoustics at the end of the twentieth century \cite{aw,Dzieciuch1993,rimLG,Sabatini2019}. Using such methods, our team has performed a series of studies in recent years focusing on solving underwater acoustic propagation models \cite{Tuhw2020a,Tuhw2020b,Tuhw2021a,Tuhw2021b,Tuhw2021c,Tuhw2021d,Tuhw2021e,Tuhw2022a}. In particular, we have developed a normal mode solver named NM-CT based on the Chebyshev--Tau spectral method and provided the code in the open-source Ocean Acoustics Library (OALIB) \cite{NM-CT}. Numerical simulations have shown that this solver achieves good robustness and accuracy \cite{Tuhw2020b,MultiLC}. 

As a useful simplification of the Helmholtz equation, the normal modes excited by a line source in a plane are also of considerable significance \cite{Jensen2011}. For example, a quasi-three-dimensional sound field can be obtained by performing a simple Fourier transform on a line source sound field \cite{Luowy2016}. Hence, many mature numerical programs include an interface for a line source sound field \cite{Kraken2001,Couple}. In this article, we employ the theory of global matrix coupled modes and the Chebyshev--Tau spectral method to design an efficient, accurate, stable and reliable algorithm to solve for the sound field in the case of a line source with a range-dependent waveguide.
 
\section{Model and Methodology}
\subsection{Stepwise approximation of a range-dependent ocean  environment}
\begin{figure}[htbp]
	\centering
	\begin{tikzpicture}[node distance=2cm,scale = 0.8]
		\node at (1.8,0){$0$};
		\fill[brown] (14,-8) rectangle (2,-6.7);
		\fill[cyan,opacity=0.7] (14,0)--(14,-1.7)--(12,-1.7)--(4,-5.7)--(2,-5.7)--(2,0)--cycle;
		\fill[orange,opacity=0.7] (14,-1.7)--(12,-1.7)--(4,-5.7)--(2,-5.7)--(2,-6.7)--(14,-6.7)--cycle;
		\draw[very thick, ->](2,0)--(14.5,0) node[right]{$x$};
		\draw[very thick, ->](2.02,0)--(2.02,-8.5) node[below]{$z$};
		\filldraw [red] (2.02,-1) circle [radius=2.5pt];
		\node at (3.2,-1){infinite};
		\node at (3.2,-1.5){line source};
		\draw[very thick](2,-5.7)--(4,-5.7);
		\draw[dashed, very thick](4,-5.7)--(5,-5.7);
		\draw[dashed, very thick](5,-5.7)--(5,-4.7);
		\draw[dashed, very thick](5,-4.7)--(7,-4.7);
		\draw[dashed, very thick](7,-4.7)--(7,-3.7);
		\draw[dashed, very thick](7,-3.7)--(9,-3.7);
		\draw[dashed, very thick](9,-3.7)--(9,-2.7);  
		\draw[dashed, very thick](9,-2.7)--(11,-2.7);
		\draw[dashed, very thick](11,-2.7)--(11,-1.7); 
		\draw[dashed, very thick](11,-1.7)--(12,-1.7);		    		
		\draw[very thick](4,-5.7)--(12,-1.7);
		\draw[very thick](12,-1.7)--(14,-1.7);
		\draw[dashed, very thick](5,0)--(5,-4.7);
		\draw[dashed, very thick](7,0)--(7,-3.7);
		\draw[dashed, very thick](9,0)--(9,-2.7); 
		\draw[dashed, very thick](11,0)--(11,-1.7); 
		\draw[dashed, very thick](2.02,-6.7)--(14,-6.7);
		\node at (4.5,-4.7){$h_{j-1}$};			
		\node at (7,0.2){$x_{j-1}$};
		\node at (6.5,-3.7){$h_{j}$};
		\node at (9,0.2){$x_{j}$};
		\node at (8.5,-2.7){$h_{j+1}$};
		\node at (11,0.2){$x_{j+1}$};
		\node at (6,-0.5){$c_w^{j-1},\rho_w^{j-1}$};
		\node at (6,-1.2){$\alpha_w^{j-1}$};    		
		\node at (8,-0.5){$c_w^j,\rho_w^j$};
		\node at (8,-1.2){$\alpha_w^j$};    	
		\node at (10,-0.5){$c_w^{j+1},\rho_w^{j+1}$};
		\node at (10,-1.2){$\alpha_w^{j+1}$};
		\node at (6,-5.5){$c_b^{j-1},\rho_b^{j-1}$};
		\node at (6,-6.3){$\alpha_b^{j-1}$};    		
		\node at (8,-4.5){$c_b^j,\rho_b^j$};
		\node at (8,-5.3){$\alpha_b^j$};     		
		\node at (10,-3.5){$c_b^{j+1},\rho_b^{j+1}$};
		\node at (10,-4.3){$\alpha_b^{j+1}$};
		\node at (8,-7.4){$c_\infty,\quad \rho_\infty,\quad \alpha_\infty$};  
	\end{tikzpicture}
	\caption{Schematic diagram of the stepwise approximation of a range-dependent ocean environment.}
	\label{Figure1}
\end{figure}
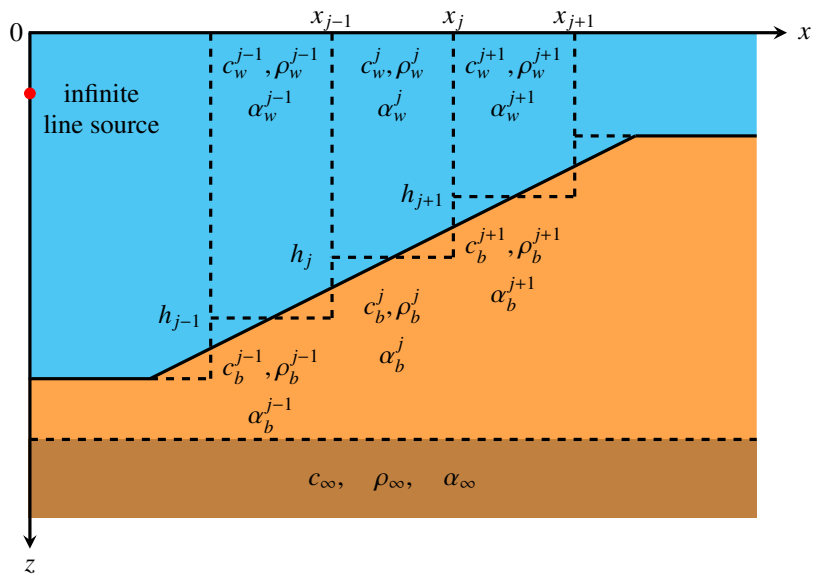
Consider a two-dimensional line source acoustic field in a planar geometric environment, where the infinite harmonic line source is located at a horizontal range of $x=x_\mathrm{s}$ and a depth of $z=z_\mathrm{s}$. Let the acoustic pressure be denoted by $p\equiv p(x,z)$, and omit the time factor $\exp(-\mathrm{i}\omega t)$. Then, the acoustic governing equation (Helmholtz equation) can be written as \cite{Jensen2011}:
\begin{equation}
	\label{eq:1}
	\rho(x,z) \frac{\partial}{\partial x}\left(\frac{1}{\rho(x,z)} \frac{\partial p}{\partial x}\right)+\rho(x,z) \frac{\partial}{\partial z}\left(\frac{1}{\rho(x,z)} \frac{\partial p}{\partial z}\right)+\frac{\omega^{2}}{c^{2}(x,z)} p=-\delta(x-x_\mathrm{s}) \delta\left(z-z_\mathrm{s}\right)
\end{equation}
where $\omega=2\pi f$, with $f$ being the frequency of the acoustic source, and $\rho(x,z)$ and $c(x,z)$ are the acoustic speed and density, respectively. Traditional normal mode theory can only be used to solve for the sound field in the case of a range-independent waveguide. For range-dependent ocean environments, the classic solution is to divide the environment into many sufficiently narrow segments \cite{Evans1983} in a stepwise pattern, as shown in Fig.~\ref{Figure1}, and then treat each segment as a range-independent environment. Once the eigenmodes and horizontal wavenumbers of each segment have been obtained, the continuity conditions at the interfaces of the $J$ segments are used to couple the acoustic subfields of all segments, thereby obtaining the acoustic field of the entire waveguide. Let us begin with the range-independent normal modes.

\subsection{Normal modes of a line source in planar geometry}
In each narrow segment, the acoustic parameters are treated as independent of range, i.e., $\rho(x,z)\equiv\rho(z)$ and $c(x,z)\equiv c(z)$; thus, the Helmholtz equation can be simplified as:
\begin{equation}
	\label{eq:2}
	\frac{\partial^{2} p}{\partial x^{2}}+\rho(z) \frac{\partial}{\partial z}\left(\frac{1}{\rho(z)} \frac{\partial p}{\partial z}\right)+\frac{\omega^{2}}{c^{2}(z)} p=-\delta(x-x_\mathrm{s}) \delta\left(z-z_\mathrm{s}\right)
\end{equation}
According to classical normal mode theory, by using the technique of the separation of variables, the acoustic pressure can be decomposed as follows:
\begin{equation}
	\label{eq:3}
	p(x, z)=X(x) \psi(z)
\end{equation}
$\psi (z)$ in Eq.~\eqref{eq:2} satisfies the following modal equation:
\begin{subequations}
	\label{eq:4}
	\begin{gather}
		\rho(z) \frac{\mathrm{d}}{\mathrm{d} z}\left(\frac{1}{\rho(z)} \frac{\mathrm{d} \psi(z)}{\mathrm{d} z}\right)+\left[k^{2}(z)-k_{x}^{2}\right]\psi(z)=0\\
		k(z)=(1+\mathrm{i}\eta\alpha)\omega/c(z),\quad \eta=(40\pi \log_{10}{\mathrm{e}})^{-1}
	\end{gather}
\end{subequations}
where $k(z)$ is called the complex wavenumber, $\alpha$ is the attenuation coefficient in units of decibels (dB) divided by $\lambda$ (dB$/\lambda$, where $\lambda$ is the wavelength), and $k_{x}$ is the horizontal wavenumber. Eq.~\eqref{eq:4} has a set of solutions $\{(k_{x,m},\psi_m)\}_{m=1}^\infty$, where $\psi_m$ is called an eigenmode. The eigenmodes of Eq.~\eqref{eq:4} should be normalized as follows:
\begin{equation}
	\label{eq:5}
	\int_{0}^{H} \frac{{\psi_m(z)}{\psi_n(z)}}{\rho(z)}\mathrm {d} z=\delta_{mn},
	\quad m,n = 1, 2, \dots
\end{equation}
where $H$ is the ocean depth and $\delta$ is the Kronecker delta function. 

$X(x)$ in Eq.~\eqref{eq:3} is related only to the horizontal range $x$ and satisfies:
\begin{equation}
	\label{eq:6}
	\frac{\mathrm{d}^{2} X(x)}{\mathrm{d} x^{2}}+k_{x}^{2} X(x)=-\frac{\delta(x-x_\mathrm{s}) \psi\left(z_\mathrm{s}\right)}{\rho\left(z_\mathrm{s}\right)}
\end{equation}
In accordance with the known properties of ordinary differential equations, the solution is:
\begin{equation}
	\label{eq:7}
	X(x)=\frac{\mathrm{i}}{2 \rho\left(z_\mathrm{s}\right)} \psi\left(z_\mathrm{s}\right) \frac{\mathrm{e}^{\mathrm{i}k_{x}\vert x\vert}}{k_{x}}  
\end{equation}
Finally, the fundamental solution to the 2D- Helmholtz equation can be approximated as:
\begin{equation}
	\label{eq:8}
	p(x, z)=\frac{\mathrm{i}}{2 \rho\left(z_\mathrm{s}\right)} \sum_{m=1}^{\infty} \psi_{m}\left(z_\mathrm{s}\right) \psi_{m}(z) \frac{\mathrm{e}^{\mathrm{i} k_{x,m}\vert x\vert}}{k_{x,m}}
\end{equation}

For a range-independent ocean environment in which the ocean floor is topped with sediment, $\rho(z)$, $c(z)$ and $\alpha(z)$ are discontinuous at the interface $z=h$. The ocean environment can be divided into a discontinuous water column layer and a bottom sediment layer. The environmental parameters in the water column and bottom sediment layers are defined as: 
\begin{equation}
	\label{eq:9}
		c(z)= \begin{cases}
			c_w(z),&0 \leq z \leq h\\
			c_b(z),&h \leq z \leq H\\
			c_\infty,&z\geq H
		\end{cases},\quad \rho(z)= \begin{cases}
			\rho_w(z),&0 \leq z \leq h\\
			\rho_b(z),&h \leq z \leq H\\
			\rho_\infty,&z\geq H
		\end{cases},\quad \alpha(z)= \begin{cases}
			\alpha_w(z),&0 \leq z \leq h\\
			\alpha_b(z),&h \leq z \leq H\\
			\alpha_\infty,&z\geq H
		\end{cases}
\end{equation}

Boundary conditions should be imposed at the sea surface ($z=0$) and the seabed ($z=H$), and interface conditions should likewise be imposed at the discontinuous interface ($z=h$) between the water column and bottom sediment layers. The sea surface is usually set as a pressure release boundary:
\begin{equation}
	\label{eq:10}
	\psi(z=0)=0
\end{equation}
The seabed can be either a pressure release boundary or a rigid boundary:
\begin{subequations}
	\label{eq:11}
	\begin{gather}
		\psi(z=H)=0\\
		\psi'(z=H)=0  
	\end{gather}
\end{subequations}
In addition, the use of an acoustic half-space is quite common in underwater acoustic modeling \cite{Jensen2011}:
\begin{equation}
	\label{eq:12}
	\psi(H)+\frac{\rho_\infty}{\rho_b(H)\gamma_\infty} \psi'(H)=0,\quad \gamma_{\infty}=\sqrt{k_x^{2}-k_{\infty}^{2}},\quad k_\infty=\omega/c_{\infty}(1+\mathrm{i}\eta\alpha)
\end{equation}
Note that for modal normalization, $z\in[H,+\infty]$ should be adopted in this case:
\begin{equation}
	\label{eq:13}
	\int_{0}^{H} \frac{\psi_{m}^{2}(z)}{\rho(z)} \mathrm{d} z +\frac{\psi_{m}^{2}(H)}{2 \rho_\infty\gamma_{\infty}} =1,\quad m=1,2,\dots
\end{equation}
The interface conditions are defined as follows:
\begin{subequations}
	\label{eq:14}
	\begin{gather}
		\psi(h^{-})=\psi(h^{+}) \\
		\frac{1}{\rho(z=h^{-})} \frac{\mathrm{d}\psi(h^{-} )}{\mathrm {d}z}= \frac{1}{\rho(z=h^{+})}\frac{\mathrm{d}\psi(h^{+} )}{\mathrm {d}z}
	\end{gather}
\end{subequations}
where the superscripts $-$ and $+$ indicate limits from above and below, respectively.

\subsection{Chebyshev--Tau spectral method}
A spectral method, which is derived from the weighted residual method \cite{Gottlieb1977}, is a numerical technique similar to the finite difference method and the finite element method that achieves high accuracy in solving differential equations \cite{Canuto1988}. Depending on the approach employed to select the weight functions, spectral methods can be divided into Galerkin-type spectral methods and spectral collocation methods. The weight functions of the former take the same form as the basis functions, while the Kronecker delta function is adopted in the latter \cite{Jshen2011,Boyd2001}. Since classical Galerkin-type spectral methods often encounter difficulties in the construction of a set of basis functions that satisfy the boundary conditions, the Tau approach is widely used as a variant of the Galerkin method for differential equations featuring complex boundary conditions because the construction of such basis functions is not required \cite{Lanczos1938}. Instead, the boundary conditions are usually transformed into the spectral space and then into constraints on the spectral coefficients of the quantity to be solved. The Chebyshev polynomials and the Legendre polynomials are the most common choices for the basis functions in spectral methods. The basis functions of the Chebyshev--Tau spectral method are the Chebyshev polynomials, which are a class of smooth orthogonal polynomials.

In our previous works \cite{Tuhw2020a,Tuhw2020b}, we comprehensively introduced the Chebyshev--Tau spectral method and its application to range-independent normal modes. We further developed the related NM-CT program, which is publicly available in the open-source library OALIB \cite{NM-CT}. Similarly, we employ and further refine the Chebyshev--Tau spectral method in this study to solve for the local modes (Eq. \eqref{eq:4}) in the range-independent segments of a range-dependent waveguide.

Since the Chebyshev polynomials $\{T_i(t)\}$ are defined on the interval $[-1,1]$, when the Chebyshev--Tau spectral method is applied to solve the modal equation, Eq.~\eqref{eq:4} needs to be scaled to $t\in[-1,1]$:
\begin{equation}
	\label{eq:15}
	\frac{4}{\vert \Delta h\vert^2}\rho(t)\frac{\mathrm{d}}{\mathrm{d}t}\left(\frac{1}{\rho(t)}\frac{\mathrm{d}\psi(t)}{\mathrm {d}t}\right) +k^2\psi(t) = k_x^2 \psi(t)
\end{equation}
where $\vert \Delta h\vert$ denotes the depth of the oceanic medium. As the variable to be solved for, the modal function $\psi(t)$ needs to be transformed into the spectral space formed by the orthogonal Chebyshev polynomials $\{T_i(t)\}$:
\begin{equation}
	\label{eq:16}
	\psi(t) \approx \sum_{i=0}^{N}\hat{\psi}_{i}T_i(t)
\end{equation}
where $\{\hat{\psi}_i\}_{i=0}^N$ denotes the spectral coefficients of $\psi(t)$. The larger $N$ is, the more accurate this approximation. Although the Chebyshev polynomials are real, they can also be used as basis functions for the spectral expansion of complex variables; the only difference is that the spectral coefficients are complex.

Due to the convenient properties of the Chebyshev polynomials, the following relations are easily derived \cite{Canuto2006}:
\begin{subequations}
	\label{eq:17}
	\begin{gather}
		\label{eq:17a}		
		\hat{\psi}'_i \approx \frac{2}{c_i}
		\sum_{\substack{j=i+1,\\ 
				j+i=\text{odd}
		}}^{N} j \hat{\psi}_j, \quad c_0=2,c_{i>1}=1
		\Longleftrightarrow \bm{\hat{\Psi}}' \approx \mathbf{D}_N \bm{\hat{\Psi}} \\
		\label{eq:17b}
		\widehat{(v\psi)}_i \approx 
		\frac{1}{2} \sum_{m+n=i}^{N} \hat{\psi}_m\hat{v}_n +
		\frac{1}{2} \sum_{\vert m-n\vert=i}^{N} \hat{\psi}_m\hat{v}_n  \Longleftrightarrow  \widehat{\bm{(v\psi)}} \approx \mathbf{C}_v \bm{\hat{\Psi}}
	\end{gather}
\end{subequations}
Eq.~\eqref{eq:17a} illustrates the relationship between the spectral coefficients of a function and those of its derivative. Likewise, Eq.~\eqref{eq:17b} describes the relationship between the spectral coefficients of a product of two functions and the spectral coefficients of one of those functions, where the right-hand side is the matrix--vector representation of this relationship.

When using the Chebyshev--Tau spectral method to solve a differential equation, we should start with the weak form of the differential equation. The weak form of the modal equation in the Chebyshev spectral space is:
\begin{equation}
	\label{eq:18}
	\begin{gathered}
		\int_{-1}^{1}\left[\frac{4}{\vert\Delta h\vert^2}\rho(t)\frac{\mathrm{d}}{\mathrm{d}t}\left(\frac{1}{\rho(t)}\frac{\mathrm{d}\psi(t)}{\mathrm {d}t}\right) +k^2\psi(t)-k_x^2 \psi(t) \right]\frac{T_i(t)}{\sqrt{1-t^2}}\mathrm{d}t=0\\
		t\in (-1,1), \quad i=0,1,\dots,N-2
	\end{gathered}
\end{equation}
By substituting Eq.~\eqref{eq:16} into Eq.~\eqref{eq:18} and considering Eq.~\eqref{eq:17}, the modal equation can be discretized into the following matrix--vector form:
\begin{equation}
	\label{eq:19}
	\left(\frac{4}{\vert\Delta h\vert^2}\mathbf{C}_{\rho}\mathbf{D}_{N}\mathbf{C}_{1/\rho}\mathbf{D}_{N}+\mathbf{C}_{k^2}\right)
	\bm{\hat{\Psi}}
	= k_x^2 \bm{\hat{\Psi}}
\end{equation}
The above clearly represents a matrix eigenvalue problem; accordingly, boundary constraints must be imposed on the actual solution. For details regarding the discretization process, please see Eq.~(29) in reference \cite{Tuhw2020b}.

For the waveguide described in Eqs.~\eqref{eq:9} through \eqref{eq:14}, the modal equation (Eq.~\eqref{eq:4}) must be established in both the water column layer and the bottom sediment layer. In a range-independent waveguide, a single set of basis functions cannot span both layers since the Chebyshev polynomials are not continuously differentiable at the interface $h$. Thus, we use the domain decomposition strategy \cite{Min2005} in Eq.~\eqref{eq:4} and split the domain interval into two subintervals:
\begin{equation}
	\label{eq:20}
	\psi(z)= \begin{cases}
		\psi_w(z) = \psi_w(t) \approx\sum_{i=0}^{N_w}\hat{\psi}_{w,i}T_i(t_w),\quad t_w=-\frac{2z}{h}+1,&
		0\leq z\leq h \\
		\psi_b(z) = \psi_b(t) \approx\sum_{i=0}^{N_b}\hat{\psi}_{b,i}T_i(t_b),\quad
		t_b=-\frac{2z}{H-h}+\frac{H+h}{H-h},&
		h\leq z\leq H
	\end{cases}
\end{equation}
where $N_w$ and $N_b$ are the spectral truncation orders in the water column and bottom sediment, respectively, and $\{\hat{\psi}_{w,i}\}_{i=0}^{N_w}$ and $\{\hat{\psi}_{b,i}\}_{i=0}^{N_b}$ are the modal spectral coefficients in these two layers. Similar to Eq.~\eqref{eq:19}, the modal equations in the water column and bottom sediment layers can be discretized into matrix--vector form:
\begin{subequations}
	\begin{align}
		\label{eq:21a}
		&\mathbf{A}
		\bm{\hat{\Psi}}_w
		= k_x^2
		\bm{\hat{\Psi}}_w,
		&\mathbf{A} 
		=\frac{4}{h^2}\mathbf{C}_{\rho_w}\mathbf{D}_{N_w}\mathbf{C}_{1/\rho_w}\mathbf{D}_{N_w}+\mathbf{C}_{k_w^2}\\
		\label{eq:21b}
		&\mathbf{B}
		\bm{\hat{\Psi}}_b
		= k_x^2
		\bm{\hat{\Psi}}_b,&\mathbf{B}= 
		\frac{4}{(H-h)^2}\mathbf{C}_{\rho_b}\mathbf{D}_{N_b}\mathbf{C}_{1/\rho_b}\mathbf{D}_{N_b}+\mathbf{C}_{k_b^2}
	\end{align}
\end{subequations}
where $\mathbf{A}$ and $\mathbf{B}$ are square matrices of order $(N_w+1)$ and $(N_b+1)$, respectively, and $\bm{\hat{\Psi}}_w$ and $\bm{\hat{\Psi}}_b$ are column vectors composed of $\{\hat{\psi}_{w,i}\}_{i=0}^{N_w}$ and $\{\hat{\psi}_{b,i}\}_{i=0}^{N_b}$, respectively. Since the interface conditions are related to both the water column and the bottom sediment, Eqs.~\eqref{eq:21a} and \eqref{eq:21b} should be solved simultaneously as follows:
\begin{equation}
	\label{eq:22}
	\left[\begin{array}{cc}
		\mathbf{A}&\mathbf{0}\\
		\mathbf{0}&\mathbf{B}\\
	\end{array}\right]
	\left[\begin{array}{c}
		\bm{\hat{\Psi}}_w\\
		\bm{\hat{\Psi}}_b\\
	\end{array}
	\right]=k_x^2\left[
	\begin{array}{c}
		\bm{\hat{\Psi}}_w\\
		\bm{\hat{\Psi}}_b\\
	\end{array}\right]
\end{equation}
The boundary conditions and interface conditions in Eqs.~\eqref{eq:10}--\eqref{eq:14} must also be expanded to the Chebyshev spectral space and explicitly added to Eq.~\eqref{eq:22}. By rearranging and combining the modified rows by means of elementary row transformations, Eq.~\eqref{eq:22} can be rewritten in the following block matrix form:
\begin{equation}
	\label{eq:23}
	\left[
	\begin{array}{cc}
		\mathbf{L}_{11}&\mathbf{L}_{12}\\
		\mathbf{L}_{21}&\mathbf{L}_{22}\\
	\end{array}
	\right]\left[
	\begin{array}{c}
		\bm{\hat{\Psi}}_1\\
		\bm{\hat{\Psi}}_2
	\end{array}
	\right]=k_x^2\left[
	\begin{array}{c}
		\bm{\hat{\Psi}}_1\\
		\mathbf{0}
	\end{array}\right],\quad \mathbf{L}=\left[
	\begin{array}{cc}
		\mathbf{L}_{11}&\mathbf{L}_{12}\\
		\mathbf{L}_{21}&\mathbf{L}_{22}\\
	\end{array}
	\right]
\end{equation}
where $\mathbf{L}_{11}$ is a square matrix of order $(N_w+N_b-2)$, $\mathbf{L}_{22}$ is a square matrix of order 4, $\mathbf{\hat{\Psi}}_1=[\hat{\psi}_{w,0},\hat{\psi}_{w,1},\cdots,\hat{\psi}_{w,N_w-2},\\ \hat{\psi}_{b,0},\hat{\psi}_{b,1},\cdots,\hat{\psi}_{b,N_b-2}]^\mathrm{T}$ and $\mathbf{\hat{\Psi}}_2=[\hat{\psi}_{w,N_w-1},\hat{\psi}_{w,N_w},\hat{\psi}_{b,N_b-1},\hat{\psi}_{b,N_b}]^\mathrm{T}$. Solving this mixed linear algebraic system can yield the horizontal wavenumbers and spectral coefficients of the eigenmodes $(k_x,\bm{\hat{\Psi}}_w^j,\bm{\hat{\Psi}}_b^j)$. For details regarding the treatment of the boundary conditions in Eqs.~\eqref{eq:10}, \eqref{eq:11} and \eqref{eq:14}, please see Eq.~(38) in reference \cite{Tuhw2020b}.

For the acoustic half-space boundary conditions expressed in Eq.~\eqref{eq:12}, since $\gamma_\infty$ contains the eigenvalue $k_x$ to be solved for, the elements of $\mathbf{L}$ on the left side of Eq.~\eqref{eq:23} contain $k_x$, so Eq.~\eqref{eq:23} is no longer a matrix eigenvalue problem and can be solved iteratively only by means of a root-finding algorithm. The greatest drawback of root-finding algorithms is that they require a reasonable initial guess concerning the eigenvalue being sought, $k_x$ \cite{Sabatini2019}. Since a prior estimate of $k_x$ is usually not available, many of the existing numerical programs following similar principles fail to converge to a specific root in some cases. To avoid the same problem when using the Chebyshev--Tau spectral method for a waveguide with acoustic half-space boundary conditions, we consider an alternative approach here: we use $k_{z,\infty}=\sqrt{k_\infty^{2}-k_{x}^{2}}$ to transform the modal equation and Eq.~\eqref{eq:12} as follows:
\begin{subequations}
	\label{eq:24}
	\begin{gather}
		\label{eq:24a}
		\rho(z) \frac{\mathrm{d}}{\mathrm{d} z}\left(\frac{1}{\rho(z)} \frac{\mathrm{d} \psi}{\mathrm{d}z}\right)+\left[k^{2}(z)-k_{\infty}^{2}+k_{z,\infty}^{2}\right] \psi=0 \\
		\label{eq:24b}		
		\frac{\mathrm{i}\rho_\infty}{\rho_b(H)}\frac{\mathrm{d} \psi(z)}{\mathrm{d} z}\bigg\vert_{z=H}+k_{z,\infty} \psi(H)=0
	\end{gather}
\end{subequations}
where $k_\infty=(1+\mathrm{i}\eta\alpha_\infty)\omega/c_\infty$ is a complex constant. Eq.~\eqref{eq:24a} can be naturally discretized into the following form:
\begin{equation}
	\label{eq:25}
	\left[\mathbf{U}+k_{z,\infty}^{2} \mathbf{I}\right] \bm{\hat{\Psi}}=\mathbf{0},\quad \mathbf{U}=\mathbf{L}-k_{\infty}^{2}\mathbf{I}
\end{equation}
where $\mathbf{I}$ is the identity matrix. However, due to the addition of Eq.~\eqref{eq:24b} including $k_{z,\infty}$, Eq.~\eqref{eq:25} finally becomes the following polynomial eigenvalue problem:
\begin{equation}
	\label{eq:26}
	\left[\mathbf{U}+k_{z,\infty} \mathbf{V}+k_{z,\infty}^{2} \mathbf{W}\right] \bm{\hat{\Psi}}=\mathbf{0}
\end{equation}
where the $\mathbf{U}$ in Eq.~\eqref{eq:26} is not exactly identical to that in Eq.~\eqref{eq:25}, as it has been modified by the boundary conditions and interface conditions; nevertheless, we still denote it by $\mathbf{U}$. In addition, $\mathbf{V}$ is a zero matrix of order $(N_w+N_b+2)$, with only the last row corresponding to the boundary condition in Eq.~\eqref{eq:24b}, and $\mathbf{W}$ is simply the identity matrix that has been modified by the boundary conditions. This polynomial eigenvalue problem can be efficiently solved via the $\mathcal{QZ}$ algorithm; alternatively, it can be transformed into a general matrix eigenvalue problem using the following trick, although the scale of the matrices is doubled:
\begin{subequations}
	\label{eq:27}
	\begin{gather}
		\tilde{\mathbf{U}} \tilde{\bm{\Psi}}=k_{z,\infty} \tilde{\mathbf{V}} \tilde{\bm{\Psi}}, \\
		\tilde{\mathbf{U}}=\left[\begin{array}{cc}
			-\mathbf{V} & -\mathbf{U} \\
			\mathbf{I} & 0
		\end{array}\right], \quad \tilde{\mathbf{V}}=\left[\begin{array}{cc}
			\mathbf{W} & 0 \\
			0 & \mathbf{I}
		\end{array}\right], \quad \tilde{\bm{\Psi}}=\left[\begin{array}{c}
			k_{z,\infty} \bm{\hat{\Psi}} \\
			\bm{\hat{\Psi}}
		\end{array}\right]
	\end{gather}
\end{subequations}

It is necessary to take the inverse Chebyshev transform of the eigenvectors $\bm{\hat{\Psi}}_w$ and $\bm{\hat{\Psi}}_b$ to $[0,h]$ and $[h,H]$, respectively, by means of Eq.~\eqref{eq:16}. The vectors $\bm{\Psi}_w$ and $\bm{\Psi}_b$ are stacked into a single column vector to form $\bm{\Psi}$, and then, Eq.~\eqref{eq:5} is used to normalize $\bm{\Psi}$. Finally, a set of modes $(k_x,\psi(x))$ is obtained.

\section{Coupled Modes in a Range-Dependent Ocean Environment}
\subsection{Line source at a horizontal range of $x_\mathrm{s}=0$}
Referring to Eq.~\eqref{eq:8}, the solution for each segment of the range-dependent environment described above can be expressed analytically in terms of exponential functions and a discrete set of unknown coefficients. The acoustic field of the $j$-th segment can generally be written as:
\begin{equation}
	\label{eq:28}
	p^{j}(x, z)\approx \sum_{m=1}^{M}\left[a_{m}^{j} E_{m}^{j}(x)+b_{m}^{j} F_{m}^{j}(x)\right] \psi_{m}^{j}(z),\quad j=1,2,\cdots,J
\end{equation}
where $M$ is the total number of normal modes needed to synthesize the acoustic field; $\{a_{m}^{j}\}_{m=1}^M$ and $\{b_{m}^{j}\}_{m=1}^M$ are the amplitudes of the forward- and backward-propagating modes, respectively, also called the coupling coefficients in the $j$-th segment; and $\psi_{m}^{j}(z)$ is the $m$-th eigenmode of the $j$-th segment. $E_{m}^{j}(x)$ and $F_{m}^{j}(x)$ are normalized range functions and are defined as follows:
\begin{gather}
	\label{eq:29}
	E_{m}^{j}(x)=\exp \left[\mathrm{i} k_{x, m}^{j}\left(x-x_{j-1}\right)\right]\\
	\label{eq:30}
	F_{m}^{j}(x)=\exp \left[-\mathrm{i} k_{x, m}^{j}\left(x-x_{j}\right)\right]
\end{gather}
where $k_{x,m}^j$ is the horizontal wavenumber of the $m$-th mode in the $j$-th segment. For the special case of $j=1$, $x_{j-1}=x_1$. It should be noted that the definition of $E_{m}^{j}(x)$ here is the same as that in COUPLE \cite{Couple}, whereas the definition of $F_{m}^{j}(x)$ is not the same. Instead, the definition of $F_{m}^{j}(x)$ in COUPLE is:
\begin{equation}
	\label{eq:31}
	F_{m}^{j}(x)=\exp \left[-\mathrm{i} k_{x, m}^{j}\left(x-x_{j-1}\right)\right]
\end{equation}
The improvement we adopt here was introduced by Luo and Yang \cite{Luowy2012a}, who called this improved approach the direct global matrix coupled mode (DGMCM) algorithm. The existence of leakage modes may cause the value of $F_{m}^{j}(x)$ defined in COUPLE to overflow; specifically, for a leakage mode, $k_{x,m}^j=\mathcal{R}+ \mathcal{I}\mathrm{i}$, where $\mathcal{R}$ and $\mathcal{I}$ are the real and imaginary parts of $k_{x,m}^j$, respectively, and $\mathcal{I}>0$. In Eq.~\eqref{eq:31}, $x-x_{j-1}>0$; therefore, Eq.~\eqref{eq:31} contains $\exp[\mathcal{I}(x-x_{j-1})]$. When $\mathcal{I}$ or $(x-x_{j-1})$ is large, using Eq.~\eqref{eq:31} may cause numerical overflow. In contrast, in Eq.~\eqref{eq:30}, since the exponential part contains $\exp[\mathcal{I}(x-x_j)]$ in this case ($x-x_j<0$), regardless of the value of $\mathcal{I}$, the value of $F_{m}^{j}(x)$ is limited, thus ensuring that no numerical overflow occurs. In general, in this improved formulation, the left boundary is used to normalize the forward acoustic field and the right boundary is used to normalize the backward acoustic field, thereby guaranteeing the numerical stability of the calculation.

In the method of coupled segments, two segment continuity conditions are explicitly imposed at the boundaries between segments. The first segment condition requires that the acoustic pressure be continuous at the $j$-th segment boundary:
\begin{equation}
	\label{eq:32}
	p^{j+1}\left(x_{j}, z\right)=p^{j}\left(x_{j}, z\right) 
\end{equation}
The second segment condition requires that the radial velocity of the acoustic pressure be continuous at the $j$-th segment boundary:
\begin{equation}
	\label{eq:33}
	\frac{1}{\rho_{j+1}(z)} \frac{\partial p^{j+1}\left(x_{j}, z\right)}{\partial x}=\frac{1}{\rho_{j}(z)} \frac{\partial p^{j}\left(x_{j}, z\right)}{\partial x}
\end{equation}
For the first segment condition, we have:
\begin{equation}
	\label{eq:34}
	\sum_{m=1}^{M}\left[a_{m}^{j+1}E_{m}^{j+1}(x_j)+b_{m}^{j+1}F_{m}^{j+1}(x_j) \right] \psi_{m}^{j+1}(z)=\sum_{m=1}^{M}\left[a_{m}^{j} E_{m}^{j}\left(x_{j}\right)+b_{m}^{j}F_{m}^{j}\left(x_{j}\right)\right] \psi_{m}^{j}(z)
\end{equation}
Next, to the above equation, we apply the following operator \cite{Jensen2011}:
\begin{equation*}
	\int_0^H (\cdot) \frac{\psi_{\ell}^{j+1}(z)}{\rho_{j+1}(z)} \mathrm{d} z
\end{equation*}
and we use the orthogonal normalization formula in Eq.~\eqref{eq:5} for the eigenmodes of the $(j+1)$-th segment, yielding the following:
\begin{subequations}
	\label{eq:35}
	\begin{gather}
		a_{\ell}^{j+1}+b_{\ell}^{j+1}F_{\ell}^{j+1}\left(x_{j}\right) =\sum_{m=1}^{M}\left[a_{m}^{j} E_{m}^{j}\left(x_{j}\right)+b_{m}^{j}\right] \tilde{c}_{\ell m}\\
		\tilde{c}_{\ell m}=\int_0^H \frac{\psi_{\ell}^{j+1}(z) \psi_{m}^{j}(z)}{\rho_{j+1}(z)} \mathrm{d} z, \quad \ell=1, \ldots, M
	\end{gather}
\end{subequations}
It is easily seen that the above formula can be written in the following matrix--vector form:
\begin{equation}
	\label{eq:36}
	\mathbf{a}^{j+1}+\mathbf{F}^{j+1}\mathbf{b}^{j+1}=\widetilde{\mathbf{C}}^{j}\left(\mathbf{E}^{j} \mathbf{a}^{j}+\mathbf{b}^{j}\right)  
\end{equation}
Eq.~\eqref{eq:28} clearly reveals:
\begin{equation}
	\label{eq:37}
	\frac{1}{\rho_{j}} \frac{\partial p^{j}(x, z)}{\partial x} \simeq \frac{1}{\rho_{j}} \sum_{m=1}^{M} k_{x,m}^{j}\left[a_{m}^{j} E_{m}^{j}(x)-b_{m}^{j} F_{m}^{j}(x)\right] \psi_{m}^{j}(z)
\end{equation}
Therefore, for the second segment condition, we have:
\begin{equation}
	\label{eq:38}
	\frac{1}{\rho_{j+1}}\sum_{m=1}^{M}k_{x,m}^{j+1}\left[a_{m}^{j+1} -b_{m}^{j+1}F_{m}^{j+1}\left(x_{j}\right)\right]\psi_{m}^{j+1}(z)=\frac{1}{\rho_{j}}\sum_{m=1}^{M}k_{x,m}^{j}\left[a_{m}^{j} E_{m}^{j}\left(x_{j}\right)-b_{m}^{j}\right] \psi_{m}^{j}(z)
\end{equation}
Similarly, we apply the following operator to the above equation:
\begin{equation*}
	\int_0^H(\cdot) \psi_{\ell}^{j+1}(z) \mathrm{d} z
\end{equation*}
and we use the orthogonal normalization formula in Eq.~\eqref{eq:5} for the eigenmodes of the $(j+1)$-th segment, yielding the following:
\begin{subequations}
	\label{eq:39}
	\begin{gather}
		a_{\ell}^{j+1}-b_{\ell}^{j+1}F_{\ell}^{j+1} =\sum_{m=1}^{M}\left[a_{m}^{j} E_{m}^{j}\left(x_{j}\right)-b_{m}^{j}\left(x_{j}\right)\right] \hat{c}_{\ell m}\\
		\hat{c}_{\ell m}=\frac{k_{x,m}^{j}}{k_{x,\ell}^{j+1}} \int_0^H \frac{\psi_{\ell}^{j+1}(z) \psi_{m}^{j}(z)}{\rho_{j}(z)} \mathrm{d} z, \quad \ell=1, \ldots, M
	\end{gather}
\end{subequations}
Then, the above formula can be naturally written in the following matrix--vector form:
\begin{equation}
	\label{eq:40}
	\mathbf{a}^{j+1}-\mathbf{F}^{j+1}\mathbf{b}^{j+1}=\widehat{\mathbf{C}}^{j}\left(\mathbf{E}^{j} \mathbf{a}^{j}-\mathbf{b}^{j}\right)
\end{equation}
Eqs.~\eqref{eq:36} and \eqref{eq:40} can be combined into the following form:
\begin{equation}
	\label{eq:41}
	\left[\begin{array}{l}
		\mathbf{a}^{j+1} \\
		\mathbf{b}^{j+1}
	\end{array}\right]=\left[\begin{array}{ll}
		\mathbf{R}_{1}^{j} & \mathbf{R}_{2}^{j} \\
		\mathbf{R}_{3}^{j} & \mathbf{R}_{4}^{j}
	\end{array}\right]\left[\begin{array}{c}
		\mathbf{a}^{j} \\
		\mathbf{b}^{j}
	\end{array}\right]
\end{equation}
where:
\begin{subequations}
	\label{eq:42}
	\begin{align}
		\mathbf{R}_{1}^{j} &=\frac{1}{2}\left(\widetilde{\mathbf{C}}^{j}+\widehat{\mathbf{C}}^{j}\right) \mathbf{E}^{j} \\
		\mathbf{R}_{2}^{j} &=\frac{1}{2}\left(\widetilde{\mathbf{C}}^{j}-\widehat{\mathbf{C}}^{j}\right) \\
		\mathbf{R}_{3}^{j} &=\frac{1}{2}\left(\mathbf{F}^{j+1}\right)^{-1}\left(\widetilde{\mathbf{C}}^{j}-\widehat{\mathbf{C}}^{j}\right) \mathbf{E}^{j} \\
		\mathbf{R}_{4}^{j} &=\frac{1}{2}\left(\mathbf{F}^{j+1}\right)^{-1}\left(\widetilde{\mathbf{C}}^{j}+\widehat{\mathbf{C}}^{j}\right)
	\end{align}
\end{subequations}
Finally, the segment boundary condition at the acoustic source ($x_\mathrm{s}=0$) and the radiation condition at $x\rightarrow \infty$ should also be considered. The segment boundary condition at $x_\mathrm{s}=0$ is:
\begin{subequations}
	\label{eq:43}
	\begin{gather}
		\mathbf{a}^{1}=\mathbf{s}\\
		s_{m}=\frac{\mathrm{i}}{2 \rho\left(z_\mathrm{s}\right)} \psi_{m}^{1}\left(z_\mathrm{s}\right) \frac{\mathrm{e}^{\mathrm{i} k_{x,m}^{1} x_{1}}}{k_{x,m}^{1}}, \quad m=1, \cdots, M  
	\end{gather}
\end{subequations}
For the radiation condition at $x\rightarrow\infty$, $\mathbf{b}^J=0$ is sufficient, where $J$ denotes the total number of horizontal segments.

By combining the continuity conditions at the boundaries of the $J$ segments with the segment boundary condition at the acoustic source and the radiation condition at infinity, the following system of linear algebraic equations can be constructed:
\begin{equation}
	\label{eq:44}
	\left[\begin{array}{ccccccccc}
		\mathbf{I} & \mathbf{0} & \mathbf{0} & & & & & \\
		\mathbf{R}_{1}^{1} & \mathbf{R}_{2}^{1} & -\mathbf{I} & \mathbf{0} & & & & \\
		\mathbf{R}_{3}^{1} & \mathbf{R}_{4}^{1} & \mathbf{0} & -\mathbf{I} & & & & \\
		& & \ddots & \ddots & \ddots & \ddots & & \\
		& & & &\mathbf{R}_{1}^{J-2} & \mathbf{R}_{2}^{J-2} & -\mathbf{I} & \mathbf{0} & \\
		& & & &\mathbf{R}_{3}^{J-2} & \mathbf{R}_{4}^{J-2} & \mathbf{0} & -\mathbf{I} & \\
		& & & & & &\mathbf{R}_{1}^{J-1} & \mathbf{R}_{2}^{J-1} & -\mathbf{I} \\
		& & & & & &\mathbf{R}_{3}^{J-1} & \mathbf{R}_{4}^{J-1} & \mathbf{0}
	\end{array}\right]\left[\begin{array}{c}
		\mathbf{a}^{1} \\
		\mathbf{b}^{1} \\
		\mathbf{a}^{2} \\
		\vdots \\
		\mathbf{b}^{J-2} \\
		\mathbf{a}^{J-1} \\
		\mathbf{b}^{J-1} \\
		\mathbf{a}^{J}
	\end{array}\right]=\left[\begin{array}{c}
		\mathbf{s} \\
		\mathbf{0} \\
		\mathbf{0} \\
		\vdots \\    
		\mathbf{0} \\
		\mathbf{0} \\
		\mathbf{0} \\
		\mathbf{0}
	\end{array}\right]
\end{equation}
The global matrix formed by a line source is clearly a sparse band matrix of order $(2J-1) \times M$; the upper bandwidth is $(2M-1)$, and the lower bandwidth is $(3M-1)$. These linear algebraic equations are then solved to obtain the coupling coefficients, and Eq.~\eqref{eq:28} is applied to synthesize the sound pressure field of the entire waveguide.

Since $x_{j-1}=x_1$ when $j=1$, in the first segment, $E_m^j(x)$ is normalized to the right side. When $\exp[\mathcal{I}(x_1-x)]$ is large, $E_m^j(x)$ will trigger numerical overflow. To avoid this problem, the superposition principle is used to solve for the acoustic field. Substituting $\mathbf{a}^1$ in Eq.~\eqref{eq:43} into Eq.~\eqref{eq:28} yields:
\begin{equation}
	\label{eq:45}
	p^{1}(x, z)=\frac{\mathrm{i}}{2 \rho\left(z_\mathrm{s}\right)} \sum_{m=1}^{M} \psi_{m}^{1}\left(z_\mathrm{s}\right) \psi_{m}^{1}(z) \frac{\mathrm{e}^{\mathrm{i} k_{x,m}^{1} x}}{k_{x,m}^{1}} +  \sum_{m=1}^{M} b_{m}^{1} F_{m}^{1}(x) \psi_{m}^{1}(z)
\end{equation}
where the first term on the right-hand side represents the range-independent acoustic field and the second term represents the backscattered acoustic field caused by range dependence.

\subsection{Generalized line source at an arbitrary position}
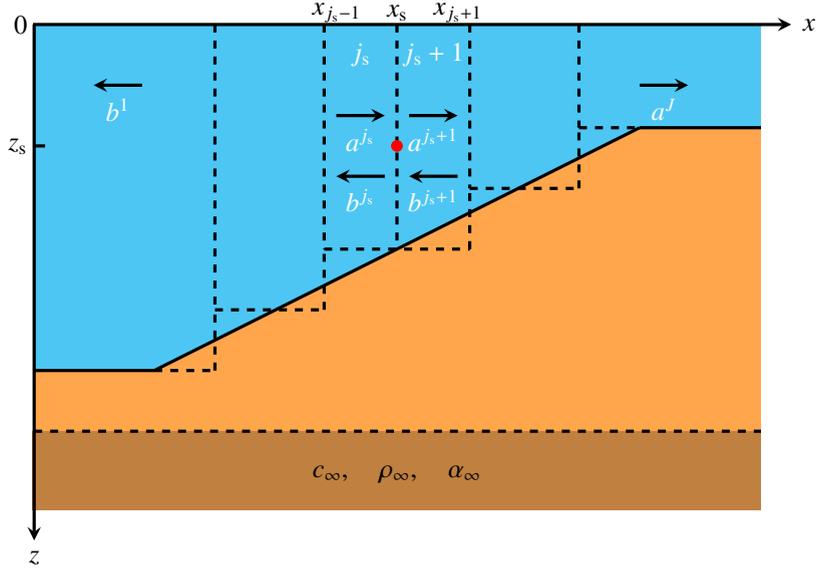
\begin{figure}[htbp]
	\centering
	\begin{tikzpicture}[node distance=2cm,scale = 0.8]
		\node at (1.8,0){$0$};
		\fill[brown] (14,-8) rectangle (2,-6.7);
		\fill[cyan,opacity=0.7] (14,0)--(14,-1.7)--(12,-1.7)--(4,-5.7)--(2,-5.7)--(2,0)--cycle;
		\fill[orange,opacity=0.7] (14,-1.7)--(12,-1.7)--(4,-5.7)--(2,-5.7)--(2,-6.7)--(14,-6.7)--cycle;
		\draw[very thick, ->](2,0)--(14.5,0) node[right]{$x$};
		\draw[very thick, ->](2.02,0)--(2.02,-8.5) node[below]{$z$};
		\draw[very thick](2,-5.7)--(4,-5.7);
		\draw[dashed, very thick](4,-5.7)--(5,-5.7);
		\draw[dashed, very thick](5,-5.7)--(5,-4.7);
		\draw[dashed, very thick](5,-4.7)--(6.8,-4.7);
		\draw[dashed, very thick](6.8,-4.7)--(6.8,-3.7);
		\draw[dashed, very thick](6.8,-3.7)--(9.2,-3.7);
		\draw[dashed, very thick](8,0)--(8,-3.7);		
		\draw[dashed, very thick](9.2,-3.7)--(9.2,-2.7);  
		\draw[dashed, very thick](9.2,-2.7)--(11,-2.7);
		\draw[dashed, very thick](11,-2.7)--(11,-1.7); 
		\draw[dashed, very thick](11,-1.7)--(12,-1.7);		    		
		\draw[very thick](4,-5.7)--(12,-1.7);
		\draw[very thick](12,-1.7)--(14,-1.7);
		\draw[dashed, very thick](5,0)--(5,-4.7);
		\draw[dashed, very thick](6.8,0)--(6.8,-3.7);
		\draw[dashed, very thick](9.2,0)--(9.2,-2.7); 
		\draw[dashed, very thick](11,0)--(11,-1.7); 
		\draw[dashed, very thick](2.02,-6.7)--(14,-6.7);
		\filldraw [red] (8,-2) circle [radius=2.5pt];
		\draw[very thick](2,-2)--(2.2,-2);
		\node at (1.75,-2){$z_\mathrm{s}$};
		\node[text=white] at (7.4,-0.5){$j_\mathrm{s}$};
		\node[text=white] at (8.6,-0.5){$j_\mathrm{s}+1$};
		\draw[very thick, ->](7,-1.5)--(7.8,-1.5);
		\node[text=white] at (7.4,-1.9){$a^{j_\mathrm{s}}$};
		\draw[very thick, ->](7.8,-2.5)--(7,-2.5);
		\node[text=white] at (7.4,-2.9){$b^{j_\mathrm{s}}$};
		\draw[very thick, ->](8.2,-1.5)--(9,-1.5);
		\node[text=white] at (8.6,-1.9){$a^{j_\mathrm{s}+1}$};
		\draw[very thick, ->](9,-2.5)--(8.2,-2.5);
		\node[text=white] at (8.6,-2.9){$b^{j_\mathrm{s}+1}$};
		\draw[very thick, ->](3.8,-1)--(3,-1);
		\node[text=white] at (3.4,-1.4){$b^{1}$};
		\draw[very thick, ->](12,-1)--(12.8,-1);
		\node[text=white] at (12.4,-1.4){$a^{J}$};							
		\node at (8,0.2){$x_\mathrm{s}$};
		\node at (7,0.2){$x_{j_\mathrm{s}-1}$};
		\node at (9,0.2){$x_{j_\mathrm{s}+1}$};
		\node at (8,-7.4){$c_\infty,\quad \rho_\infty,\quad \alpha_\infty$};    		
	\end{tikzpicture}
	\caption{Schematic diagram of the stepwise coupled modes of a line source at an arbitrary position.}
	\label{Figure2}
\end{figure}

In the above environment, the line source must be located at $x_\mathrm{s}=0$. However, after a simple improvement, the sound source can be located at any horizontal range $x_\mathrm{s}$ \cite{Luowy2012a}. This improvement is valuable for solving for the acoustic propagation when the line source is located on a slope, as shown in Fig.~\ref{Figure2}. Suppose that a line source is located at $(x_\mathrm{s}, z_\mathrm{s})$, and a virtual boundary $j_\mathrm{s}$ is introduced at the horizontal range $x_\mathrm{s}$. All boundaries except $j_\mathrm{s}$ still satisfy the boundary conditions of Eqs.~\eqref{eq:32} and \eqref{eq:33}. However, due to the existence of the sound source, the conditions at boundary $j_\mathrm{s}$ need to be modified. The sound fields in the $j_\mathrm{s}$-th and ($j_\mathrm{s}+1$)-th segments can be expressed as:
\begin{subequations}
	\label{eq:46}
	\begin{gather}
		p^{j_\mathrm{s}}(x, z)=\sum_{m}^M\left[a_{m}^{j_\mathrm{s}} E_{m}^{j_\mathrm{s}}(x)+b_{m}^{j_\mathrm{s}} F_{m}^{j_\mathrm{s}}(x)\right] \psi_{m}^{j_\mathrm{s}}(z) \\
		p^{j_\mathrm{s}+1}(x, z)=\sum_{m}^M\left[a_{m}^{j_\mathrm{s}+1} E_{m}^{j_\mathrm{s}+1}(x)+b_{m}^{j_\mathrm{s}+1} F_{m}^{j_\mathrm{s}+1}(x)\right] \psi_{m}^{j_\mathrm{s}+1}(z)
	\end{gather}
\end{subequations}
By applying segment boundary condition \eqref{eq:32} at the virtual boundary $j_\mathrm{s}$ and taking into account $\psi_m^{j_\mathrm{s}}(z)=\psi_m^{j_\mathrm{s}+1}(z)$, $F_m^{j_\mathrm{s}}(x_{j_\mathrm{s}})=1$ and $E_m^{j_\mathrm{s}+1}(x_{j_\mathrm{s}})=1$, we can obtain:
\begin{subequations}
	\label{eq:47}
	\begin{gather}
		a_{m}^{j_\mathrm{s}+1}+b_{m}^{j_\mathrm{s}+1} F_{m}^{j_\mathrm{s}+1}=a_{m}^{j_\mathrm{s}} E_{m}^{j_\mathrm{s}}+b_{m}^{j_\mathrm{s}}\\
		\mathbf{a}^{j_\mathrm{s}+1}+\mathbf{F}^{j_\mathrm{s}+1} \mathbf{b}^{j_\mathrm{s}+1}=\mathbf{E}^{j_\mathrm{s}} \mathbf{a}^{j_\mathrm{s}}+\mathbf{b}^{j_\mathrm{s}}    
	\end{gather}
\end{subequations}
These equations are consistent with the form of Eq.~\eqref{eq:36} except that the acoustic properties on both sides of boundary $j_\mathrm{s}$ are completely consistent, resulting in $\widetilde{\mathbf{C}}=\mathbf{I}$.

The integral of Eq.~\eqref{eq:6} in an infinitesimal neighborhood yields the following:
\begin{equation}
	\label{eq:48}
	\frac{\mathrm{d} X_{m}}{\mathrm{~d} x}\bigg\vert_{x_\mathrm{s}^{-}} ^{x_\mathrm{s}^{+}}=-\frac{\psi_{m}^{j_\mathrm{s}}\left(z_\mathrm{s}\right)}{\rho\left(z_\mathrm{s}\right)}
\end{equation}
Considering the boundary conditions and $k_{x,m}^{j_\mathrm{s}}=k_{x,m}^{j_\mathrm{s}+1}$, we can obtain:
\begin{subequations}
	\label{eq:49}
	\begin{gather}
		\left(\mathrm{i} k_{x,m}^{j_\mathrm{s}+1} a_{m}^{j_\mathrm{s}+1}-\mathrm{i} k_{x,m}^{j_\mathrm{s}+1} b_{m}^{j_\mathrm{s}+1} F_{m}^{j_\mathrm{s}+1}\right)-\left(\mathrm{i} k_{x,m}^{j_\mathrm{s}} a_{m}^{j_\mathrm{s}} E_{m}^{j_\mathrm{s}}-\mathrm{i} k_{x,m}^{j_\mathrm{s}} b_{m}^{j_\mathrm{s}}\right)=-\frac{\psi_{m}^{j_\mathrm{s}}\left(z_\mathrm{s}\right)}{\rho\left(z_\mathrm{s}\right)}\\
		\mathbf{a}^{j_\mathrm{s}+1}-\mathbf{F}^{j_\mathrm{s}+1} \mathbf{b}^{j_\mathrm{s}+1}=\mathbf{E}^{j_\mathrm{s}} \mathbf{a}^{j_\mathrm{s}}-\mathbf{b}^{j_\mathrm{s}}-\mathbf{s}\\
		s_{m}=-\frac{\mathrm{i}}{\rho\left(z_\mathrm{s}\right)} \frac{\psi_{m}^{j_\mathrm{s}}\left(z_\mathrm{s}\right)}{k_{x,m}^{j_\mathrm{s}}}, \quad m=1,2, \cdots, M
	\end{gather}
\end{subequations}
Eq.~\eqref{eq:49} similarly corresponds to \eqref{eq:40}. Because the acoustic properties on both sides of the virtual boundary $j_\mathrm{s}$ are exactly the same, $\widehat{\mathbf{C}}=\mathbf{I}$. Eqs. \eqref{eq:47} and \eqref{eq:49} indicate that the relationship between the coupling coefficients of the $j_\mathrm{s}$-th and $(j_\mathrm{s}+1)$-th segments is:
\begin{equation}
	\label{eq:50}
	\left[\begin{array}{l}
		\mathbf{a}^{j_\mathrm{s}+1} \\
		\mathbf{b}^{j_\mathrm{s}+1}
	\end{array}\right]=\left[\begin{array}{ll}
		\mathbf{E}^{j_\mathrm{s}} & \mathbf{0} \\
		\mathbf{0} & (\mathbf{F}^{j_\mathrm{s}+1})^{-1}
	\end{array}\right]\left[\begin{array}{c}
		\mathbf{a}^{j_\mathrm{s}} \\
		\mathbf{b}^{j_\mathrm{s}}
	\end{array}\right]+\left[\begin{array}{c}
		-\frac{1}{2}\mathbf{s} \\
		\frac{1}{2}\left(\mathbf{F}^{j_\mathrm{s}+1}\right)^{-1}\mathbf{s}
	\end{array}\right]
\end{equation}
Since the line source is located at $x_\mathrm{s} \neq 0$, the boundary condition at $x=0$ becomes the radiation condition $\mathbf{a}^1=0$. In this case, the global matrix used to solve for the coupling coefficients can be constructed as follows:
\begin{equation}
	\label{eq:51}
	\left[\begin{array}{ccccccccccccc}
		\mathbf{R}_{2}^{1} & -\mathbf{I} & \mathbf{0} & & & & & \\
		\mathbf{R}_{4}^{1} & \mathbf{0} & -\mathbf{I} & & & & & \\
		& \mathbf{R}_{1}^{2} &\mathbf{R}_{2}^{2} &-\mathbf{I} &\mathbf{0} & & & & \\
		& \mathbf{R}_{3}^{2} &\mathbf{R}_{4}^{2} &\mathbf{0}  &-\mathbf{I} & & & & \\
		& & &\ddots & \ddots & \ddots & \ddots& & \\
		& & & & & \mathbf{E}^{j_\mathrm{s}} & \mathbf{0} & -\mathbf{I} & \mathbf{0} \\
		& & & & & \mathbf{0} & (\mathbf{F}^{j_\mathrm{s}+1})^{-1} & \mathbf{0} & -\mathbf{I}\\
		& & & & & & &  \ddots & \ddots & \ddots & \ddots&  \\
		& & & & & & & & &\mathbf{R}_{1}^{J-1} & \mathbf{R}_{2}^{J-1} & -\mathbf{I} \\
		& & & & & & & & &\mathbf{R}_{3}^{J-1} & \mathbf{R}_{4}^{J-1} & \mathbf{0}
	\end{array}\right]\left[\begin{array}{c}
		\mathbf{b}^{1} \\
		\mathbf{a}^{2} \\
		\mathbf{b}^{2} \\
		\mathbf{a}^{3} \\
		\vdots \\
		\mathbf{a}^{j_\mathrm{s}} \\
		\mathbf{b}^{j_\mathrm{s}} \\
		\vdots \\
		\mathbf{b}^{J-1}\\
		\mathbf{a}^{J}
	\end{array}\right]=\left[\begin{array}{c}
		\mathbf{0} \\
		\mathbf{0} \\
		\mathbf{0} \\
		\mathbf{0} \\
		\vdots \\
		\frac{1}{2}\mathbf{s} \\
		-\frac{1}{2}(\mathbf{F}^{j_\mathrm{s}+1})^{-1}\mathbf{s} \\
		\vdots \\
		\mathbf{0}\\
		\mathbf{0} \\
	\end{array}\right]
\end{equation}
The coupling coefficients can be obtained by solving this system of linear equations. Similar to the previous analysis, the global matrix formed by the line source is a sparse band matrix of order $(2J-2) \times M$; the upper bandwidth is $(2M-1)$, and the lower bandwidth is $(3M-1)$.

\section{Algorithm}
Based on the above derivation, the proposed algorithm is devised as follows.

\textbf{Input:} Data from the ocean environment and the program parameters.

\textbf{Output:} The sound pressure field.
\begin{enumerate}
	\item
	Set the parameters.
	
	The parameters include the frequency $f$ and location $(x_\mathrm{s},z_\mathrm{s})$ of the line source, the total ocean depth $H$, the topography of the seabed, the number of acoustic profiles, and the specific information of each group of acoustic profiles. In addition, the parameters should include the spectral truncation orders ($N_w$ and $N_b$), the horizontal and vertical resolutions, and the number of coupled modes $M$. If the lower boundary is the upper interface of an acoustic half-space, then the speed $c_\infty$, density $\rho_\infty$ and attenuation $\alpha_\infty$ of the half-space must also be specified.
	\item
	Segment the ocean environment based on the seabed topography and acoustic profiles.
	
	Stair-step discretization criteria have been established to accurately represent a smoothly varying bathymetry in numerical models. Jensen confirmed that the strictest segmentation criterion is $\Delta x \le \lambda/4$, where $\lambda=\min\{c_w,c_b\}/f$ \cite{Jensen1998}. We assume that the entire waveguide is divided into $J$ segments.
	\item
	Apply the Chebyshev--Tau spectral method to solve for the horizontal wavenumbers and eigenmodes $(k_x,\psi(z))$ of the $J$ range-independent segments.
	
	The modal spectral coefficients $\bm{\hat{\Psi}}$ obtained for the $J$ segments must be transformed to a uniform vertical resolution. Otherwise, $\tilde{c}_{\ell m}$ in Eq.~\eqref{eq:36} and $\hat{c}_{\ell m}$ in Eq.~\eqref{eq:40} cannot be calculated. The uniform vertical resolution is specified in the first step by the environment file. This step exhibits natural parallelism because the solution processes for the individual segments are independent.
	
	\item
	Calculate the coupling submatrices $\{\mathbf{R}_{1}^{j}\}_{j=1}^{J-1}$, $\{\mathbf{R}_{2}^{j}\}_{j=1}^{J-1}$, $\{\mathbf{R}_{3}^{j}\}_{j=1}^{J-1}$, and $\{\mathbf{R}_{4}^{j}\}_{j=1}^{J-1}$. This step also exhibits natural parallelism.
	\item
	Depending on whether $x_\mathrm{s}$ is equal to 0, calculate $\mathbf{s}$ in the boundary conditions, construct the global matrix of coupling coefficients according to Eq.~\eqref{eq:44} or Eq.~\eqref{eq:51}, and solve the system of linear equations to obtain the coupling coefficients $(\{\mathbf{a}^j\}_{j=1}^J,\{\mathbf{b}^j\}_{j=1}^J)$ of the $J$ segments.
	\item
	Calculate the entire sound field.
	
	The sound pressure subfields of the segments are calculated according to Eq.~\eqref{eq:28}, and the subfield in the first segment is corrected according to Eq.~\eqref{eq:45} if $x_\mathrm{s}=0$. The subfields of the $J$ segments are then spliced to form the sound pressure field of the entire waveguide.
\end{enumerate}

\section{Numerical Simulations}
\begin{figure}
	\subfigure[]{
		\begin{minipage}[t]{0.5\linewidth}
			\begin{tikzpicture}[node distance=2cm,scale=0.5]
				\tikzstyle{every node}=[font=\small]
				\node at (1.4,0){0 $\text{m}$};
				\node at (1.1,-3.5){100 $\text{m}$};
				\node at (1.1,-6.9){200 $\text{m}$};
				\node at (2.3,0.4){0 $\text{km}$};
				\node at (14.3,0.4){4 $\text{km}$};
				\fill[cyan,opacity=0.7] (14,0)--(2,0)--(2,-7)--cycle;
				\draw[very thick, ->](2,0)--(15,0)   node[right]{$x$};
				\draw[very thick, ->](2.02,0.2)--(2.02,-7.5)  node[below]{$z$};
				\filldraw [red] (2.02,-3.5) circle [radius=2.5pt];
				\node at (3.2,-3.5){source};
				\node at (3.2,-3){$f$=25 Hz};
				\node at (7.5,-1.0){$c$=1500 m/s};
				\node at (7.5,-1.6){$\rho$=1.0  g/cm$^3$};
				\draw[very thick](14,0)--(14,0.2);  				
				\draw[very thick](2,-7)--(14,0);
				\draw[dashed, very thick](2.02,-7)--(6.02,-7);
				\draw[very thick](3,-7) arc (0:30:1);
				\node at (3.7,-6.7){2.86°};
				\pgftext[rotate=30,left,x=3 cm,y=-7.5 cm]{\LARGE Pressure release bottom};
			\end{tikzpicture}
	\end{minipage}}
	\subfigure[]{
		\begin{minipage}[t]{0.5\linewidth}
			\begin{tikzpicture}[node distance=2cm,scale=0.5]
				\tikzstyle{every node}=[font=\small]
				\node at (1.4,0){0 $\text{m}$};
				\node at (1.1,-1.75){100 $\text{m}$};
				\node at (1.1,-6.9){400 $\text{m}$};
				\node at (2.3,0.4){0 $\text{km}$};
				\node at (14.3,0.4){8 $\text{km}$};
				\fill[cyan,opacity=0.7] (14,0)--(2,0)--(2,-7)--cycle;
				\draw[very thick, ->](2,0)--(15,0)   node[right]{$x$};
				\draw[very thick, ->](2.02,0.2)--(2.02,-7.5)  node[below]{$z$};
				\draw[very thick](2,-1.75)--(2.2,-1.75);
				\filldraw [red] (8,-1.75) circle [radius=2.5pt];
				\node at (7,-1.5){$f$=25 Hz};
				\node at (7,-2){source};
				\node at (4,-3.0){$c$=1500 m/s};
				\node at (4,-3.6){$\rho$=1.0 g/cm$^3$};	
				\draw[very thick](8,0)--(8,0.2);
				\node at (8,0.4){4 $\text{km}$}; 
				\draw[very thick](14,0)--(14,0.2);  				
				\draw[very thick](2,-7)--(14,0);
				\draw[dashed, very thick](2.02,-7)--(6.02,-7);
				\draw[very thick](3,-7) arc (0:30:1);
				\node at (3.7,-6.7){2.86°};
				\pgftext[rotate=30,left,x=3 cm,y=-7.5 cm]{\LARGE Pressure release bottom};
			\end{tikzpicture}
	\end{minipage}}
	\subfigure[]{
		\begin{minipage}[t]{0.5\linewidth}
			\begin{tikzpicture}[node distance=2cm,scale=0.5]
				\tikzstyle{every node}=[font=\small]
				\node at (1.4,0){0 $\text{m}$};
				\node at (1.1,-3.5){100 $\text{m}$};
				\node at (1.1,-6.9){200 $\text{m}$};
				\node at (2.3,0.4){0 $\text{km}$};
				\node at (14.3,0.4){4 $\text{km}$};
				\fill[cyan,opacity=0.7] (14,0)--(2,0)--(2,-7)--cycle;
				\draw[very thick, ->](2,0)--(15,0)   node[right]{$x$};
				\draw[very thick, ->](2.02,0.2)--(2.02,-7.5)  node[below]{$z$};
				\filldraw [red] (2.02,-3.5) circle [radius=2.5pt];
				\node at (3.2,-3.5){source};
				\node at (3.2,-3){$f$=25 Hz};
				\node at (7.5,-1.0){$c$=1500 m/s};
				\node at (7.5,-1.6){$\rho$=1.0  g/cm$^3$};
				\draw[very thick](14,0)--(14,0.2);  				
				\draw[very thick](2,-7)--(14,0);
				\draw[dashed, very thick](2.02,-7)--(6.02,-7);
				\draw[very thick](3,-7) arc (0:30:1);
				\node at (3.7,-6.7){2.86°};
				\pgftext[rotate=30,left,x=4 cm,y=-7.5 cm]{\LARGE Rigid bottom};
			\end{tikzpicture}
	\end{minipage}}
	\subfigure[]{
		\begin{minipage}[t]{0.5\linewidth}
			\begin{tikzpicture}[node distance=2cm,scale=0.5]
				\tikzstyle{every node}=[font=\small]
				\node at (1.4,0){0 $\text{m}$};		
				\node at (1.1,-6.9){400 $\text{m}$};
				\node at (1.1,-1.75){100 $\text{m}$};
				\node at (2.3,0.4){0 $\text{km}$};
				\node at (14.3,0.4){8 $\text{km}$};
				\fill[cyan,opacity=0.7] (14,0)--(2,0)--(2,-7)--cycle;
				\draw[very thick, ->](2,0)--(15,0) 	node[right]{$x$};
				\draw[very thick, ->](2.02,0.2)--(2.02,-7.5) 	node[below]{$z$};
				\draw[very thick](2,-1.75)--(2.2,-1.75);
				\filldraw [red] (8,-1.75) circle [radius=2.5pt];
				\node at (7,-1.5){$f$=25 Hz};
				\node at (7,-2){source};
				\node at (4,-3.0){$c$=1500 m/s};
				\node at (4,-3.6){$\rho$=1.0 g/cm$^3$};	
				\draw[very thick](8,0)--(8,0.2);
				\node at (8,0.4){4 $\text{km}$}; 
				\draw[very thick](14,0)--(14,0.2); 
				\draw[very thick](2,-7)--(14,0);
				\draw[dashed, very thick](2.02,-7)--(6.02,-7);
				\draw[very thick](3,-7) arc (0:30:1);
				\node at (3.7,-6.7){2.86°};
				\pgftext[rotate=30,left,x=4 cm,y=-7.5 cm]{\LARGE Rigid bottom};
			\end{tikzpicture}
	\end{minipage}}
	\subfigure[]{
		\begin{minipage}[t]{0.5\linewidth}
			\begin{tikzpicture}[node distance=2cm,scale=0.5]
				\tikzstyle{every node}=[font=\small]
				\node at (1.4,0){0 $\text{m}$};
				\node at (1.1,-6.9){400 $\text{m}$};
				\node at (2.3,0.4){0 $\text{km}$};
				\node at (14.3,0.4){8 $\text{km}$};
				\fill[cyan,opacity=0.7] (14,0)--(2,0)--(2,-3.5)--(8,-3.5)--cycle;
				\draw[very thick, ->](2,0)--(15,0) 	node[right]{$x$};
				\draw[very thick, ->](2.02,0.1)--(2.02,-7.5) 		node[below]{$z$};
				\filldraw [red] (8.02,-1.75) circle [radius=2.5pt];
				\node at (9,-1.75){source};
				\node at (9,-1.25){$f$=25 Hz};
				\node at (1.1,-1.75){100 $\text{m}$};
				\node at (1.1,-3.5){200 $\text{m}$};
				\draw[very thick](2,-1.75)--(2.2,-1.75);
				\draw[very thick](8,0)--(8,0.2);
				\draw[very thick](14,0)--(14,0.2); 
				\node at (8,0.4){4 $\text{km}$}; 
				\node at (4.5,-1.0){$c$=1500 m/s};
				\node at (4.5,-1.6){$\rho$=1.0 g/cm$^3$};	
				\draw[very thick](2,-3.5)--(8.02,-3.5);		
				\draw[very thick](8,-3.5)--(14,0);
				\draw[dashed, very thick](2.02,-7)--(8.02,-3.5);
				\draw[dashed, very thick](2.02,-7)--(6.02,-7);		
				\draw[very thick](3,-7) arc (0:30:1);
				\node at (3.7,-6.7){2.86°};
				\pgftext[rotate=30,left,x=3 cm,y=-7.5 cm]{\LARGE Pressure release bottom};
			\end{tikzpicture}
	\end{minipage}}
	\subfigure[]{
		\begin{minipage}[t]{0.5\linewidth}
			\begin{tikzpicture}[node distance=2cm,scale=0.5]
				\tikzstyle{every node}=[font=\small]
				\node at (1.4,0){0 $\text{m}$};
				\node at (1.1,-6.9){400 $\text{m}$};
				\node at (2.3,0.4){0 $\text{km}$};
				\node at (14.3,0.4){8 $\text{km}$};
				\fill[cyan,opacity=0.7] 	(2,-7)--(2,0)--(14,0)--(14,-3.5)--(8,-3.5)--cycle;
				\draw[very thick, ->](2,0)--(15,0) node[right]{$x$};
				\draw[very thick, ->](2.02,0.1)--(2.02,-7.5) 	node[below]{$z$};
				\filldraw [red] (8.02,-1.75) circle [radius=2.5pt];
				\node at (9,-1.75){source};
				\node at (9,-1.25){$f$=25 Hz};
				\node at (1.1,-1.75){100 $\text{m}$};
				\draw[very thick](2,-1.75)--(2.2,-1.75);
				\draw[very thick](8,0)--(8,0.2);
				\draw[very thick](14,0)--(14,0.2); 
				\node at (8,0.4){4 $\text{km}$}; 
				\node at (4.5,-1.0){$c$=1500 m/s};
				\node at (4.5,-1.6){$\rho$=1.0 g/cm$^3$};	
				\draw[very thick](2,-7)--(8.02,-3.5);
				\draw[very thick](8,-3.5)--(14,-3.5);
				\draw[dashed, very thick](8.02,-3.5)--(14.02,0);			
				\draw[dashed, very thick](2.02,-7)--(6.02,-7);
				\draw[very thick](3,-7) arc (0:30:1);
				\node at (3.7,-6.7){2.86°};
				\pgftext[rotate=30,left,x=3 cm,y=-7.5 cm]{\LARGE Pressure};
				\pgftext[rotate=0,left,x=8.5 cm,y=-4 cm]{\LARGE release bottom};
			\end{tikzpicture}
	\end{minipage}}
	\caption{ASA benchmark problems (a)--(d) and modified wedge-shaped waveguides (e)--(f).}
	\label{Figure3}
\end{figure}
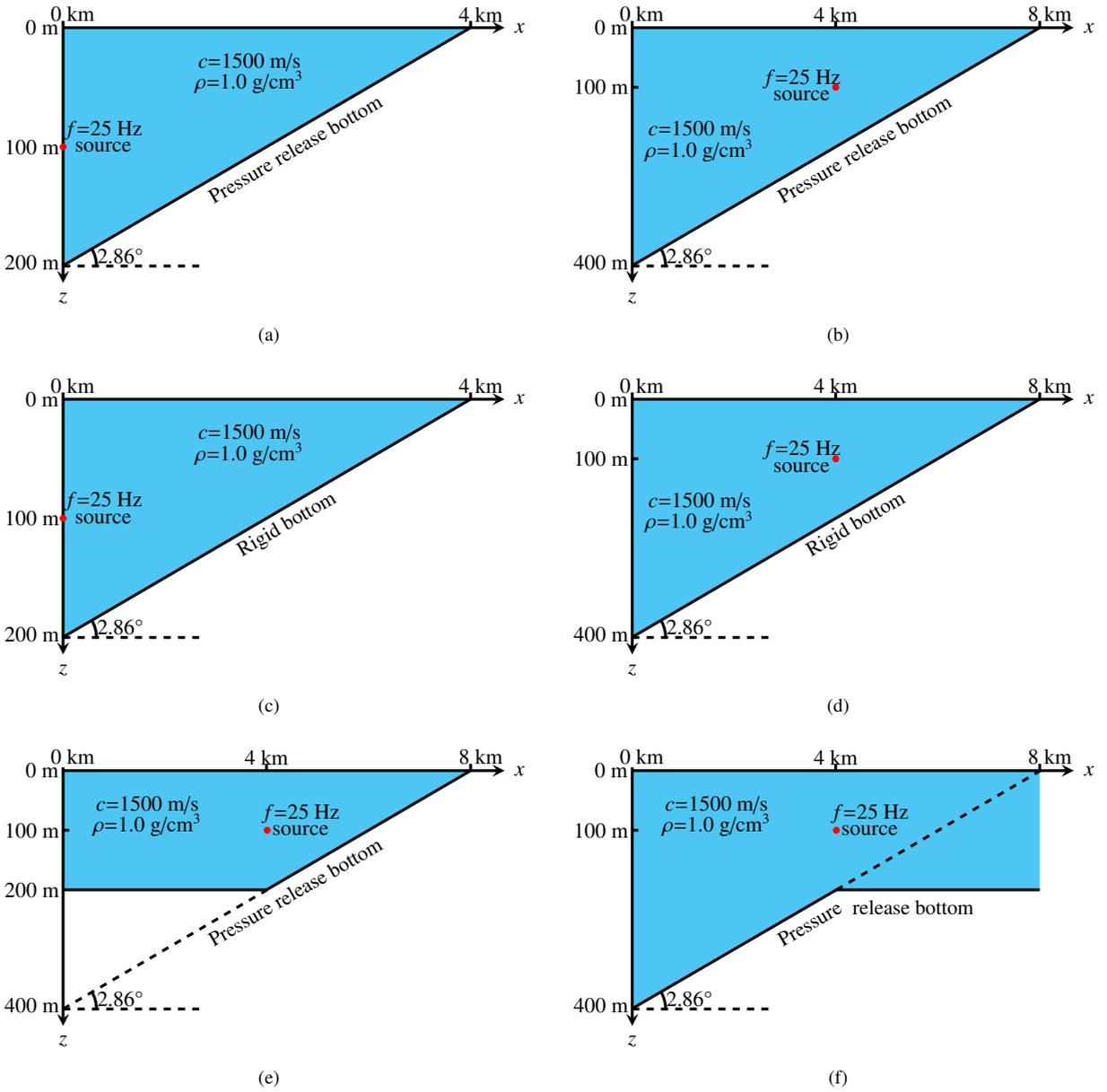

To verify the correctness of the algorithm and to test and demonstrate the developed code, the following ten numerical experiments were conducted. We have named the program developed based on the proposed algorithm SPEC. To present the acoustic field results, the transmission loss (TL) of the acoustic pressure is defined as $\text{TL}=-20\log_{10}(\vert p\vert/\vert p_0\vert)$ in dB, where $p_0=\mathrm{i}\mathcal{H}_0^{(1)}(k_\mathrm{s})/4$ is the acoustic pressure 1 m from the line source and $k_\mathrm{s}$ is the wavenumber of the medium at the location of the line source. The TL field is often used in actual displays to compare and analyze sound fields \cite{Jensen2011}. For convenience, the spectral truncation orders were set to $N_w=N_b=30$ in SPEC for all examples; in actual calculations, users can arbitrarily specify the spectral truncation orders in the SPEC input file.

\subsection{Analytical examples: simple wedge-shaped waveguide}

Consider the ideal wedge-shaped waveguide shown in Fig.~\ref{Figure3}(a), which is a primary benchmark problem for range-dependent waveguides. Both the sea surface and the seabed are pressure release boundaries. The line source is perpendicular to the plane and intersects the plane at $(x_\mathrm{s}, z_\mathrm{s})$. This benchmark problem has been proposed and discussed at two consecutive conferences of the Acoustic Society of America (ASA), and Buckingham presented an analytical solution to this problem \cite{Buckingham1990}. Here, we briefly introduce this problem. The velocity potential in the water body is:
\begin{subequations}
	\label{eq:52}
	\begin{gather}
		\label{eq:52a}
		\Phi\left(r, r^{\prime}, \theta, \theta^{\prime}\right)= \frac{\mathrm{i} \pi}{\theta_{0}} \sum_{m=1}^{M} \mathcal{J}_{\nu_m}\left(k r_{<}\right) \mathcal{H}_{\nu_m}^{(1)}\left(k r_{>}\right)\sin (\nu_m \theta) \sin \left(\nu_m \theta^{\prime}\right) \\
		\nu_m = \frac{m\pi}{\theta_0}, \quad m=1,2,\cdots,M
	\end{gather}		
\end{subequations}
where $r$ and $r^{\prime}$ are the distances from the receiver and line source, respectively, to the apex of the wedge; $\theta$ and $\theta'$ are the angles to the depths of the receiver and source, respectively, measured about the apex; $\theta_0$ is the wedge angle; $k$ is the wavenumber of the seawater; $r_< = \min(r,r')$ and $r_> = \max(r,r')$; and $\mathcal{J}_{\nu_m}(\cdot)$ and $\mathcal{H}_{\nu_m}^{(1)}(\cdot)$ are the Bessel and Hankel functions, respectively, of the first kind of order $\nu_m$. The TL field is calculated via the following formula:
\begin{subequations}
	\label{eq:53}
	\begin{gather}
		\mathrm{TL}=-20 \log _{10}\bigg\vert\frac{\Phi\left(r, r^{\prime}, \theta, \theta^{\prime}\right)}{\Phi_{0}(1)}\bigg\vert \\
		\Phi_{0}(R)=\frac{\mathrm{i}}{4} \mathcal{H}_{0}^{(1)}(k R)
	\end{gather}		
\end{subequations}
where $R$ is the radial distance from the line source to a point in the field.

\begin{figure}[htbp]
	\centering
	\subfigure[]{\includegraphics[width=0.49\linewidth]{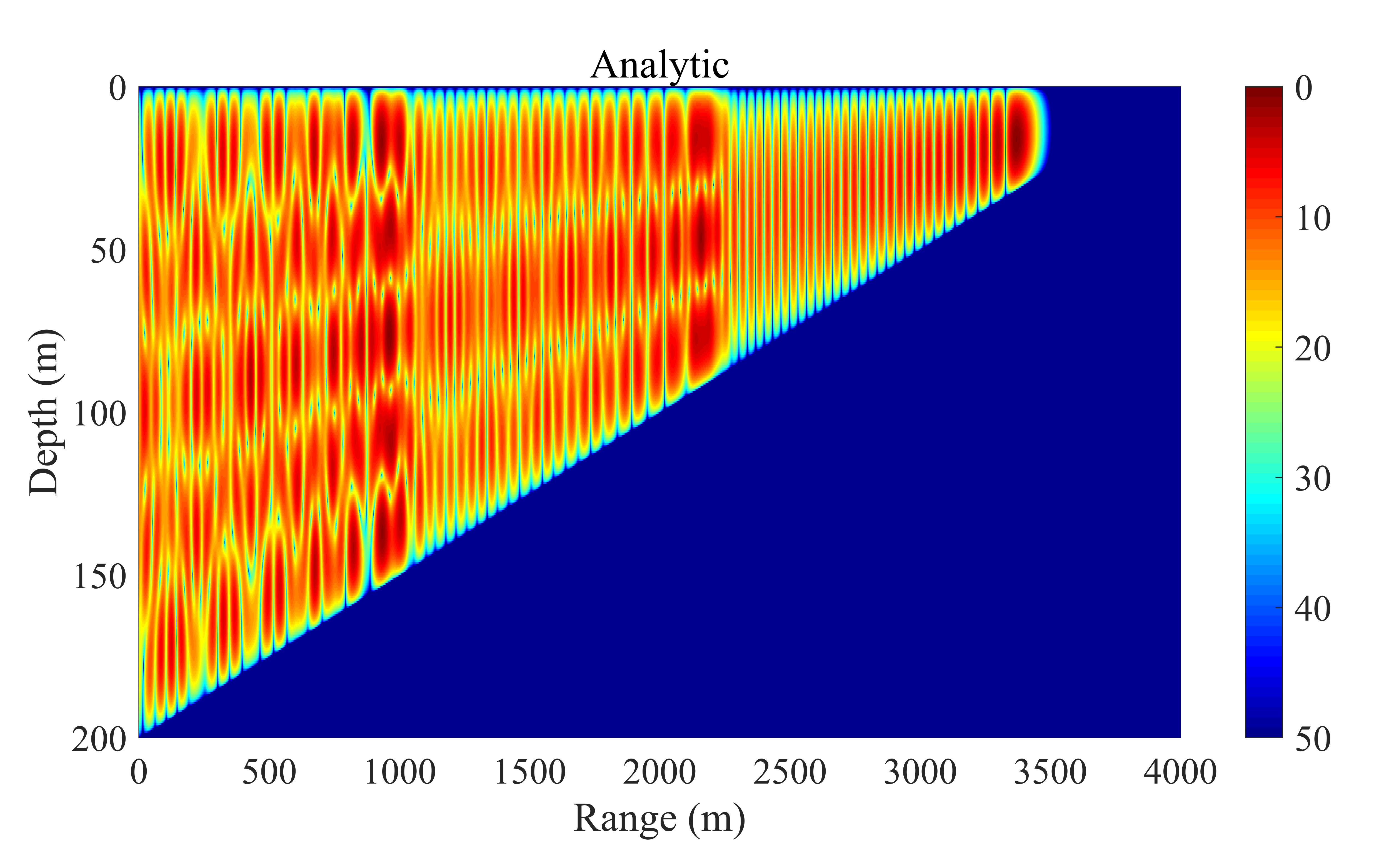}}
	\subfigure[]{\includegraphics[width=0.49\linewidth]{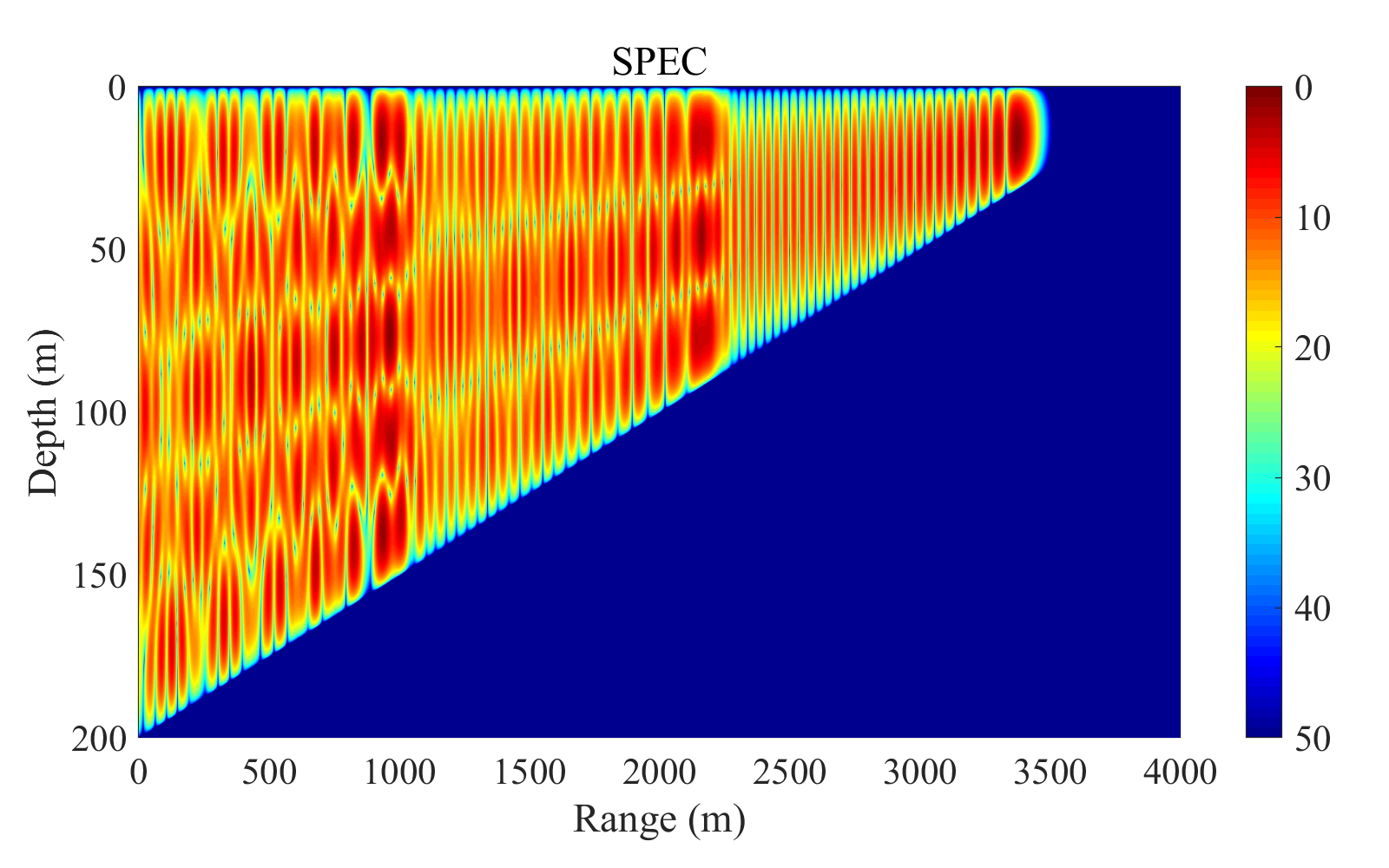}}\\
	\subfigure[]{\includegraphics[width=\linewidth]{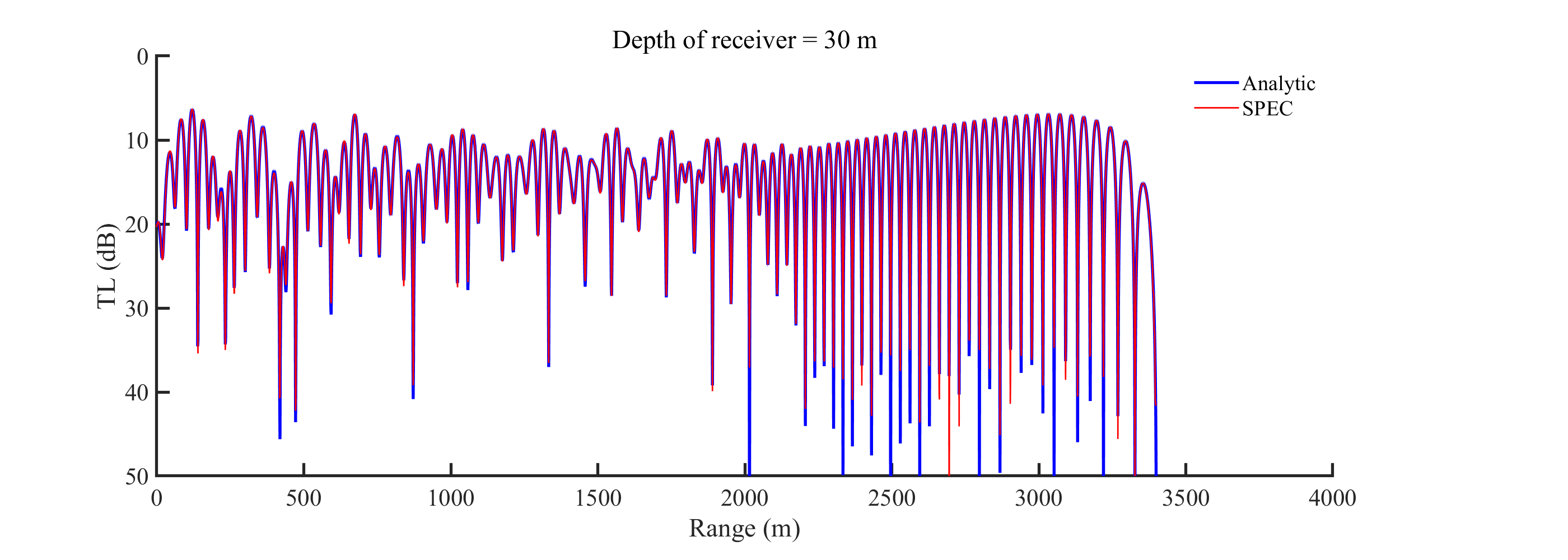}}
	\caption{Sound fields for example 1 calculated using the analytical solution (a) and SPEC (b) and the TL curves at a depth of 30 m (c).}
	\label{Figure4}
\end{figure}

\begin{figure}[htbp]
	\centering
	\subfigure[]{\includegraphics[width=0.8\linewidth]{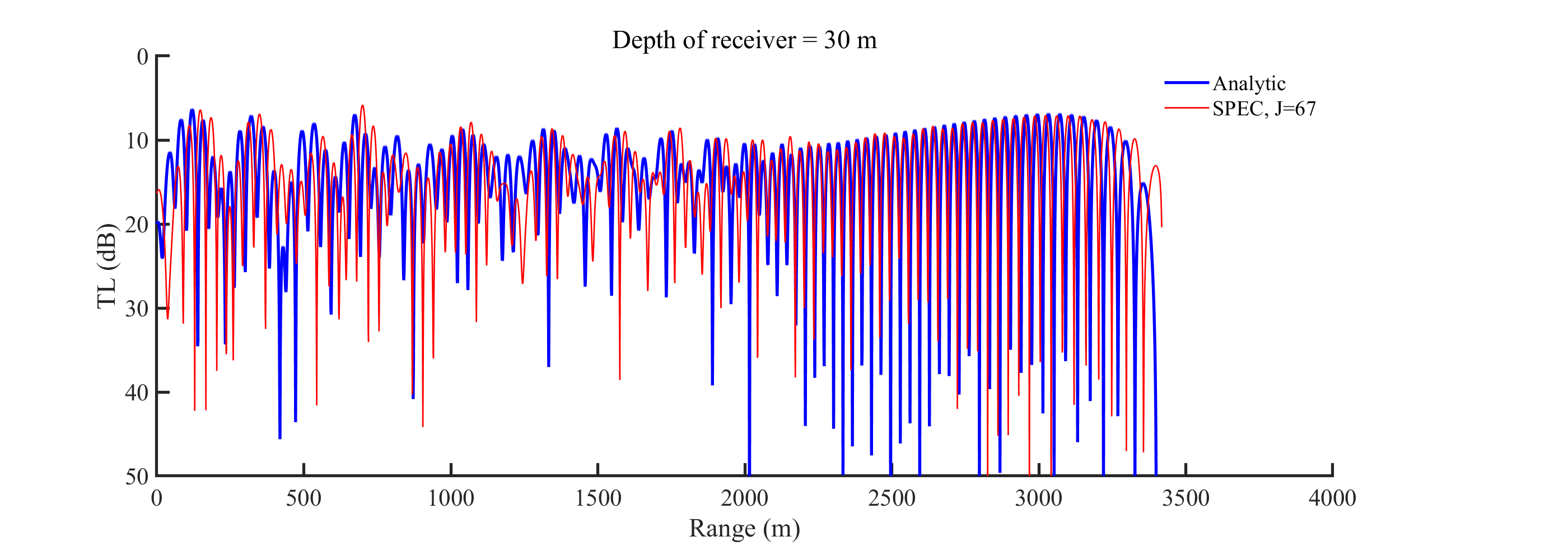}}
	\subfigure[]{\includegraphics[width=0.8\linewidth]{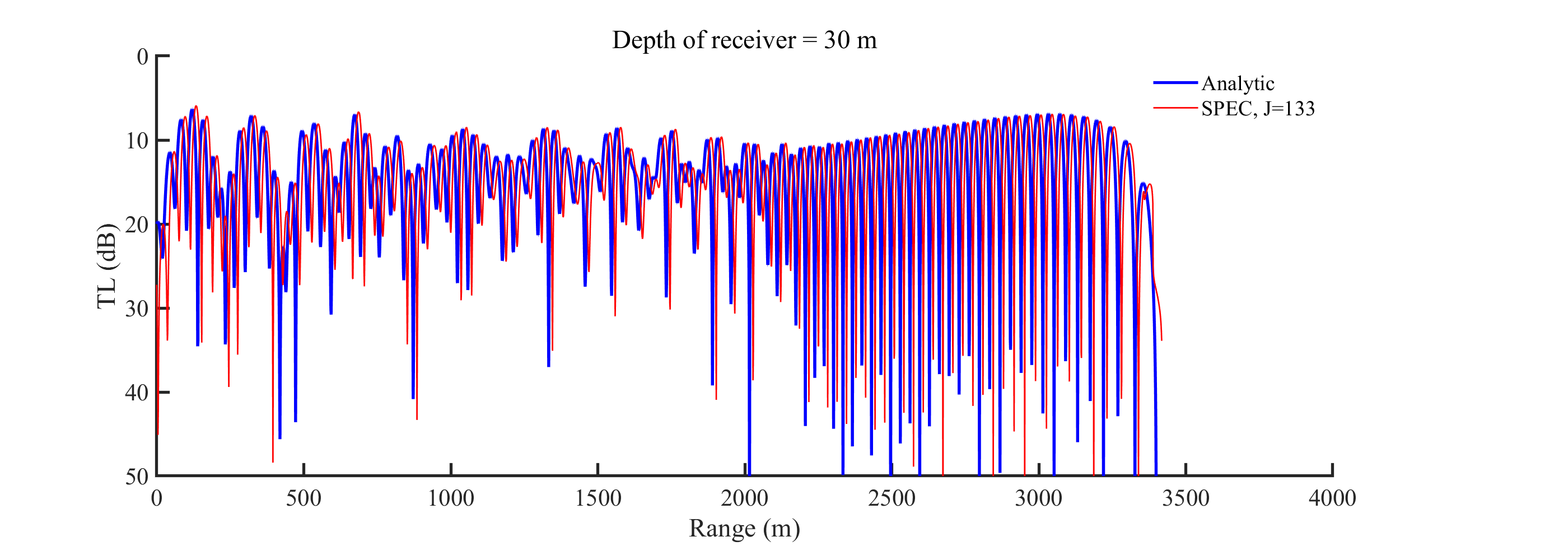}}
	\subfigure[]{\includegraphics[width=0.8\linewidth]{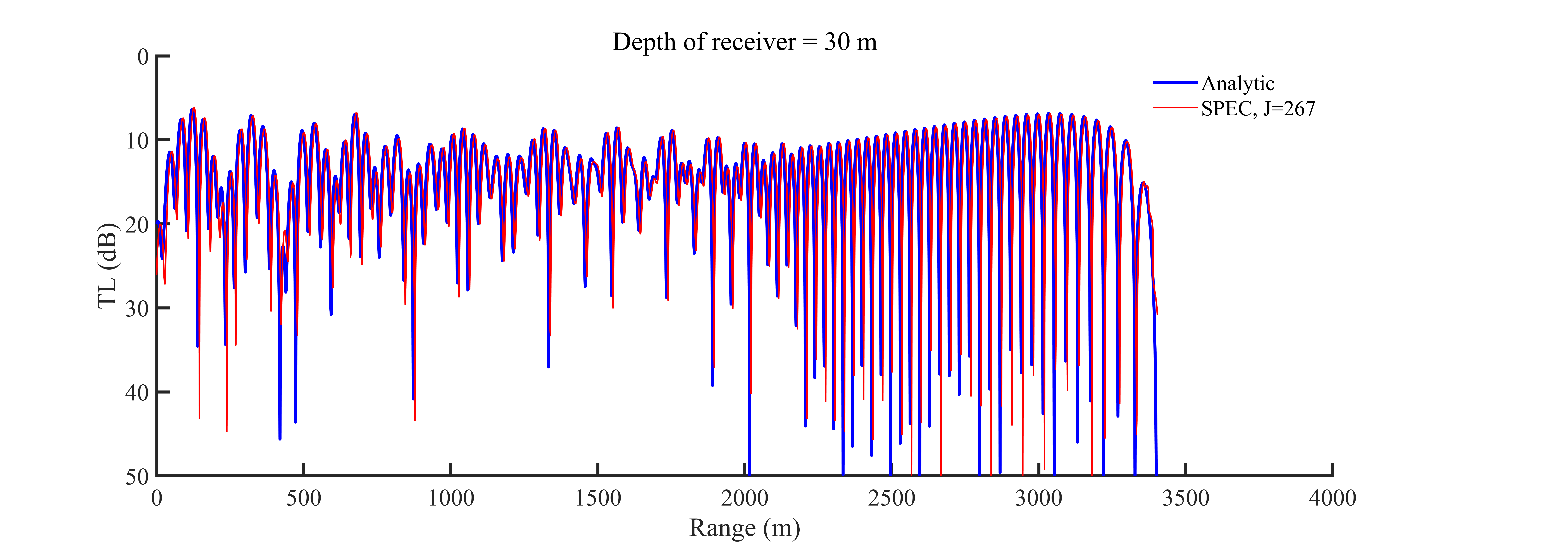}}
	\subfigure[]{\includegraphics[width=0.8\linewidth]{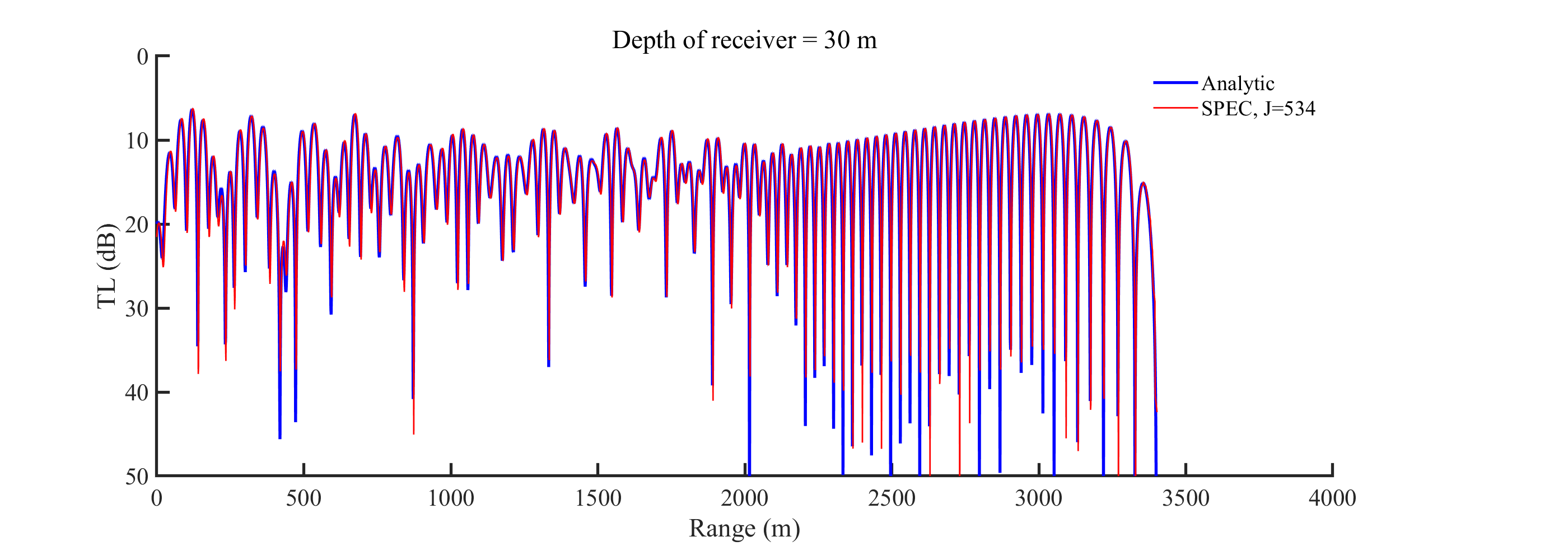}}
	\caption{TL curves for example 1 at a depth of 30 m as calculated by SPEC with different numbers of segments.}
	\label{Figure5}
\end{figure}

For example 1, Fig.~\ref{Figure4} shows the sound fields of the wedge-shaped waveguide as calculated using the analytical solution and SPEC. To ensure the accuracy of the results, we divided the SPEC model into 800 horizontal segments; this segmentation fully satisfies the step approximation discretization criterion presented by Jensen \cite{Jensen1998}. The number of modes is taken to be $M=6$. Fig.~\ref{Figure4} clearly confirms that the SPEC results are almost perfectly consistent with the analytical solution regardless of whether the TL is calculated throughout the whole field or at a certain receiving depth.
Fig.~\ref{Figure5} shows the SPEC-calculated TL curves for comparison with the analytical solution for different numbers of segments. As the number of segments increases, the sound field calculated by SPEC gradually approaches the analytical solution, and the error continuously decreases.

\begin{figure}[htbp]
	\centering
	\subfigure[]{\includegraphics[width=0.49\linewidth]{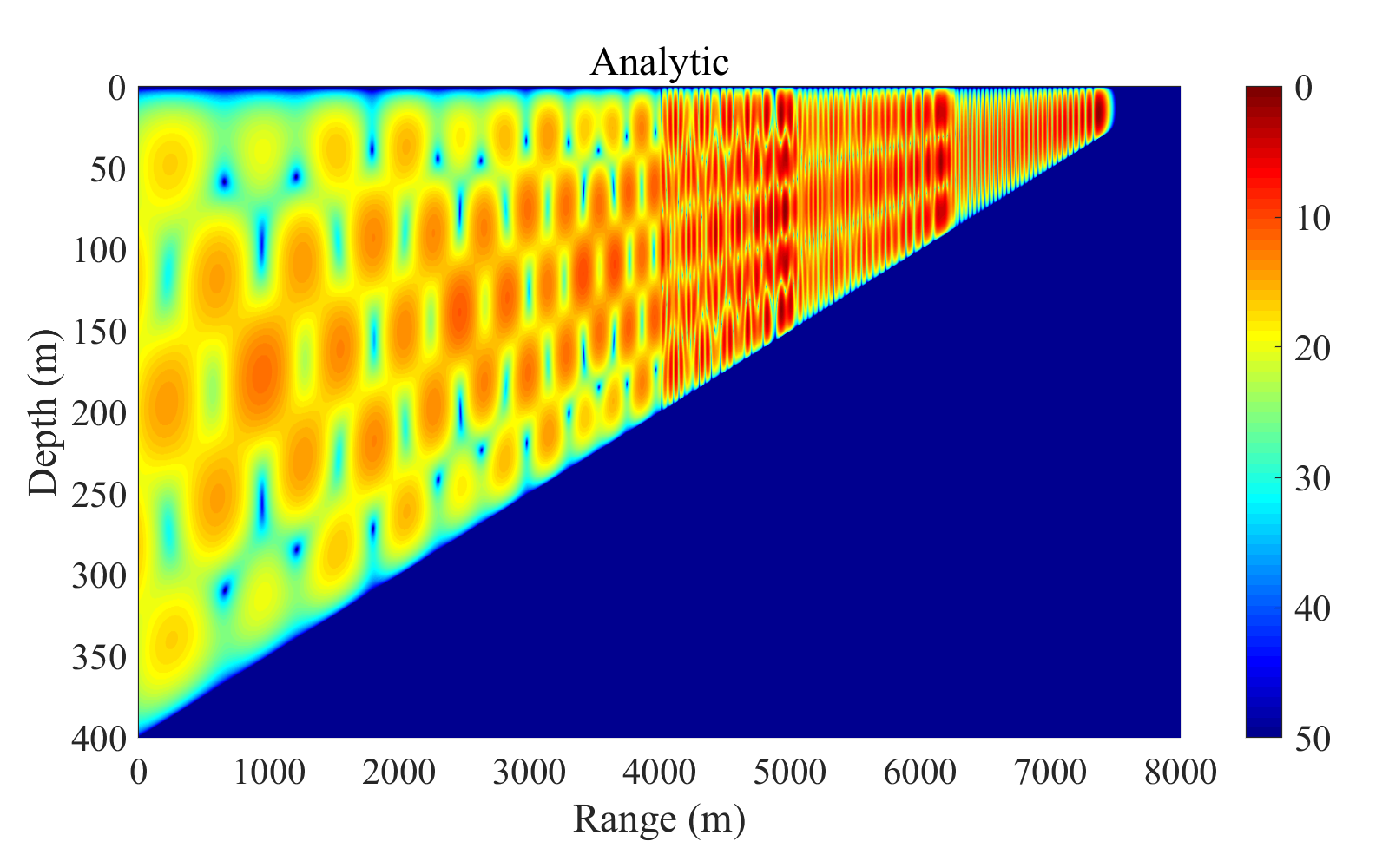}}
	\subfigure[]{\includegraphics[width=0.49\linewidth]{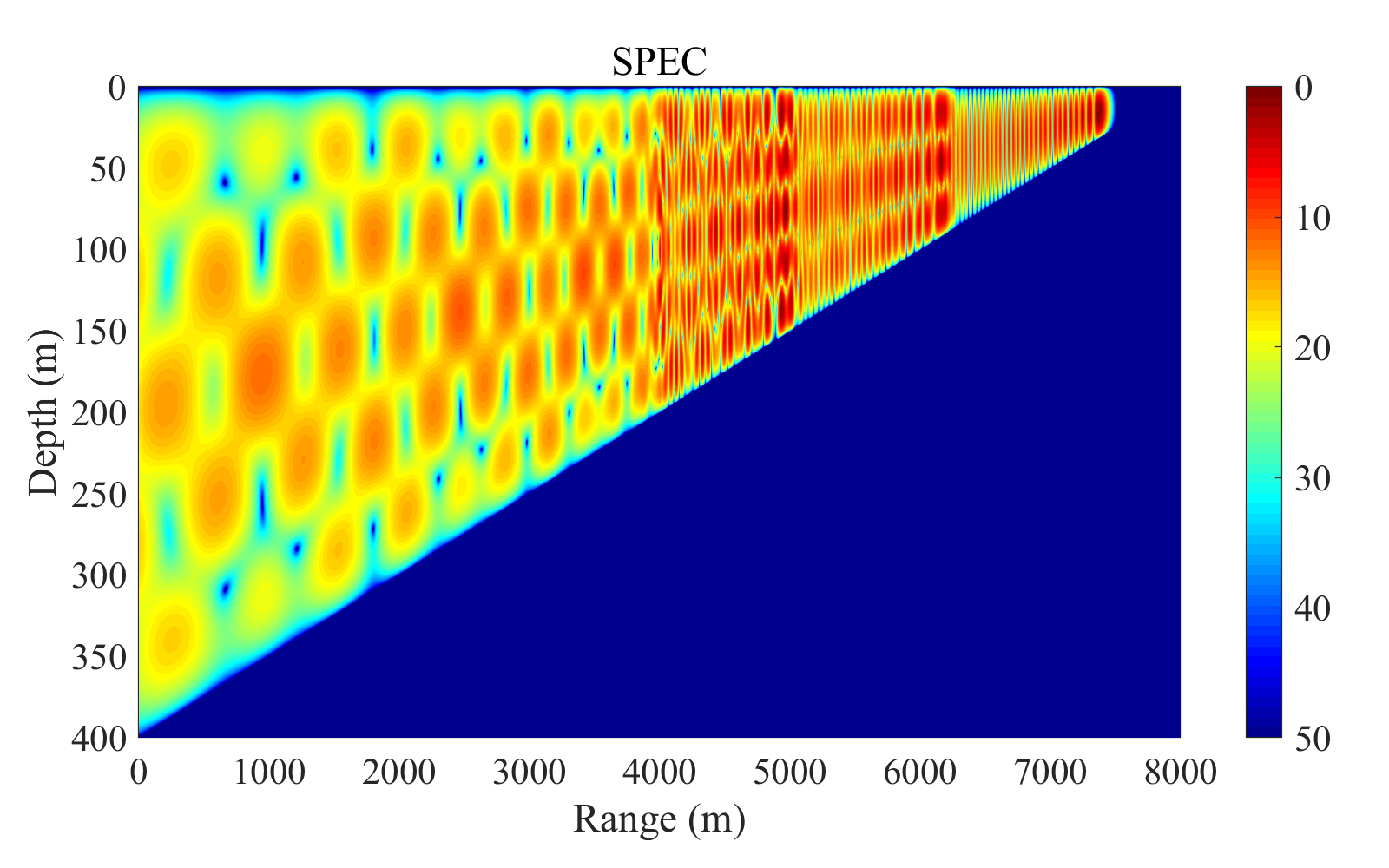}}\\
	\subfigure[]{\includegraphics[width=\linewidth]{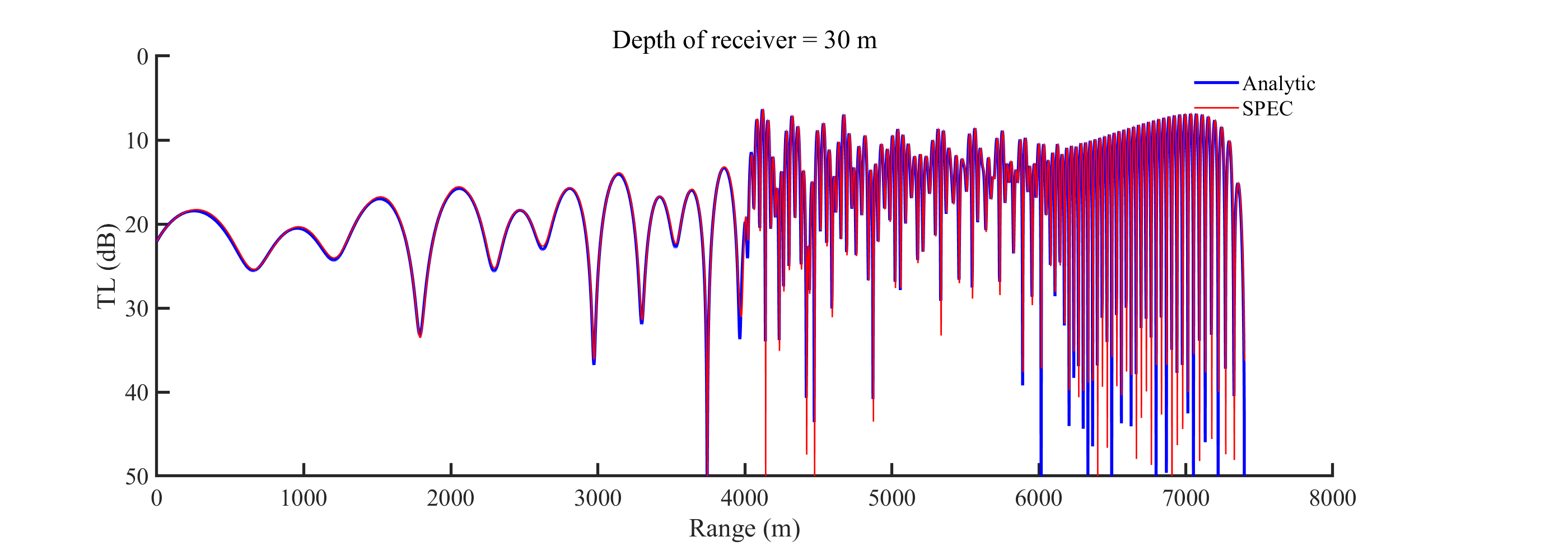}}
	\caption{Sound fields for example 2 calculated using the analytical solution (a) and SPEC (b) and the TL curves at a depth of 30 m (c).}
	\label{Figure6}
\end{figure}
For example 2, the configuration in Fig.~\ref{Figure3}(b) is the same as that in Fig.~\ref{Figure3}(a). The main differences are that the line source in Fig.~\ref{Figure3}(b) is located in the center of the oceanic wedge, and the maximum depth of the water column is twice as large as that in Fig.~\ref{Figure3}(a). In this case, the analytical solution shown in Eqs.~\eqref{eq:52} and \eqref{eq:53} is still valid, as illustrated in Fig.~\ref{Figure6}. For this example, the horizontal range was divided into 1600 segments, and the number of modes was taken to be $M=12$. The SPEC solution shown in Fig.~\ref{Figure6}(b) demonstrates excellent overall agreement with the analytical solution shown in Fig.~\ref{Figure6}(a). The positions of the peaks and troughs are the same, with only occasional, slight discrepancies in the heights of the interference peaks. Moreover, the absolute levels of the TL curves are in good agreement.

\begin{figure}[htbp]
	\centering
	\subfigure[]{\includegraphics[width=0.49\linewidth]{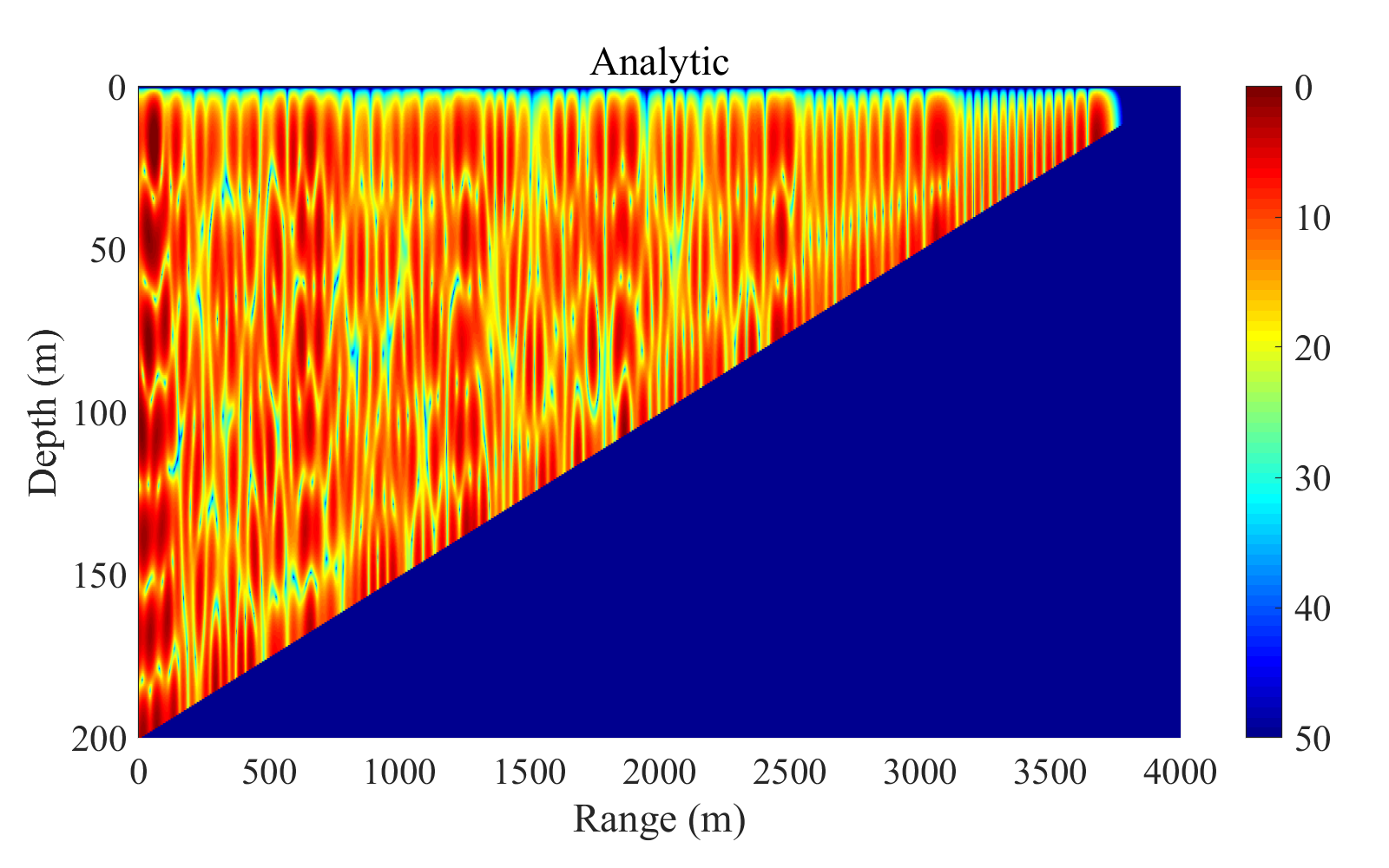}}
	\subfigure[]{\includegraphics[width=0.49\linewidth]{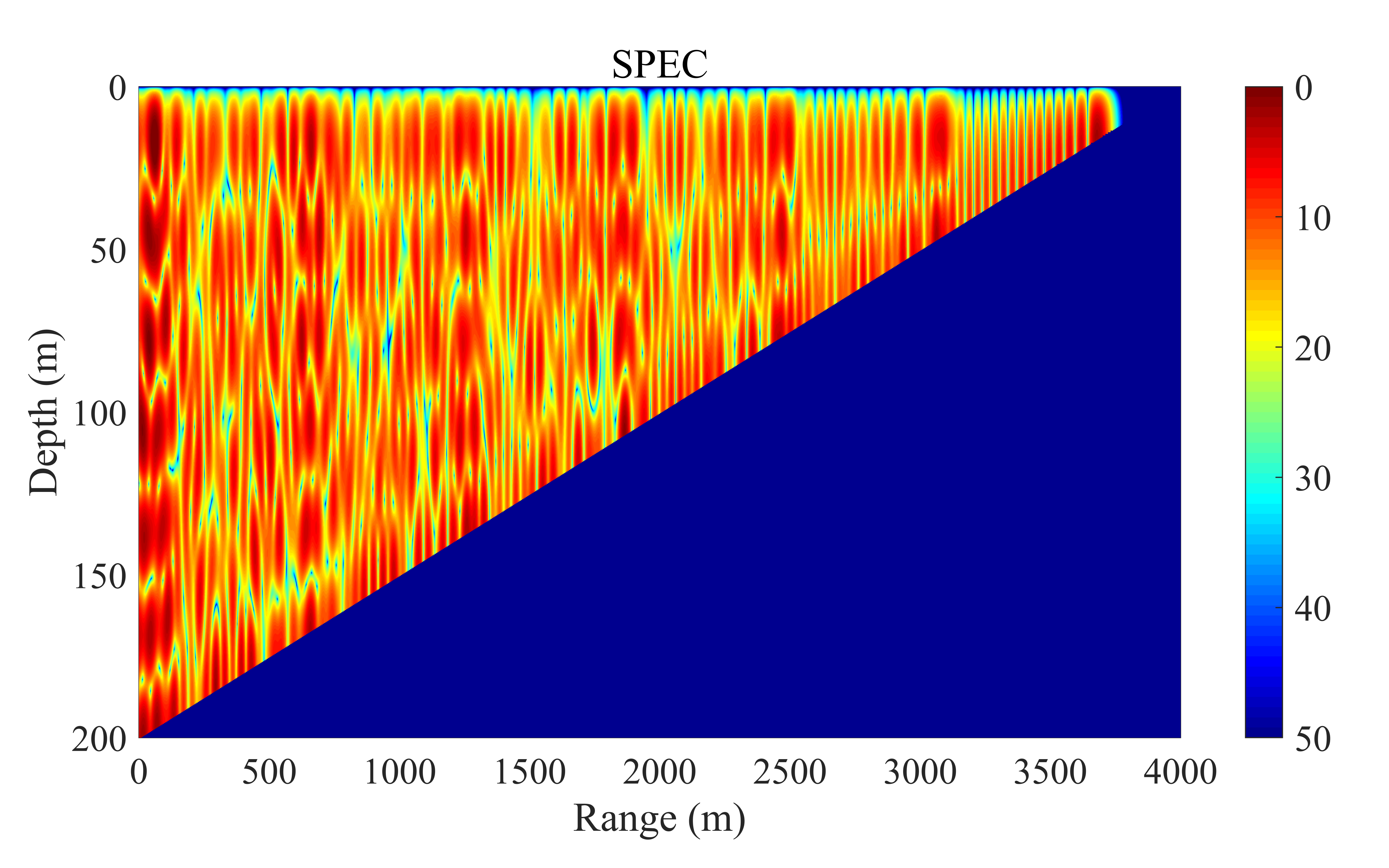}}\\
	\subfigure[]{\includegraphics[width=\linewidth]{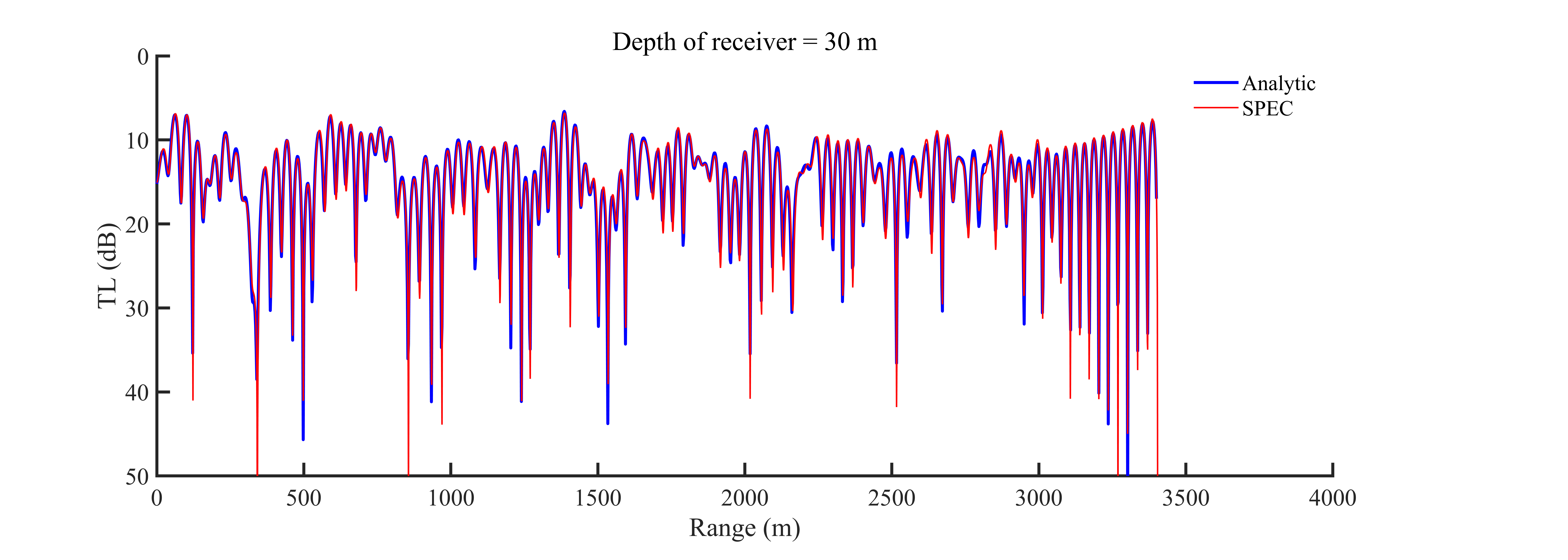}}
	\caption{Sound fields for example 3 calculated using the analytical solution (a) and SPEC (b) and the TL curves at a depth of 30 m (c).}
	\label{Figure7}
\end{figure}

\begin{figure}[htbp]
	\centering
	\subfigure[]{\includegraphics[width=0.8\linewidth]{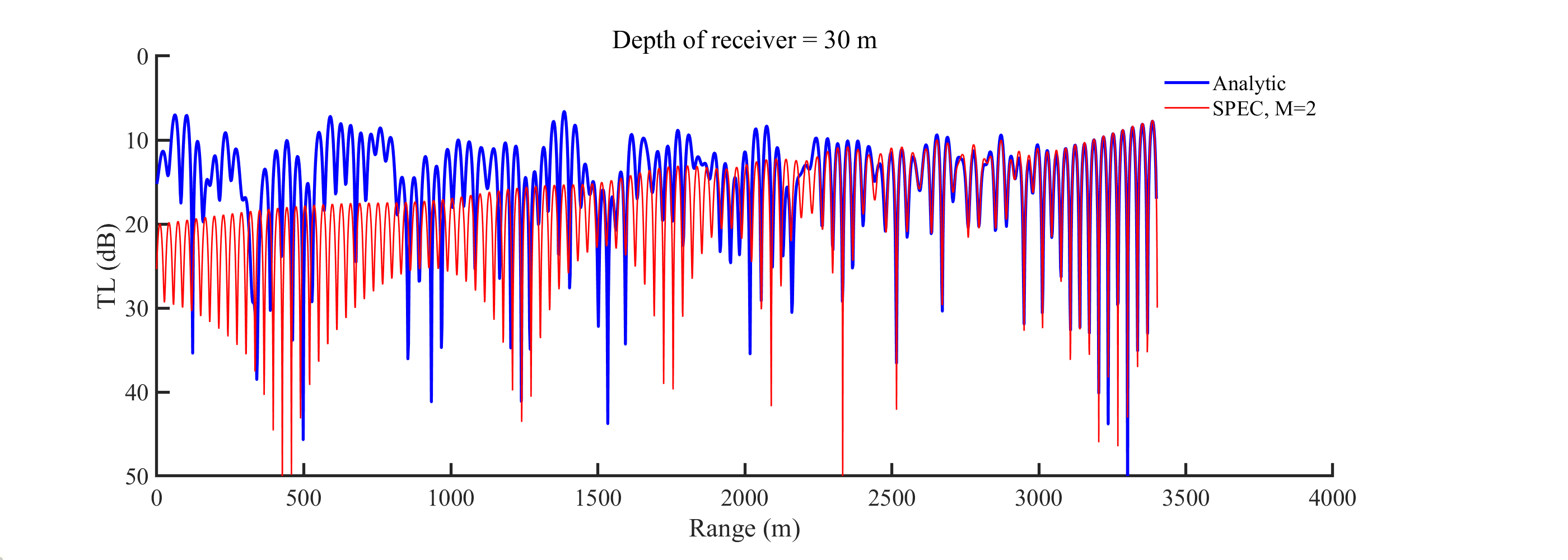}}
	\subfigure[]{\includegraphics[width=0.8\linewidth]{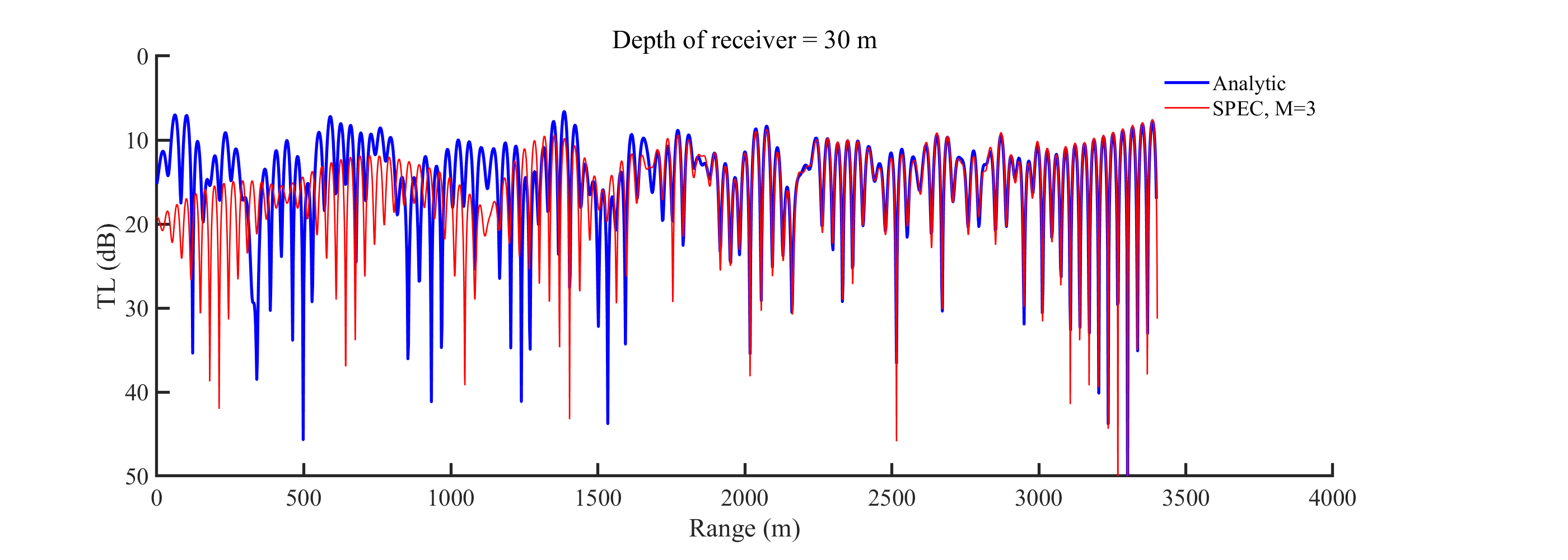}}
	\subfigure[]{\includegraphics[width=0.8\linewidth]{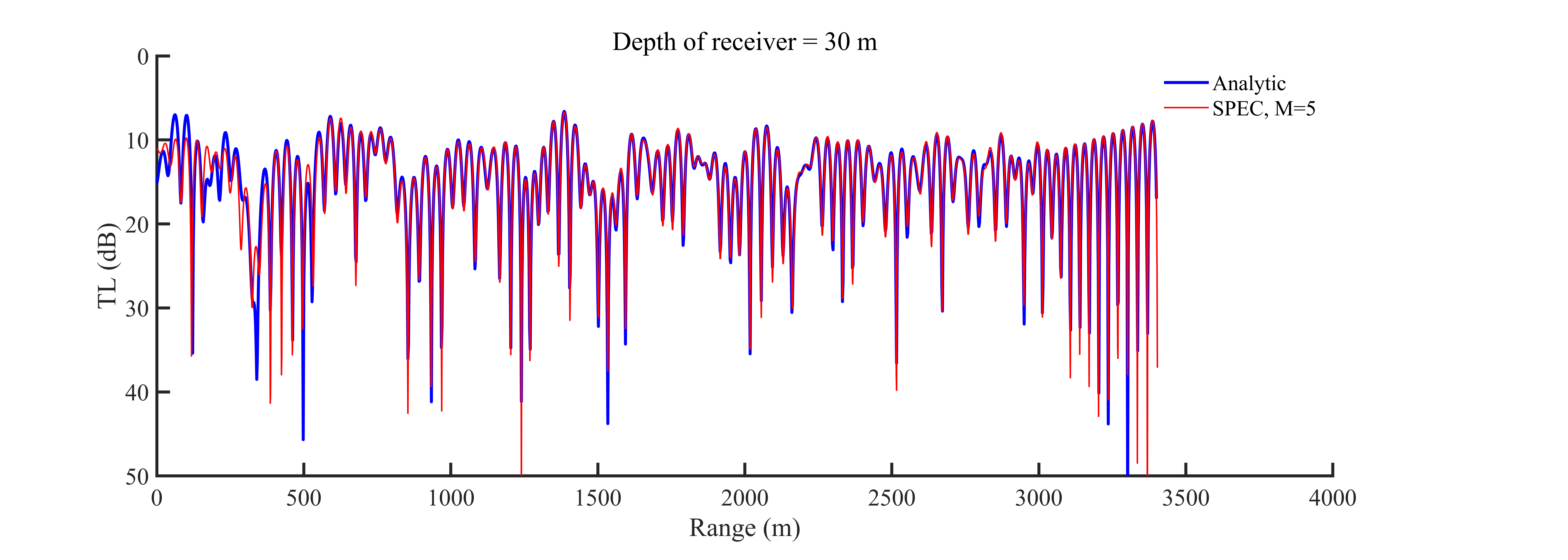}}
	\subfigure[]{\includegraphics[width=0.8\linewidth]{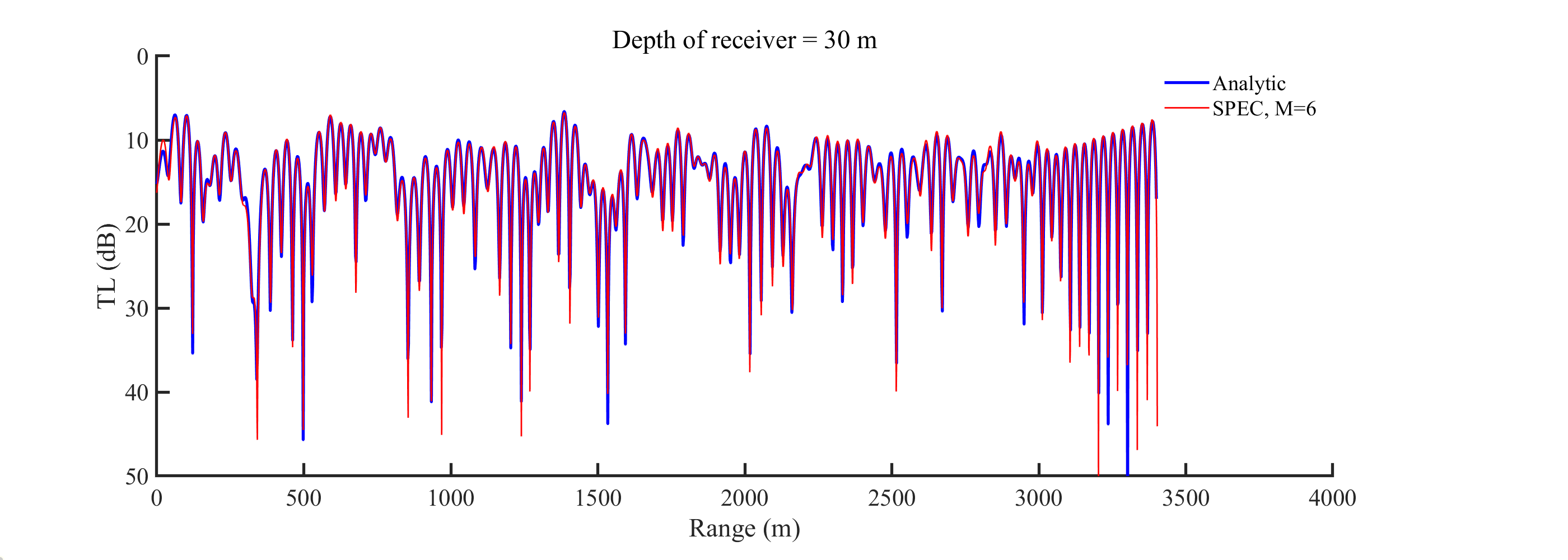}}
	\caption{TL curves for example 3 at a depth of 30 m as calculated by SPEC with different numbers of modes.}
	\label{Figure8}
\end{figure}

Fig.~\ref{Figure3}(c) and Fig.~\ref{Figure3}(d) illustrate wedge-shaped waveguides on a rigid seabed in configurations that are otherwise equivalent to Fig.~\ref{Figure3}(a) and Fig.~\ref{Figure3}(b), respectively. Except for the nature of the ocean bottom, these configurations are exactly the same as those in examples 1 and 2. In these cases, an analytical solution is also available, and the expression is the same as Eq.~\eqref{eq:52a}; however, the following usually holds for a rigid seabed:
\begin{equation}
	\nu_m = \left(m-\frac{1}{2}\right)\frac{\pi}{\theta_0}, \quad m=1,2,\cdots
\end{equation}
Fig.~\ref{Figure7} shows the analytical solution for the first wedge-shaped waveguide configuration atop a rigid seabed (example 3) and the corresponding SPEC-calculated sound field. The number of segments in SPEC was set to the same value as in the case of the free seabed, and the number of modes was taken to be $M=7$. Given the perfect agreement between the analytical and SPEC solutions, the same conclusion as that from example 1 can naturally be drawn with confidence.
Fig.~\ref{Figure8} compares the TL curves calculated by SPEC for different numbers of coupled modes against the analytical solution. As the number of coupled modes increases, the sound field calculated by SPEC gradually approaches the analytical solution, and the error rapidly decreases. Another obvious phenomenon is that the wedge angle region on the right tends to match the analytical solution earlier than the one on the left. This is because the depth of this wedge angle region is shallower and the number of normal modes that can be excited is smaller.

\begin{figure}[htbp]
	\centering
	\subfigure[]{\includegraphics[width=0.49\linewidth]{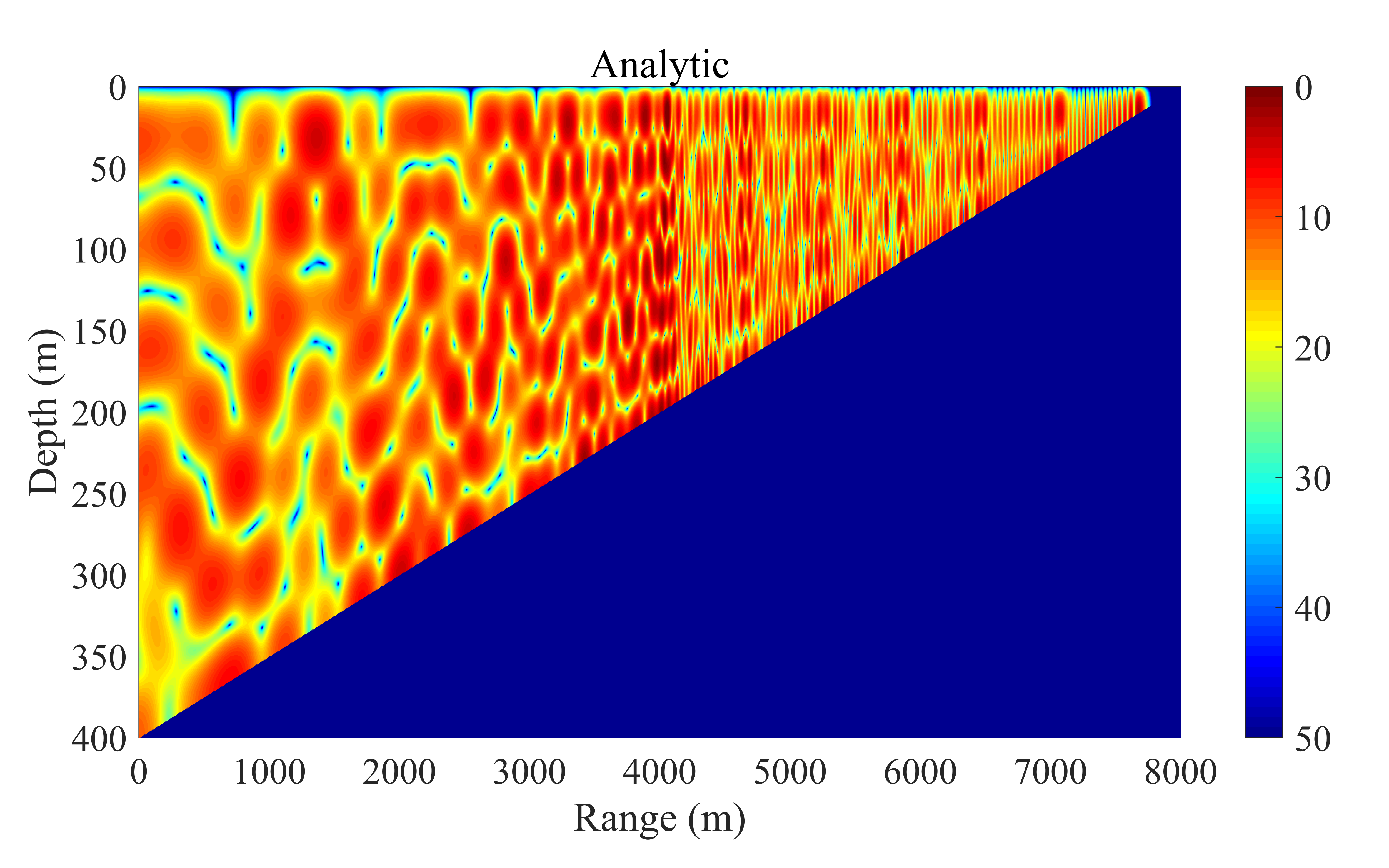}}
	\subfigure[]{\includegraphics[width=0.49\linewidth]{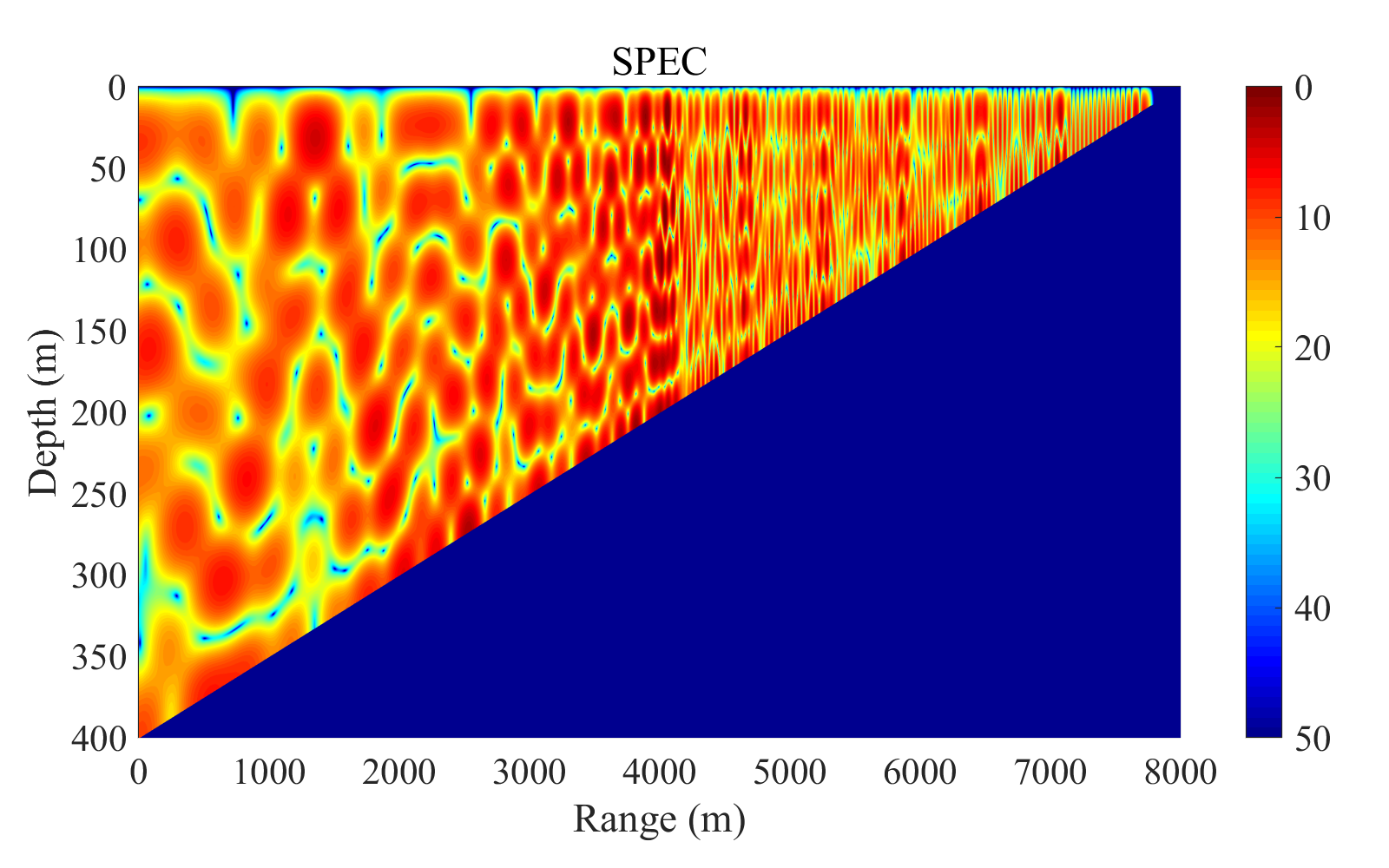}}\\
	\subfigure[]{\includegraphics[width=\linewidth]{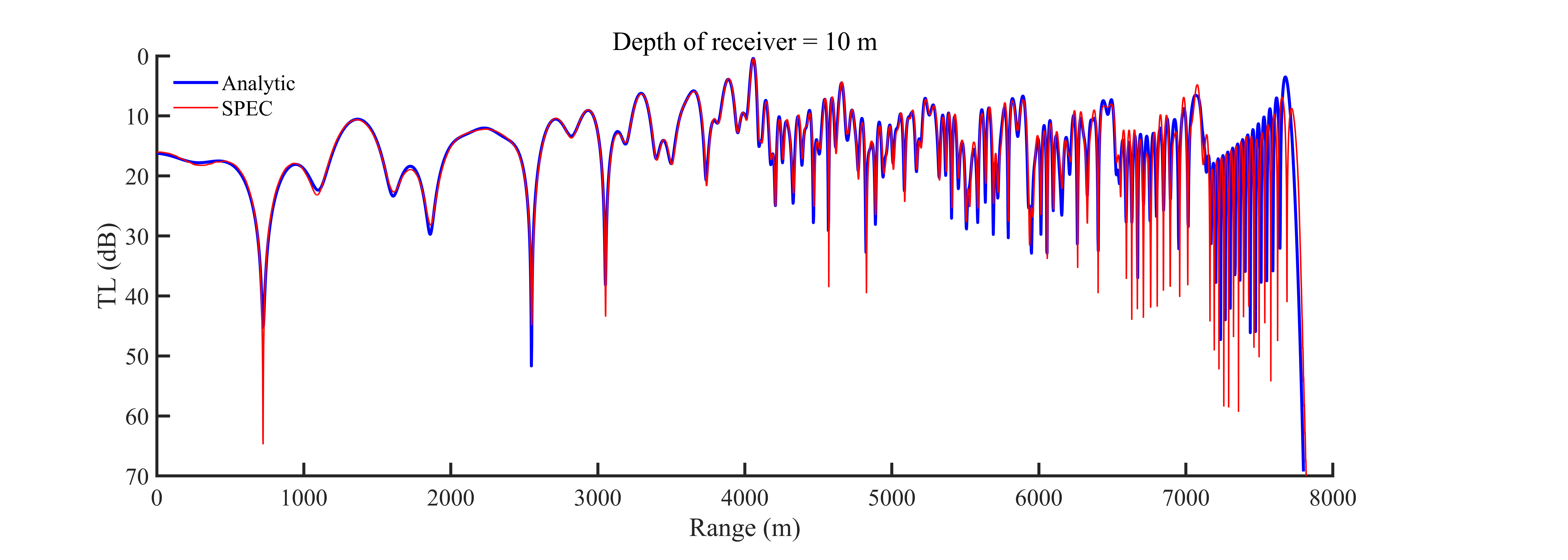}}
	\caption{Sound fields for example 4 calculated using the analytical solution (a) and SPEC (b) and the TL curves at a depth of 10 m (c).}
	\label{Figure9}
\end{figure}

Similarly, we calculated the sound field of the rigid-bottomed waveguide when the line source is in the center of the wedge (example 4), and the results are shown in Fig.~\ref{Figure9}. Again, the number of segments in SPEC was set the same as in the case of the free seabed, and the number of modes was taken to be $M=13$. As in the previous examples, Fig.~\ref{Figure9} demonstrates that the SPEC results still match the analytical solution very well, although certain differences are observed in areas far from the line source.

The above comparisons of the SPEC results with the analytical solutions clearly reflect the excellent accuracy achieved by SPEC, which is sufficient to verify the reliability of the proposed algorithm (regardless of whether the line source is classical or generalized) and the correctness of the code's implementation. The observed stability and accuracy provide motivation to use this code for the numerical simulation of more practical ocean acoustic problems in future research.
\subsection{Complicated examples without analytical solutions}
\subsubsection{Modified wedge-shaped ocean environment}
\begin{figure}[htbp]
	\centering
	\subfigure[]{\includegraphics[width=0.49\linewidth]{Figure6a}}
	\subfigure[]{\includegraphics[width=0.49\linewidth]{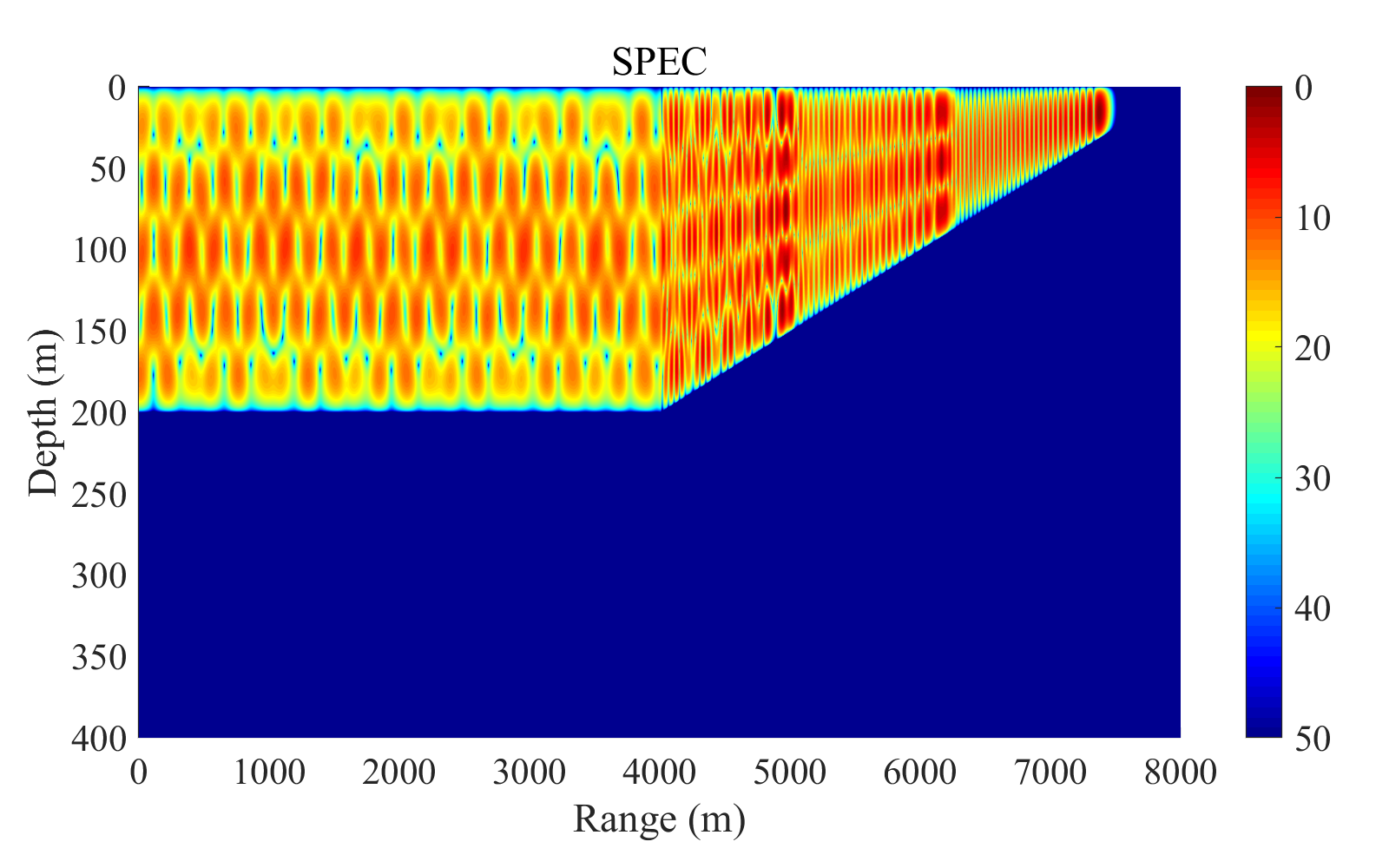}}\\
	\subfigure[]{\includegraphics[width=\linewidth]{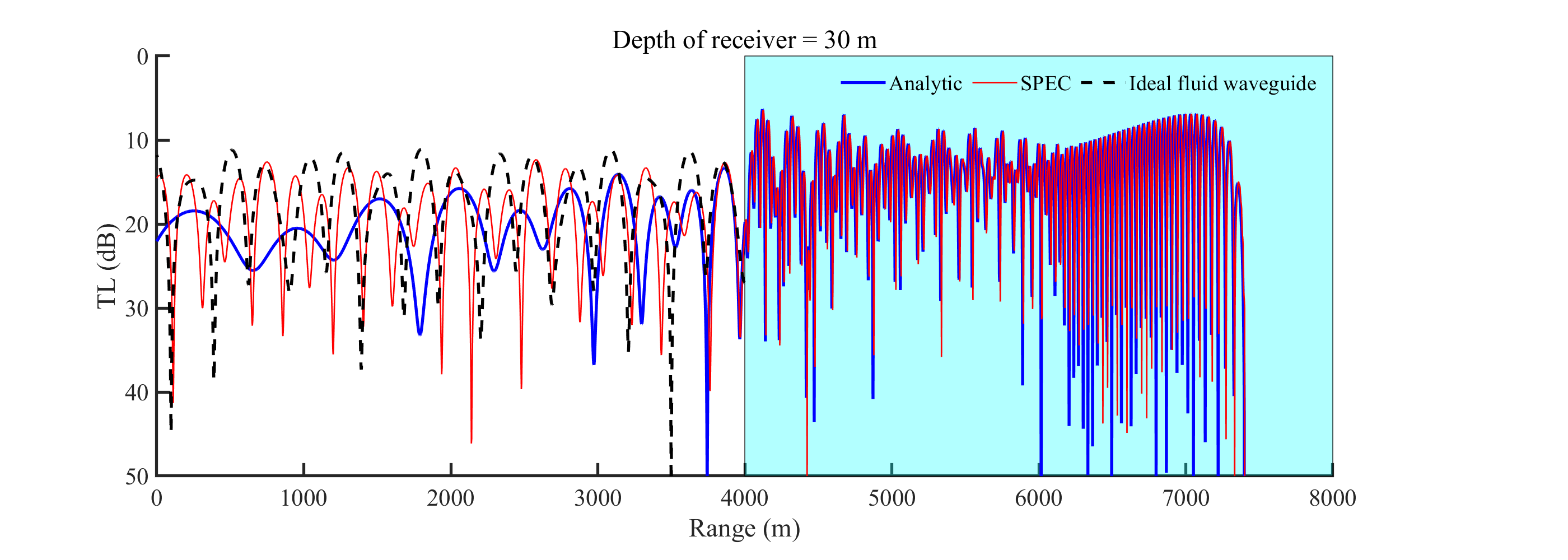}}
	\caption{Analytical sound field for example 2 (a), the sound field for example 5 as calculated by SPEC (b) and the TL curves at a depth of 30 m (c).}
	\label{Figure10}
\end{figure}

As depicted in Fig.~\ref{Figure3}(e), we consider two modifications to the ideal wedge-shaped waveguide, leading to some interesting conclusions. The first modification is to flatten the downward slope to the left of the line source in Fig.~\ref{Figure3}(b). Please note that this modification involves only half of the sound field, i.e., the region to the left of the sound source. The same number of segments was used as for the problem depicted in Fig.~\ref{Figure3}(b), but since the maximum ocean depth in this case is 200 m, the number of modes was taken to be $M=6$. For comparison, we show the analytical sound field and TL curves for Fig.~\ref{Figure3}(b) in Fig.~\ref{Figure10}. Remarkably, for the unmodified right half of the waveguide (from 4 km to 8 km), the sound field generated by the line source is exactly the same as in the configuration depicted in Fig.~\ref{Figure3}(b). This indicates that the uphill sound field excited by the line source is not affected by the downhill sound field; that is, the backscatter occurring in the downhill area has a negligible impact on the uphill area. The TL curve for an ideal fluid waveguide with $H=200$ m and $z_\mathrm{s}=100$ m is also shown in Fig.~\ref{Figure10}(c). Obviously, it is not consistent with the left half of Fig.~\ref{Figure10}(b), which illustrates that the backscattering of the uphill sound field excited by the line source has a nonnegligible influence on the flat-slope sound field.

\begin{figure}[htbp]
	\centering
	\subfigure[]{\includegraphics[width=0.49\linewidth]{Figure6a}}
	\subfigure[]{\includegraphics[width=0.49\linewidth]{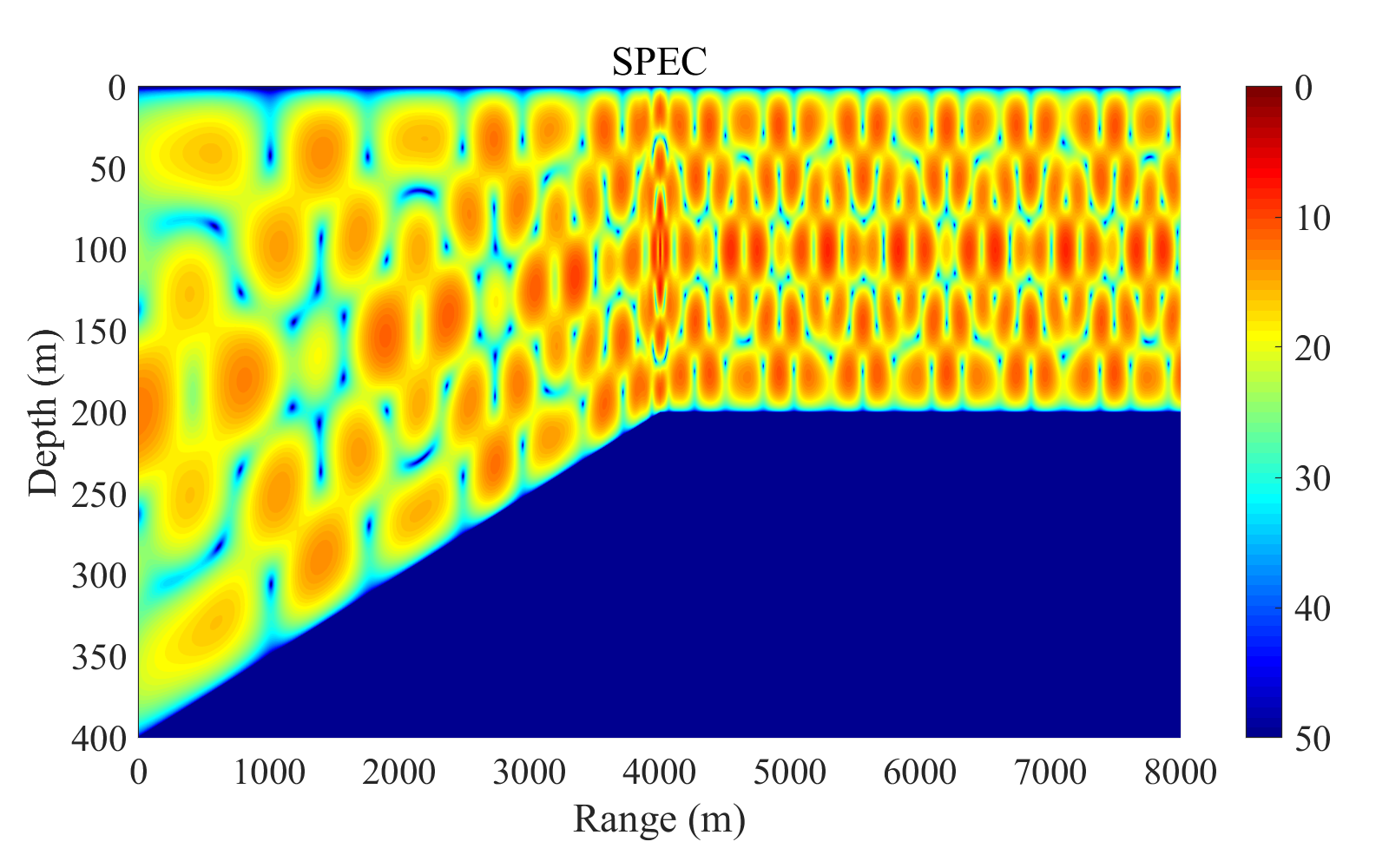}}\\
	\subfigure[]{\includegraphics[width=\linewidth]{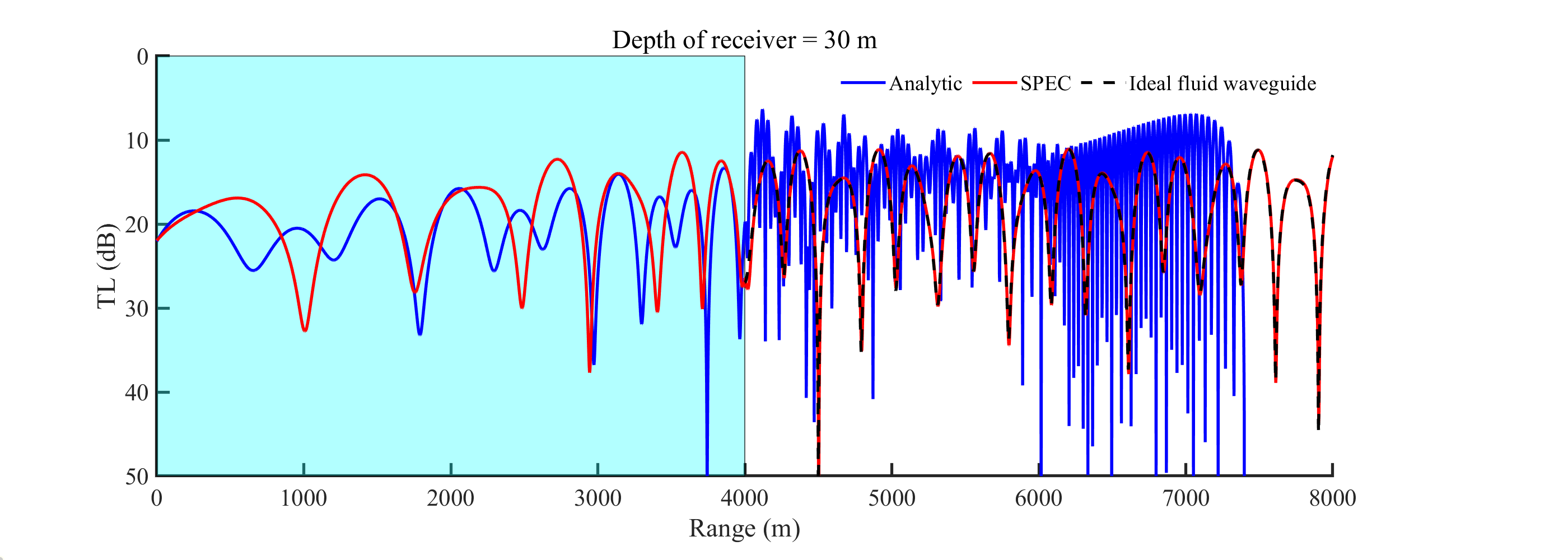}}
	\caption{Analytical sound field for example 2 (a), the sound field for example 6 as calculated by SPEC (b) and the TL curves at a depth of 30 m (c).}
	\label{Figure11}
\end{figure}

Similarly, the second modification is to flatten the upward slope to the right of the line source in Fig.~\ref{Figure3}(b). Please note that this modification again involves only half of the sound field, this time to the right of the sound source. The same numbers of segments and modes were used as for the problem depicted in Fig.~\ref{Figure3}(b). For comparison, we show the analytical sound field and TL curves for Fig.~\ref{Figure3}(b) in Fig.~\ref{Figure11}. Compared to the results of the first modification, we do not observe a similar phenomenon; that is, for the unmodified left half of the sound field (from 0 km to 4 km), the sound field generated by the line source does not match the analytical solution. This indicates that the downhill sound field excited by the line source is affected by the backscattering occurring in the uphill area, indicating that backscattering in the uphill area cannot be ignored because it is sufficient to induce completely different behavior in the downhill sound field. The TL curves for an ideal fluid waveguide with the same configuration as in Fig.~\ref{Figure10}(c) are also shown in Fig.~\ref{Figure11}(c). As seen from the figure, the sound field of the flat-slope segment in Fig.~\ref{Figure11}(b) is in good agreement with the sound field of the ideal fluid waveguide, indicating that the backscattering of the downhill sound field excited by the line source has a negligible effect on the flat-slope sound field.

These two modifications not only demonstrate the characteristics of the sound field excited by a line source but also further illustrate that the proposed algorithm for a generalized line source is completely correct. Moreover, these examples verify the robustness of the SPEC program when dealing with line sources located exactly above changes in the seabed topography.

\subsubsection{Penetrable slope and seamount problems}
\begin{figure}
	\subfigure[]{
		\begin{minipage}[t]{0.5\linewidth}
			\begin{tikzpicture}[node distance=2cm,scale=0.5]
				\tikzstyle{every node}=[font=\small]
				\node at (1.4,0){0 $\text{m}$};
				\node at (1.2,-4.8){80 $\text{m}$};
				\node at (1.1,-6){100 $\text{m}$};
				\node at (2.2,0.4){0 $\text{m}$};
				\node at (14.3,0.4){2500 $\text{m}$};
				\node at (14.6,-1.8){30 $\text{m}$};
				\fill[cyan,opacity=0.7] 	(14,0)--(14,-1.8)--(11.2,-1.8)--(4.8,-4.8)--(2,-4.8)--(2,0)--cycle;
				\fill[orange,opacity=0.7] (2,-4.8)--(4.8,-4.8)--(11.2,-1.8)--(14,-1.8)--(14,-6)--(2,-6)--cycle;
				\draw[very thick, ->](2,0)--(14.5,0) 	node[right]{$x$};
				\draw[very thick, ->](2.02,0.2)--(2.02,-7.5) 	node[below]{$z$};
				\filldraw [red] (2.02,-2.16) circle [radius=2.5pt];
				\node at (3.5,-2.16){$f$=50 Hz};
				\node at (1.2,-2.16){36 m};
				\draw[very thick](2,-4.8)--(4.8,-4.8);
				\draw[very thick](4.8,-4.8)--(11.2,-1.8);
				\draw[very thick](11.2,-1.8)--(14,-1.8);
				\draw[dashed, very thick](14.02,0.2)--(14.02,0);
				\draw[very thick](2.02,-6)--(14.02,-6);
				\draw[dashed, very thick](4.82,-4.8)--(4.82,-6);
				\draw[dashed, very thick](11.22,-1.8)--(11.22,-6);
				\node at (4.8,-6.4){500 $\text{m}$};
				\node at (11.2,-6.4){2000 $\text{m}$};
				\node at (4.5,-0.5){$c_w$=1500 m/s};
				\node at (4.5,-1.1){$\rho_w$=1.0 g/cm$^3$};
				\node at (8,-4.4){$c_b$=1800 m/s};
				\node at (8,-5){$\rho_b$=1.5 g/cm$^3$};
				\node at (8,-5.6){$\alpha_b$=2.0 dB/$\lambda$};
				\node at (8,-6.9){Pressure release bottom};
			\end{tikzpicture}
	\end{minipage}}
	\subfigure[]{
		\begin{minipage}[t]{0.5\linewidth}
			\begin{tikzpicture}[node distance=2cm,scale=0.5]
				\tikzstyle{every node}=[font=\small]
				\node at (1.4,0){0 $\text{m}$};
				\node at (1.2,-4.8){80 $\text{m}$};
				\node at (1.1,-6){100 $\text{m}$};
				\node at (2.2,0.4){0 $\text{m}$};
				\node at (14.3,0.4){2500 $\text{m}$};
				\node at (14.6,-1.8){30 $\text{m}$};
				\fill[brown] (14,-7) rectangle (2,-6);
				\fill[cyan,opacity=0.7] 	(14,0)--(14,-1.8)--(11.2,-1.8)--(4.8,-4.8)--(2,-4.8)--(2,0)--cycle;
				\fill[orange,opacity=0.7] 	(2,-4.8)--(4.8,-4.8)--(11.2,-1.8)--(14,-1.8)--(14,-6)--(2,-6)--cycle;
				\draw[very thick, ->](2,0)--(14.5,0) 	node[right]{$x$};
				\draw[very thick, ->](2.02,0.2)--(2.02,-7.5) 	node[below]{$z$};
				\filldraw [red] (2.02,-2.16) circle [radius=2.5pt];
				\node at (3.5,-2.16){$f$=50 Hz};
				\node at (1.2,-2.16){36 m};
				\draw[very thick](2,-4.8)--(4.8,-4.8);
				\draw[very thick](4.8,-4.8)--(11.2,-1.8);
				\draw[very thick](11.2,-1.8)--(14,-1.8);
				\draw[dashed, very thick](14.02,0.2)--(14.02,0);
				\draw[dashed, very thick](2.02,-6)--(14.02,-6);
				\draw[dashed, very thick](4.82,-4.8)--(4.82,-6);
				\draw[dashed, very thick](11.22,-1.8)--(11.22,-6);
				\node at (4.5,-0.5){$c_w$=1500 m/s};
				\node at (4.5,-1.1){$\rho_w$=1.0 g/cm$^3$};
				\node at (8,-4.4){$c_b$=1800 m/s};
				\node at (8,-5.0){$\rho_b$=1.5 g/cm$^3$};
				\node at (8,-5.6){$\alpha_b$=2.0 dB/$\lambda$};
				\node at (8,-6.5){$c_\infty$=2000 m/s $\rho_\infty$=1.5 g/cm$^3$ $\alpha_\infty$=2 dB/$\lambda$};
			\end{tikzpicture}
	\end{minipage}}
	\subfigure[]{
		\begin{minipage}[t]{0.5\linewidth}
			\begin{tikzpicture}[node distance=2cm,scale=0.5]
				\tikzstyle{every node}=[font=\small]
				\node at (1.4,0){0 $\text{m}$};
				\node at (1.1,-3){150 $\text{m}$};
				\node at (1.1,-2){100 $\text{m}$};
				\node at (1.1,-6){300 $\text{m}$};
				\node at (1.1,-6.9){350 $\text{m}$};
				\node at (2.3,0.4){0 $\text{km}$};
				\node at (14,0.4){10 $\text{km}$};
				\draw[very thick](5,0)--(5,0.2);
				\node at (5,0.4){2.5 $\text{km}$};
				\fill[cyan,opacity=0.7] 	 (2,0)--(2,-6)--(3.2,-6)--(4.4,-1)--(5.6,-6)--(14,-6)--(14,0)--cycle;
				\fill[orange,opacity=0.7] 	 (2,-6)--(3.2,-6)--(4.4,-1)--(5.6,-6)--(14,-6)--(14,-7)--(2,-7)--cycle;
				\draw[very thick, ->](2,0)--(14.5,0) 	node[right]{$x$};
				\draw[very thick, ->](2.02,0.2)--(2.02,-7.5) 	 node[below]{$z$};
				\filldraw [red]  (5.0,-2) circle [radius=2.5pt];		
				\filldraw [blue] (5.0,-3) circle [radius=2.5pt];
				\node at (6.5,-2.5){$f$=23 Hz};
				\draw[very thick](2,-3)--(2.2,-3);
				\draw[very thick](2,-2)--(2.2,-2);
				\draw[very thick](2,-6)--(3.2,-6);
				\draw[very thick](3.2,-6)--(4.4,-1);
				\draw[very thick](4.4,-1)--(5.6,-6);
				\draw[very thick](5.6,-6)--(14,-6);
				\draw[dashed, very thick](14.02,0.2)--(14.02,0);
				\draw[very thick](2.02,-7)--(14.02,-7);
				\draw[dashed, very thick](3.2,-6)--(3.2,-7);
				\draw[dashed, very thick](4.4,-1)--(4.4,-7);
				\draw[dashed, very thick](5.6,-6)--(5.6,-7);
				\node at (3.2,-7.4){1 $\text{km}$};
				\node at (5.6,-7.4){3 $\text{km}$};
				\node at (10.5,-2.5){$c_w$=1500 m/s};
				\node at (10.5,-3.1){$\rho_w$=1.0 g/cm$^3$};
				\node[text=white] at (4.5,-6.5){$c_b$=1800 m/s};
				\node[text=white] at (8.5,-6.5){$\rho_b$=1.5 g/cm$^3$};
				\node[text=white] at (12.3,-6.5){$\alpha_b$=5 dB/$\lambda$};
				\node at (10,-7.4){Pressure release bottom};
			\end{tikzpicture}
	\end{minipage}}
	\subfigure[]{
		\begin{minipage}[t]{0.5\linewidth}
			\begin{tikzpicture}[node distance=2cm,scale=0.5]
				\tikzstyle{every node}=[font=\small]
				\node at (1.4,0){0 $\text{m}$};
				\node at (1.2,-1.4){40 $\text{m}$};
				\node at (1.1,-3.5){100 $\text{m}$};
				\node at (1.1,-5.2){150 $\text{m}$};
				\node at (1.1,-6.9){200 $\text{m}$};
				\node at (2.67,0.4){0.5 $\text{km}$};
				\node at (14,0.4){9 $\text{km}$};
				\node at (4.66,0.4){2 $\text{km}$};
				\node at (11.34,0.4){7 $\text{km}$};
				\fill[cyan,opacity=0.7] 	  (2,0)--(2,-5.2)--(4.66,-1.4)--(11.34,-1.4)--(14,-5.2)--(14,0)--cycle;
				\fill[orange,opacity=0.7] 	 (2,-5.2)--(4.66,-1.4)--(11.34,-1.4)--(14,-5.2)--(14,-7)--(2,-7)--cycle;
				\draw[very thick, ->](2,0)--(14.5,0) 	node[right]{$x$};
				\draw[very thick, ->](2.02,0.2)--(2.02,-7.5) 	 node[below]{$z$};
				\filldraw [red] (2.67,-3.5) circle [radius=2.5pt];	
				\filldraw [blue] (13.34,-3.5) circle [radius=2.5pt];	
				\node at (13.34,-0.4){8.5 $\text{km}$};
				\draw[very thick](13.34,0)--(13.34,-0.2);
				\node at (3.4,-1.2){$f$=25 Hz};
				\draw[very thick](2.66,0)--(2.66,0.2);
				\draw[very thick](4.66,0)--(4.66,0.2);
				\draw[very thick](11.34,0)--(11.34,0.2);		
				\draw[very thick](2,-1.4)--(2.2,-1.4);
				\draw[very thick](2,-3.5)--(2.2,-3.5);
				\draw[very thick](2,-5.2)--(2.2,-5.2);
				\draw[very thick](2,-5.2)--(4.67,-1.4);
				\draw[very thick](4.66,-1.4)--(11.34,-1.4);
				\draw[very thick](11.33,-1.4)--(14,-5.2);
				\draw[dashed, very thick](14.02,0.2)--(14.02,0);
				\draw[very thick](2.02,-7)--(14.02,-7);
				\draw[dashed, very thick](4.66,-1.4)--(4.66,-7);
				\draw[dashed, very thick](11.34,-1.4)--(11.34,-7);
				\node at (8,-0.4){$c_w$=1500 m/s};
				\node at (8,-1.0){$\rho_w$=1.0 g/cm$^3$};
				\node at (8,-3.5){$c_b$=1700 m/s};
				\node at (8,-4.1){$\rho_b$=1.5 g/cm$^3$};
				\node at (8,-4.8){$\alpha_b$=0.5 dB/$\lambda$};
				\node at (8,-7.5){Rigid bottom};
			\end{tikzpicture}
	\end{minipage}}
	\caption{Penetrable slope ocean environments (a)--(b) and seamount problems (c)--(d).}
	\label{Figure12}
\end{figure}
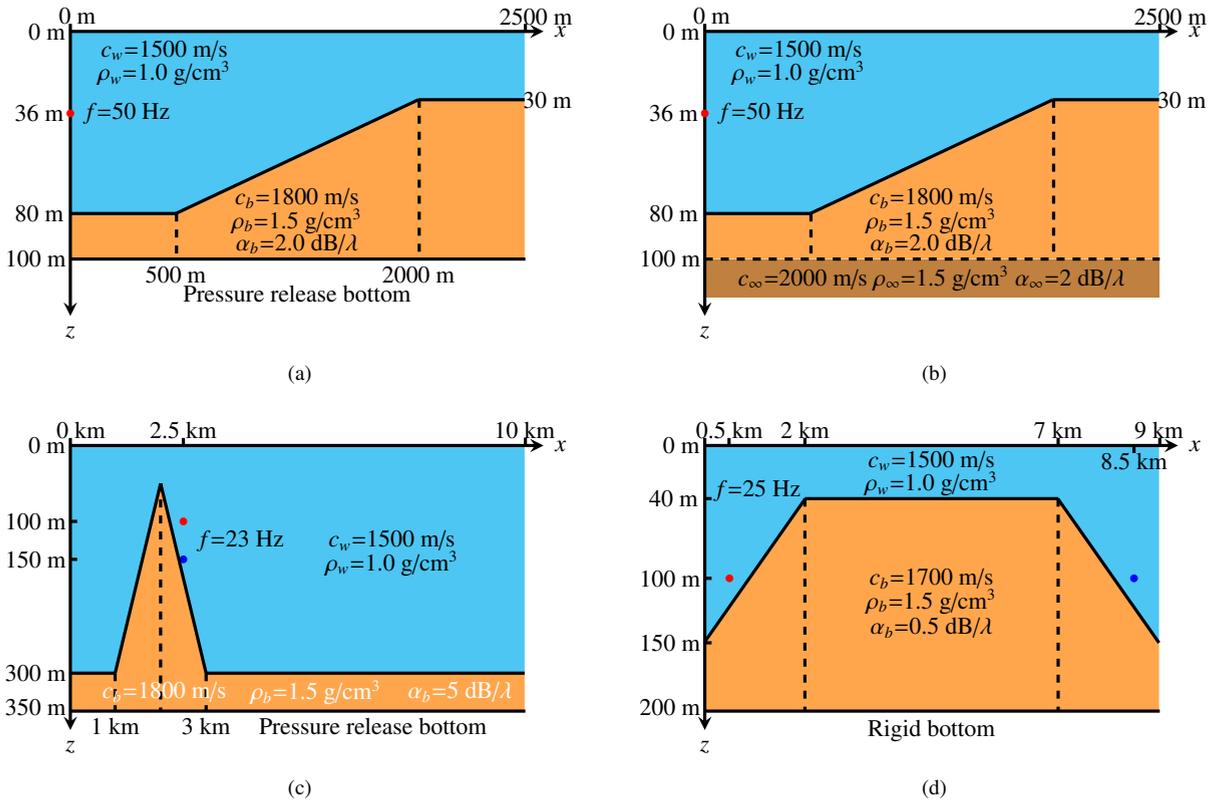

\begin{figure}[htbp]
	\centering
	\subfigure[]{\includegraphics[width=0.49\linewidth]{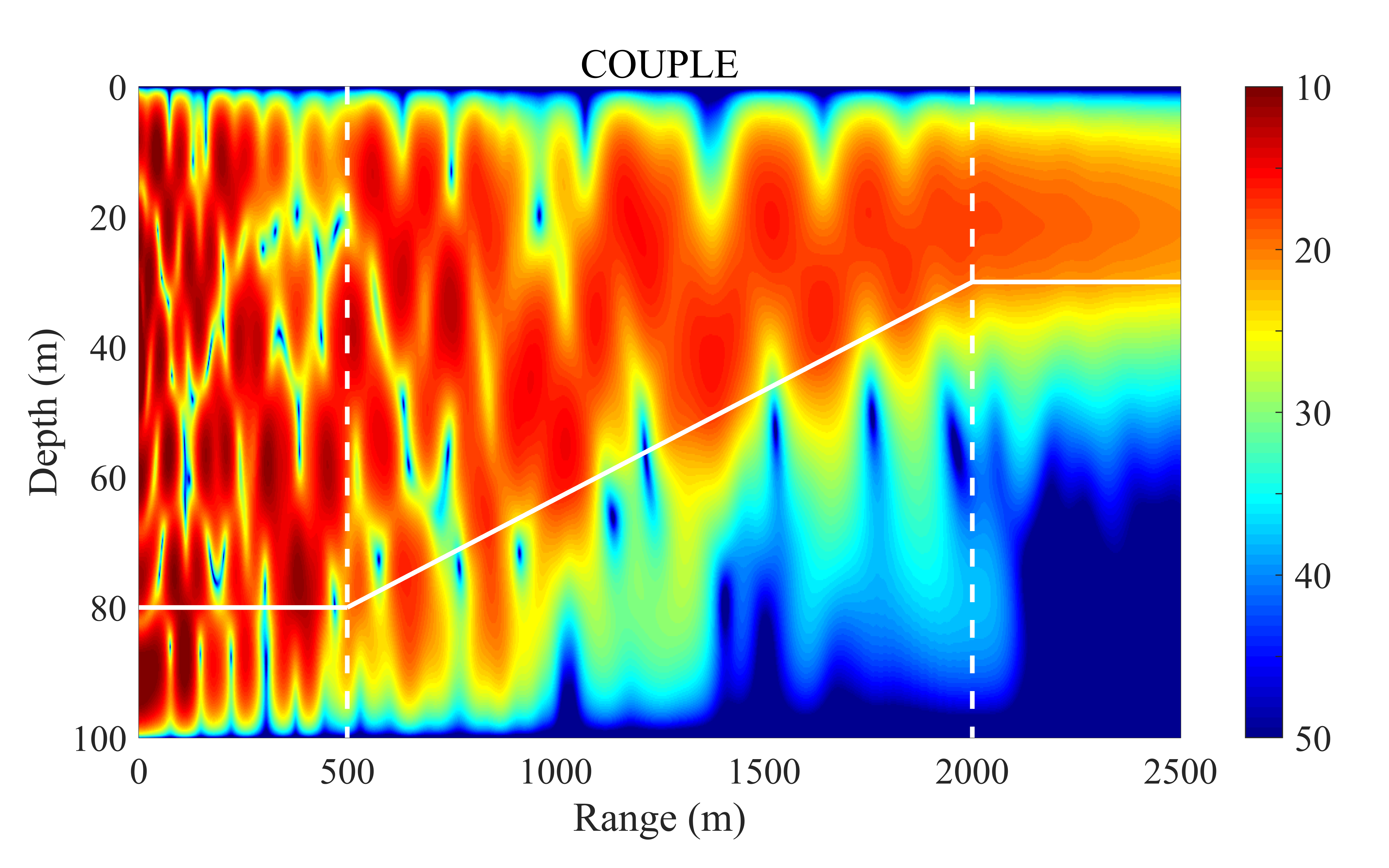}}
	\subfigure[]{\includegraphics[width=0.49\linewidth]{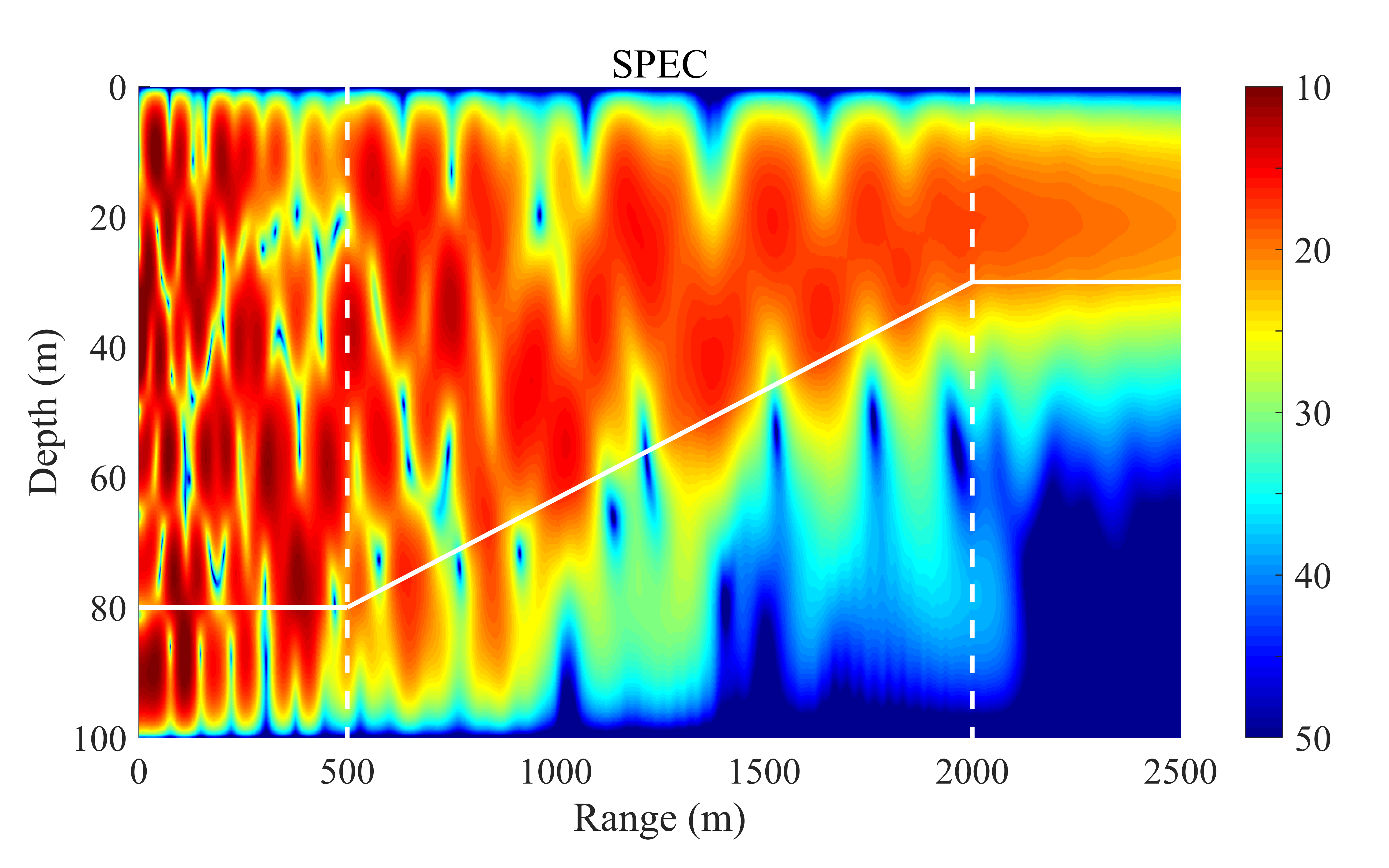}}\\
	\subfigure[]{\includegraphics[width=\linewidth]{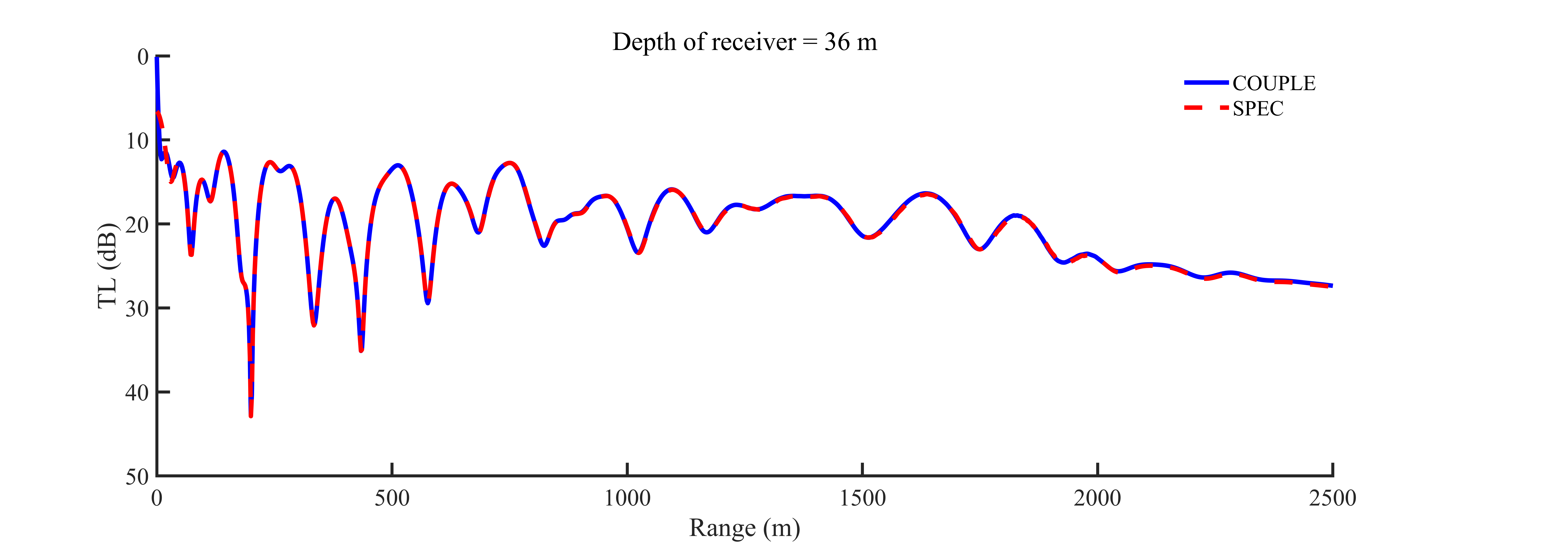}}
	\caption{Sound fields for example 7 as calculated by COUPLE (a) and SPEC (b) and the TL curves at a depth of 36 m (c).}
	\label{Figure13}
\end{figure}

Consider a penetrable slope in an ocean environment as shown in Fig.~\ref{Figure12}(a), which is a very common scenario of range dependency. The sound field in this case does not have an analytical solution, so we present the results of COUPLE for comparison. First, we consider the situation in which the sound source is placed at $x_\mathrm{s}=0$. Fig.~\ref{Figure13} displays the sound fields calculated by COUPLE and SPEC. In the simulation, the number of coupled modes was taken to be $M=6$, and the slope was divided into 200 segments. The sound fields confirm that the results of the two programs match fairly well.

\begin{figure}[htbp]
	\centering
	\subfigure[]{\includegraphics[width=0.49\linewidth]{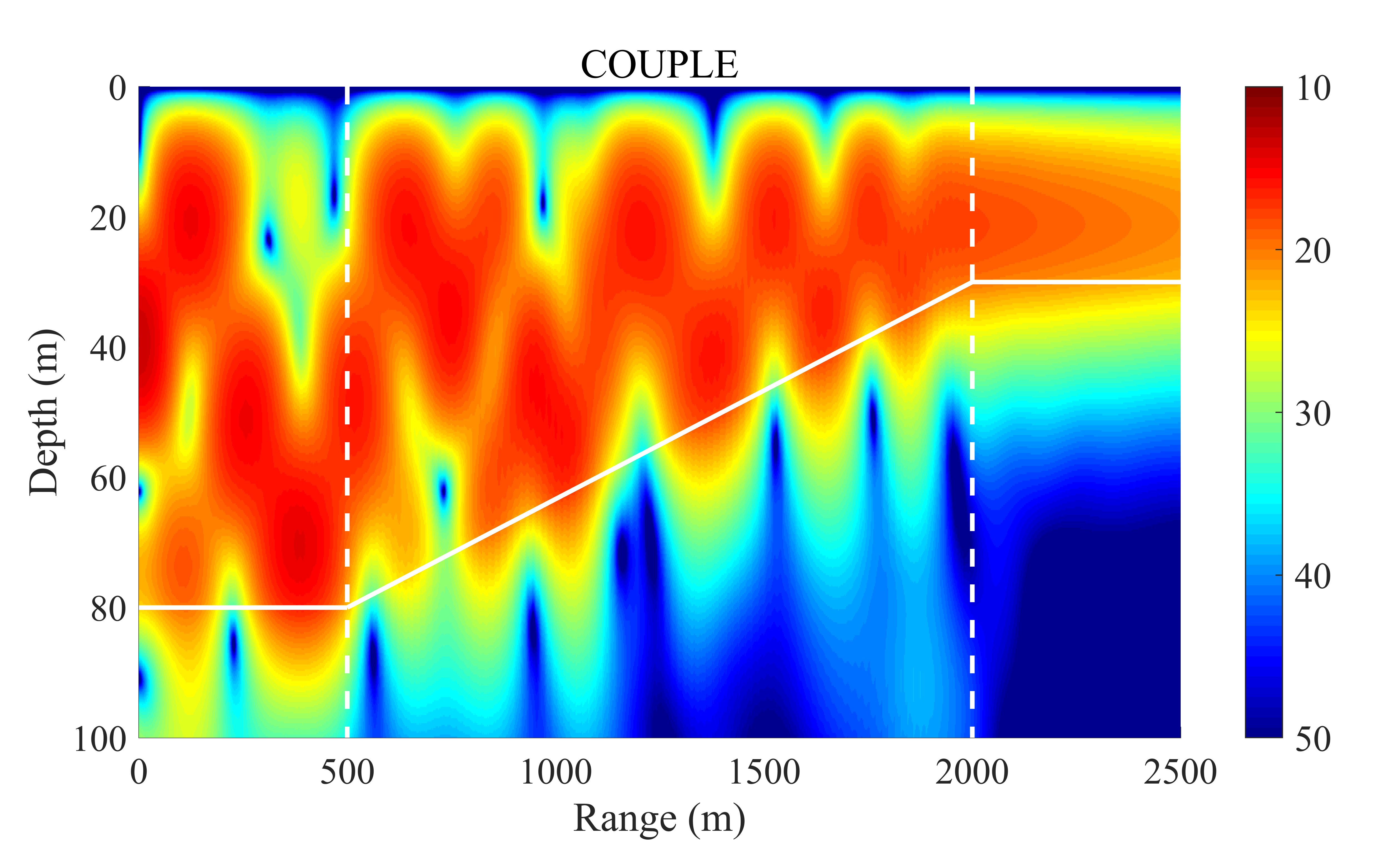}}
	\subfigure[]{\includegraphics[width=0.49\linewidth]{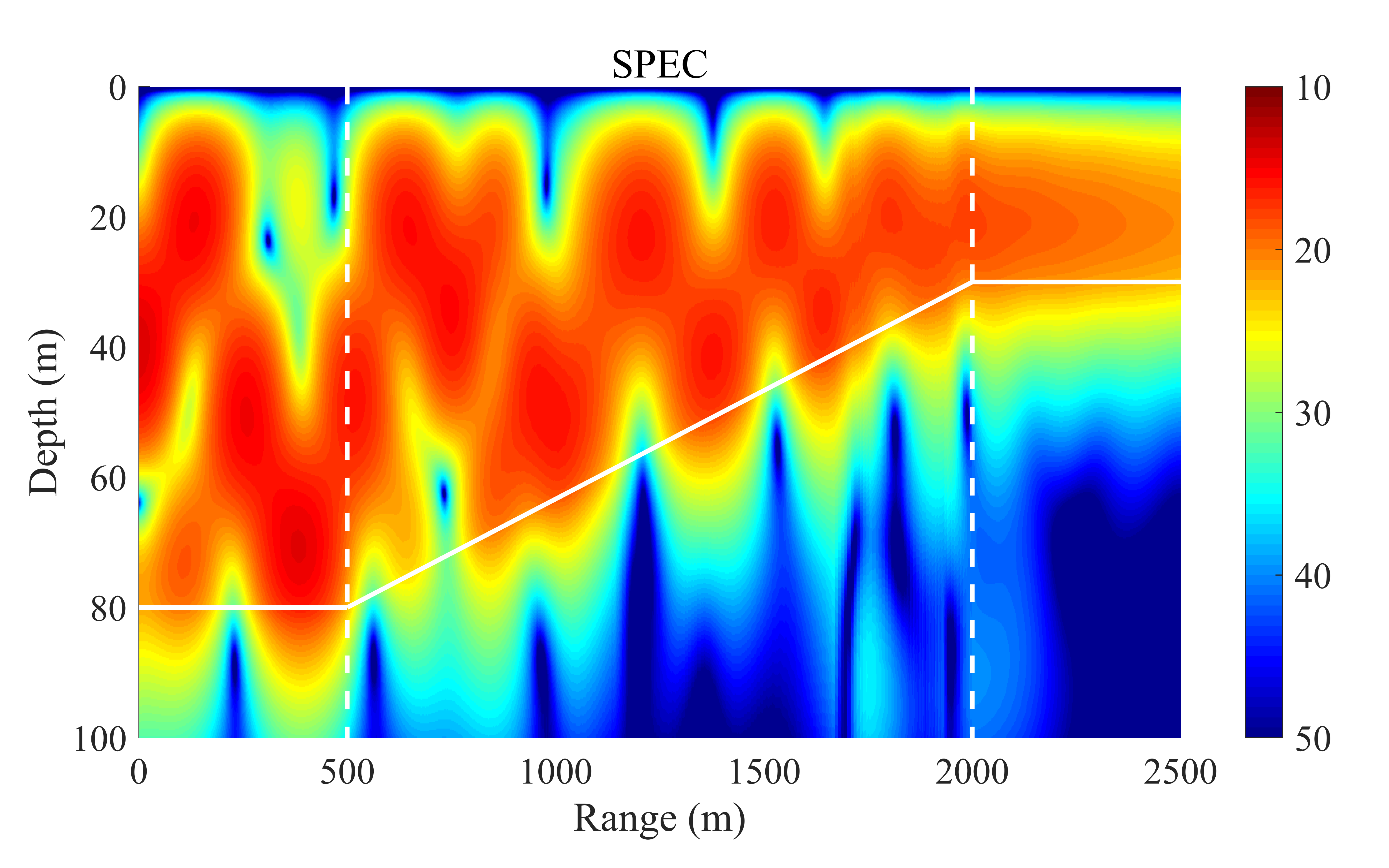}}
	\subfigure[]{\includegraphics[width=0.49\linewidth]{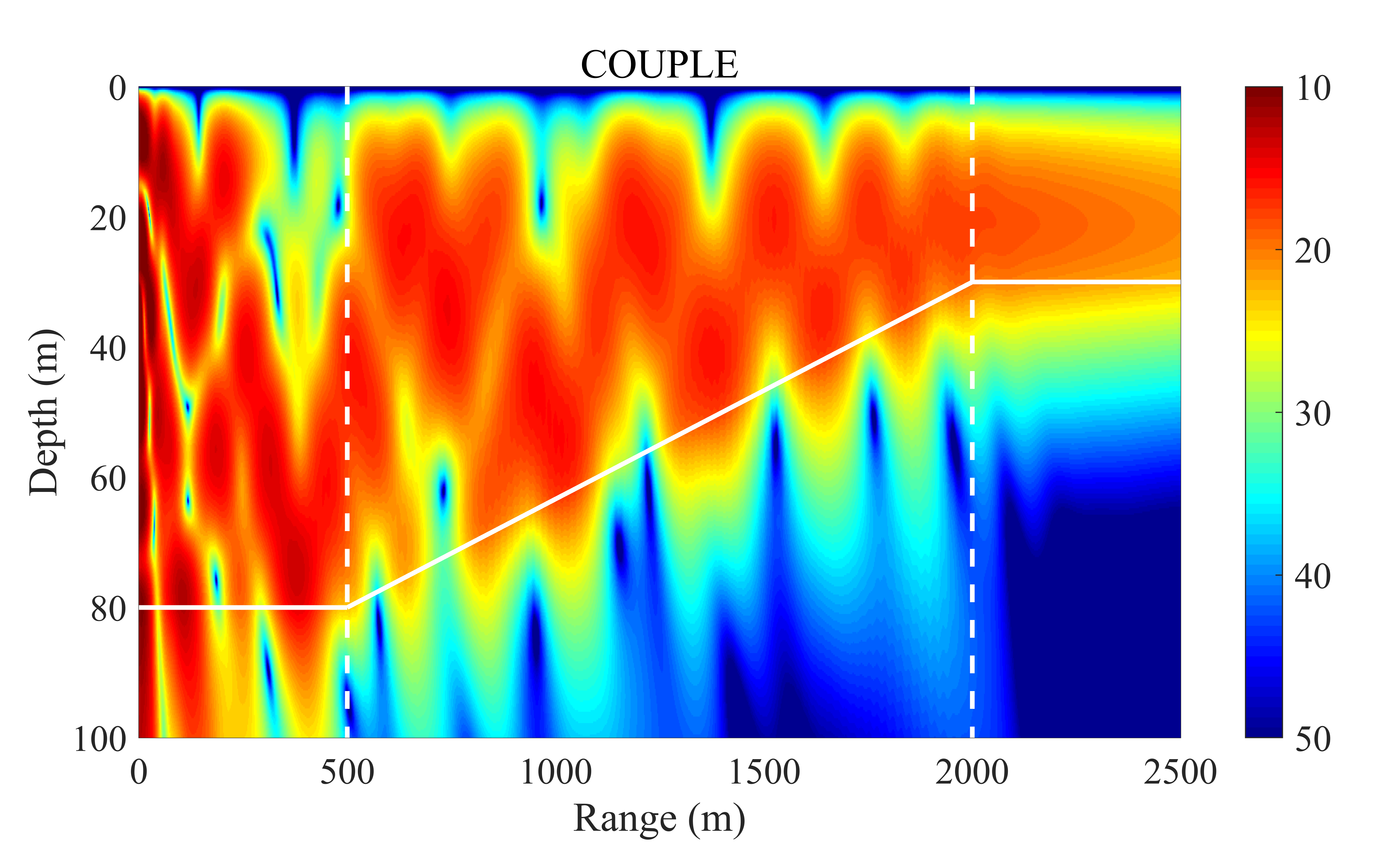}}
	\subfigure[]{\includegraphics[width=0.49\linewidth]{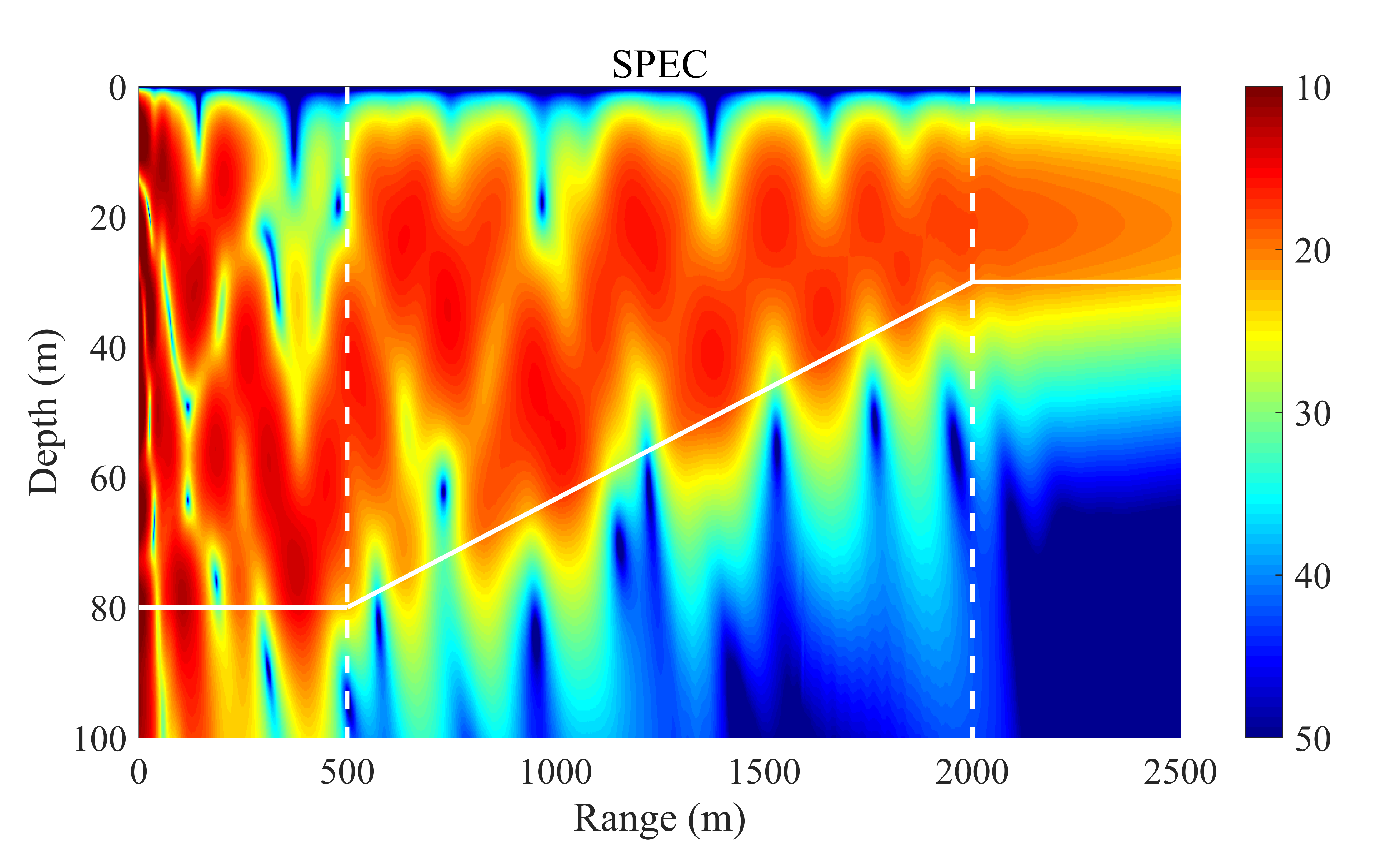}}
	\caption{Sound fields for example 8 as calculated by COUPLE ((a) and (c)) and SPEC ((b) and (d)); in (a) and (b), only the coupling of 3 propagating modes is considered, and in (c) and (d), an absorbing layer is adopted to contain the leaky modes.}
	\label{Figure14}
\end{figure}

We also show the case in which the lower boundary of the waveguide is an acoustic half-space in Fig.~\ref{Figure12}(b). The configurations in Fig.~\ref{Figure12}(b) and Fig.~\ref{Figure12}(a) are exactly the same; the only difference is that in the latter, the lower boundary is taken to be an acoustic half-space. Fig.~\ref{Figure14} shows the sound fields under the configuration of Fig.~\ref{Figure12}(b), where the range-dependent region is discretized into 200 flat segments. It is worth explaining that when the Chebyshev--Tau spectral method is used to solve for local modes, the continuous spectrum of the wavenumbers can be obtained, but instability often occurs when the leakage modes are coupled. Therefore, in actual simulation, SPEC usually adopts one of two methods to simulate a waveguide with half-space characteristics. One is to exclude the leakage modes in the coupling analysis because they mainly affect the near field. If the user is not greatly concerned about the near field, then this method is effective. The second is to set a sufficiently deep absorbing layer on the seabed that the energy entering the absorbing layer will be completely absorbed and the sound field above the absorbing layer will not be affected by the absorbing layer, while a free seabed is still used at the bottom. Fig.~\ref{Figure14}(b) and \ref{Figure14}(d) display the sound fields calculated by SPEC using these two strategies. The results of both methods are in good agreement with the results of COUPLE, and it can be seen that the influence of the leakage modes is mainly reflected in the near field.

\begin{figure}[htbp]
	\centering
	\subfigure[]{\includegraphics[width=0.49\linewidth]{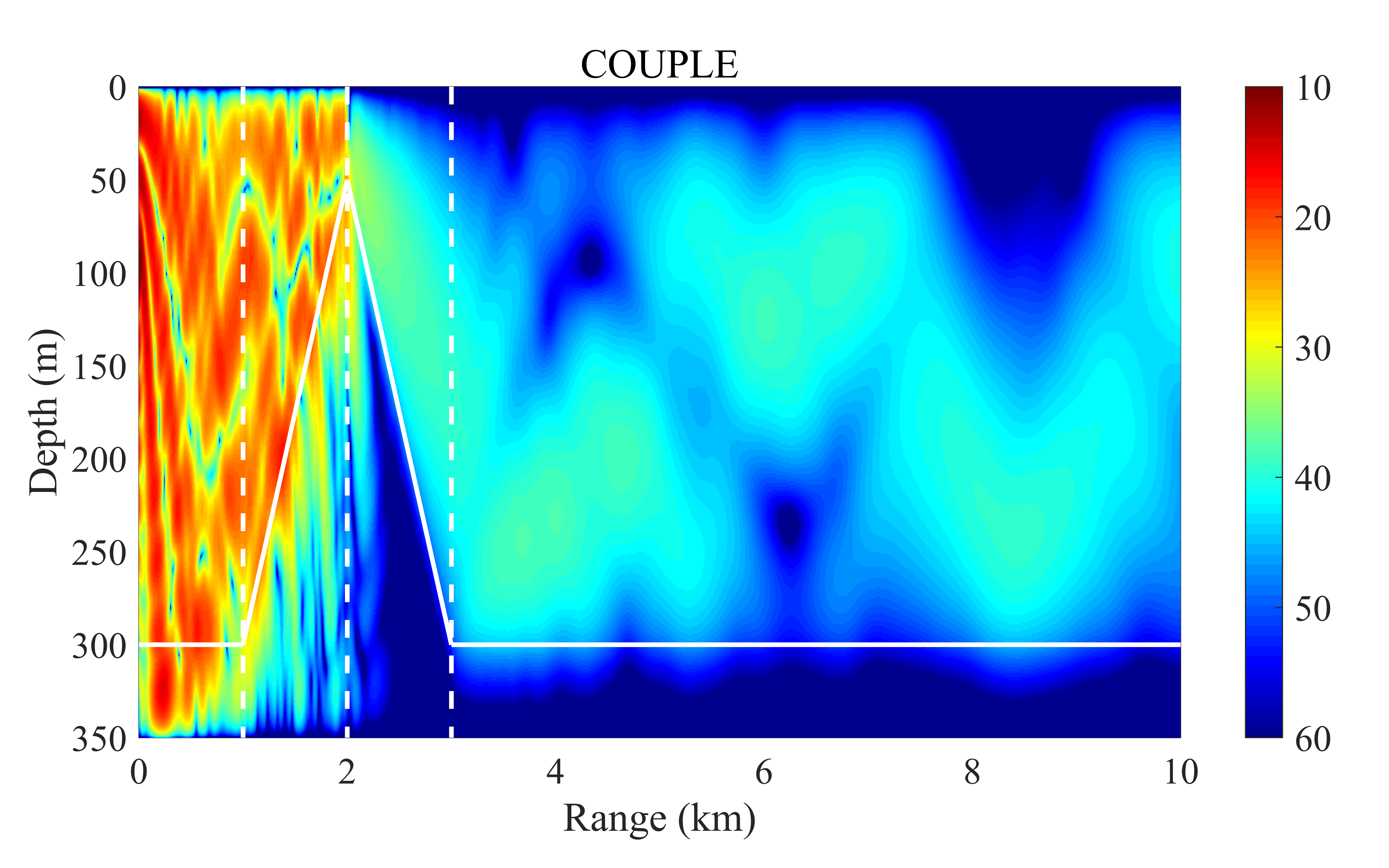}}
	\subfigure[]{\includegraphics[width=0.49\linewidth]{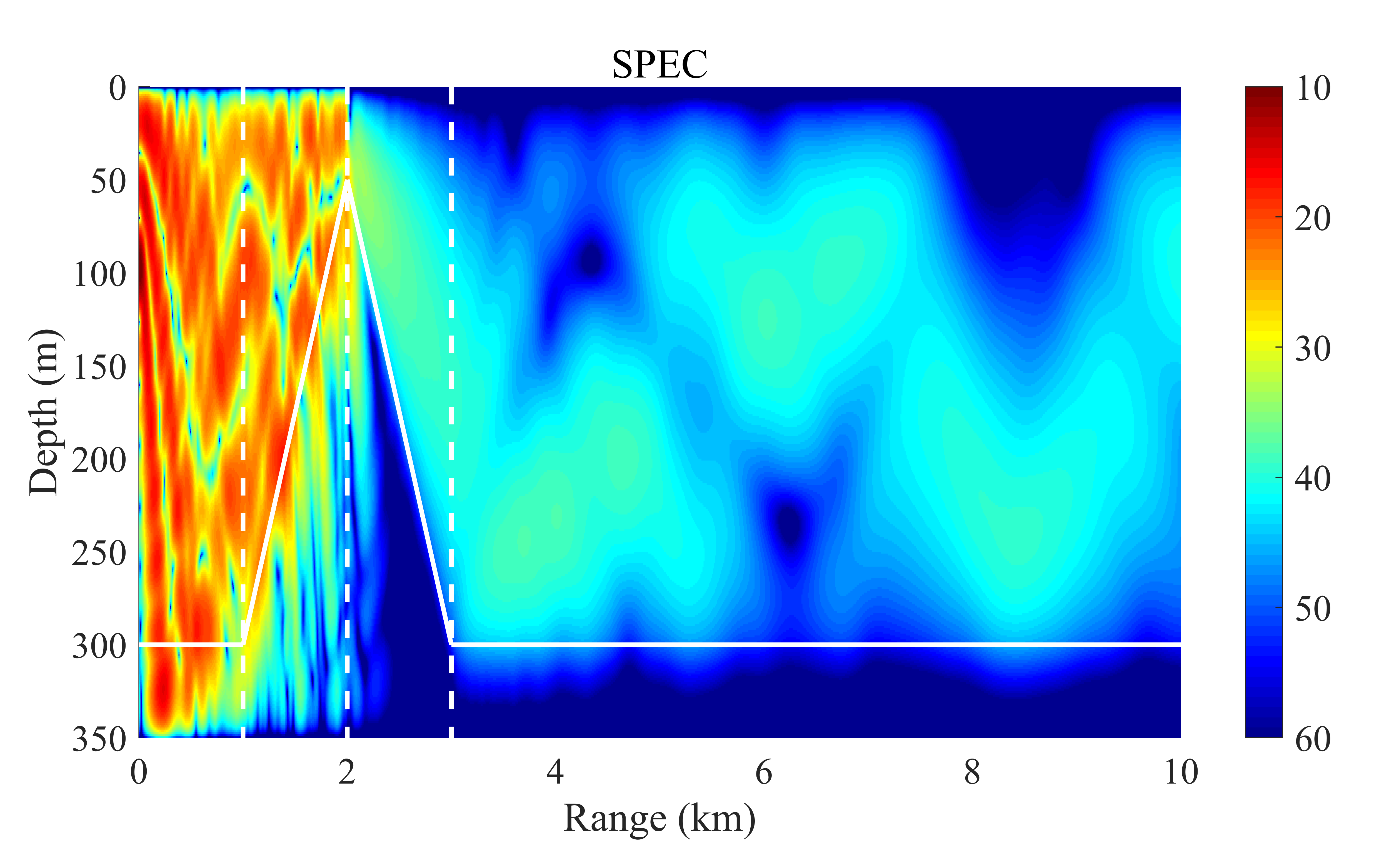}}\\
	\subfigure[]{\includegraphics[width=\linewidth]{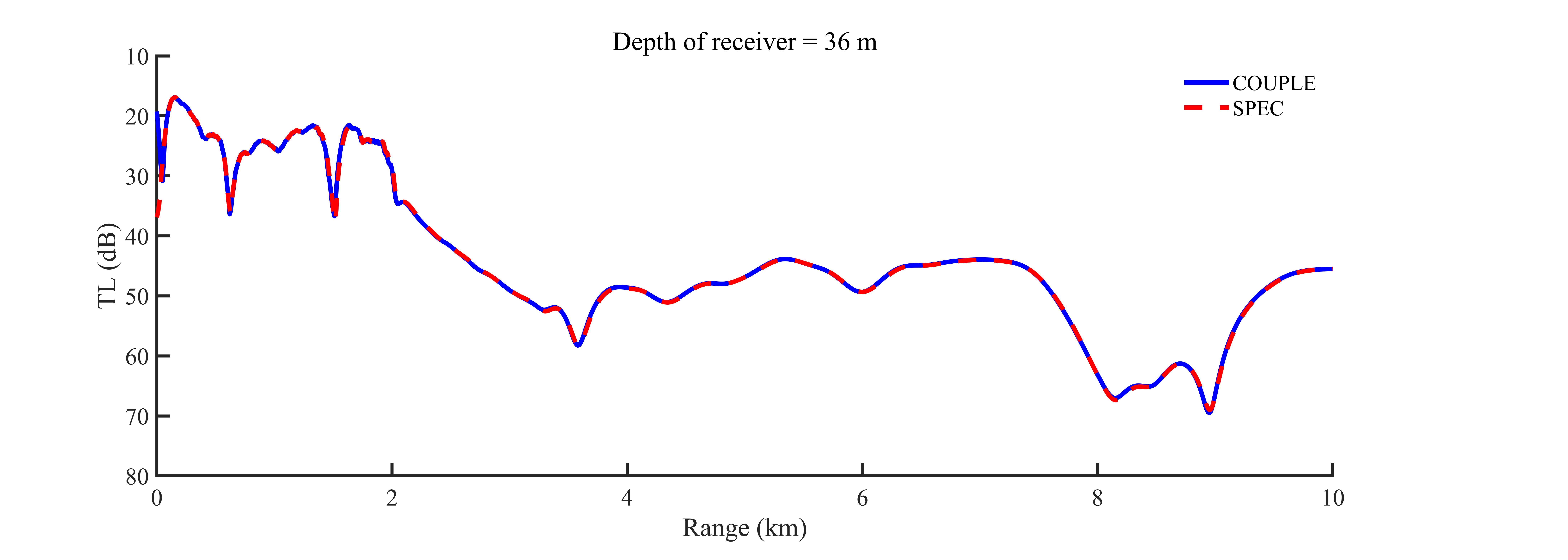}}
	\caption{Sound fields for example 9 as calculated by COUPLE (a) and SPEC (b) and the TL curves at a depth of 36 m (c); the line source is located at $x_\mathrm{s}=0$ m and $z_\mathrm{s}=100$ m.}
	\label{Figure15}
\end{figure}

\begin{figure}[htbp]
	\centering
	\subfigure[]{\includegraphics[width=0.49\linewidth]{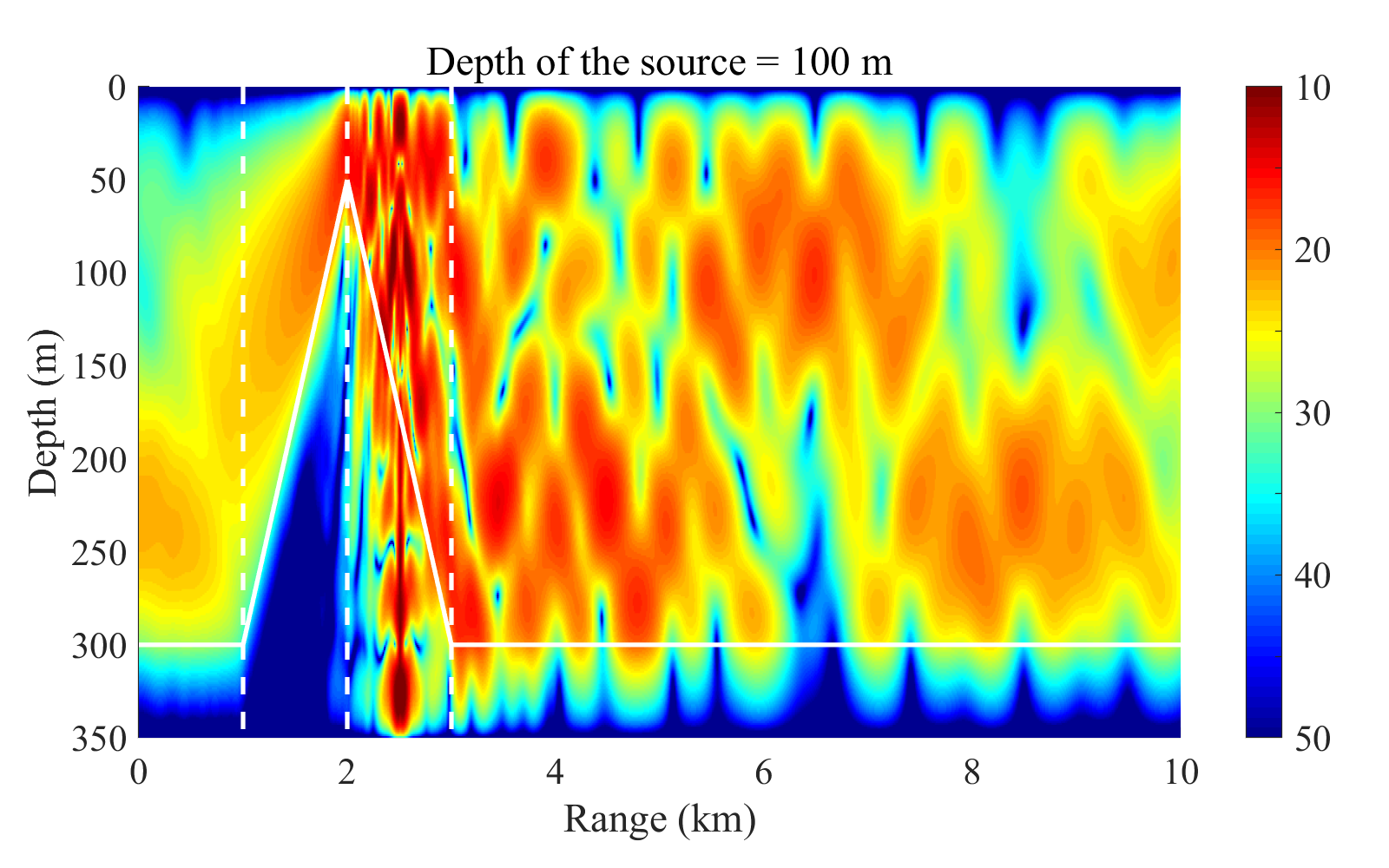}}
	\subfigure[]{\includegraphics[width=0.49\linewidth]{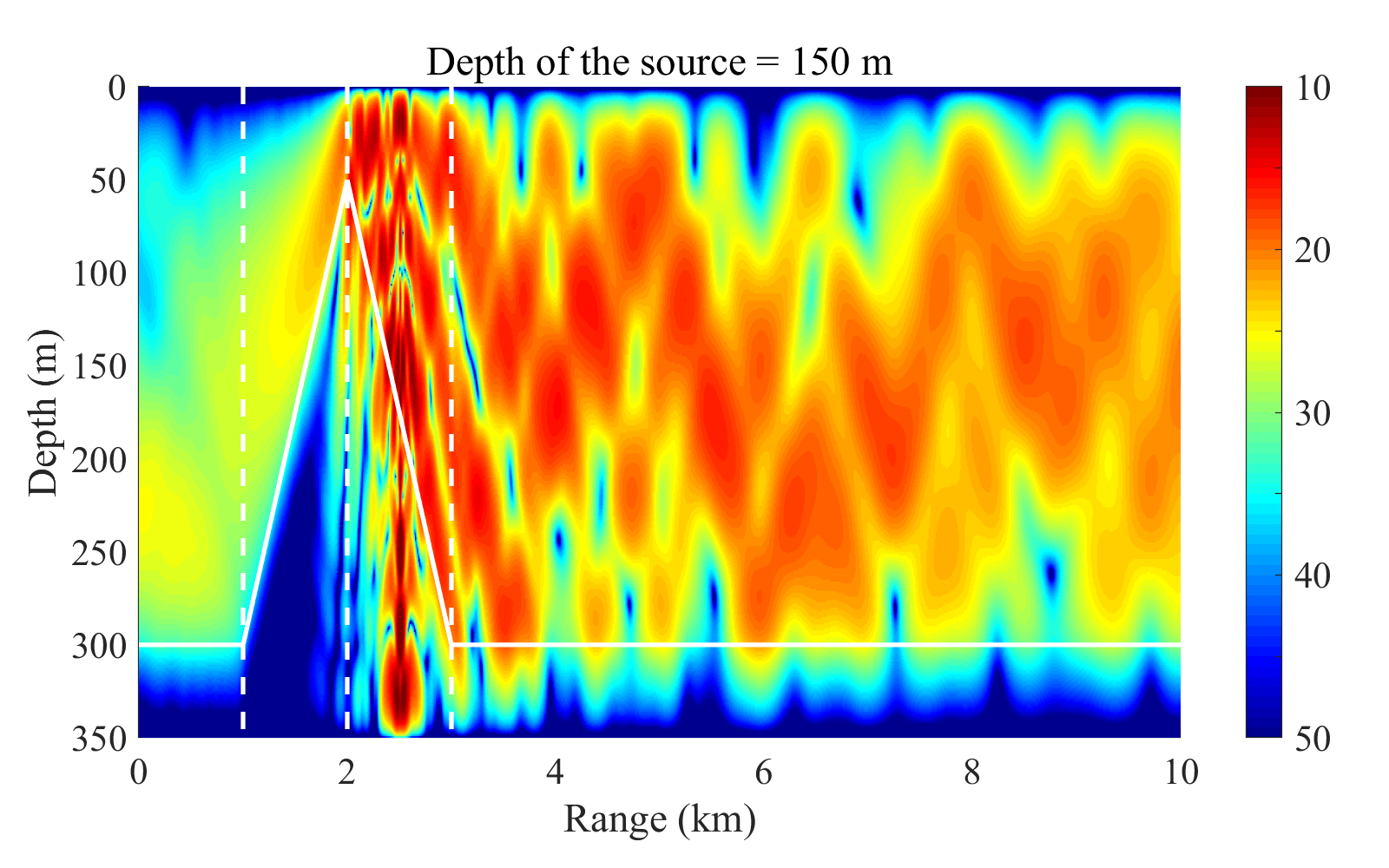}}
	\caption{Sound fields excited by sound sources at different depths in example 9; the sound source in (a) is located at $x_\mathrm{s}=100$ m and $z_\mathrm{s}=100$ m, and the sound source in (b) is located at $x_\mathrm{s}=100$ m and $z_\mathrm{s}=150$ m.}
	\label{Figure16}
\end{figure}

Example 9 is a classic seamount problem, with the configuration shown in Fig.~\ref{Figure12}(c). We first consider the sound field when the sound source is located at $x_\mathrm{s}$=0 and $z_\mathrm{s}$=100 m, and we compare the results with those of COUPLE. Fig.~\ref{Figure15} shows the sound fields calculated by COUPLE and SPEC when the same numbers of segments and coupled modes are used, confirming that the results of the two programs are in good agreement.
In addition, we consider the situation in which the sound source is placed on a slope. Here, we consider two different sound source depths above the slope: the sound sources are located at the same horizontal range of $x_\mathrm{s}=2500$ m but at different depths of $z_\mathrm{s}=50$ m and $z_\mathrm{s}=100$ m. Because COUPLE cannot handle this situation, we visualize only the results of SPEC to demonstrate its functionality. The SPEC-calculated sound fields produced by these sound sources at two different depths are shown in Fig.~\ref{Figure16}. The deeper sound source produces more energy that is distributed in the center of the waveguide, whereas less energy reaches the other side of the seamount.

\begin{figure}[htbp]
	\centering
	\subfigure[]{\includegraphics[width=0.49\linewidth]{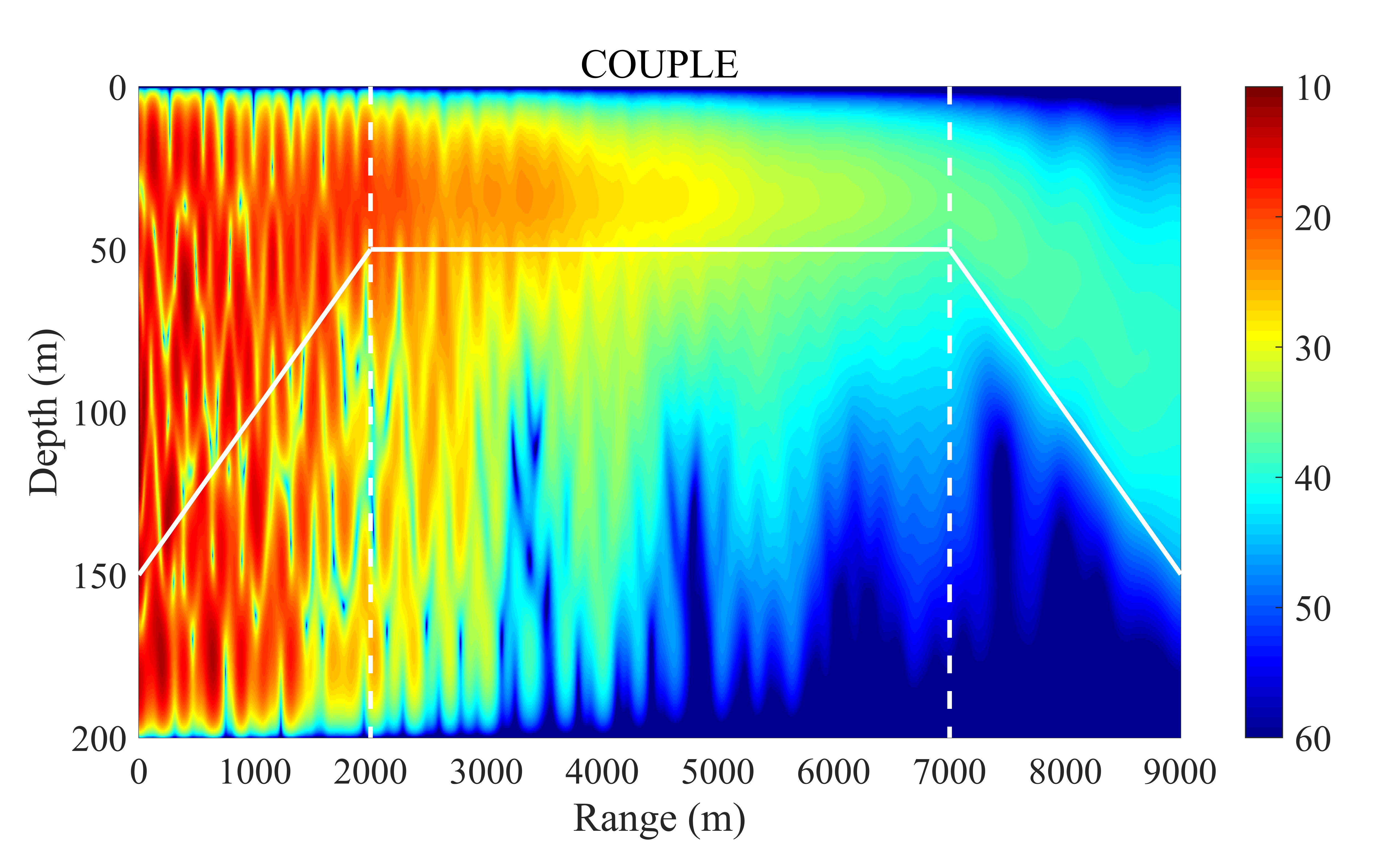}}
	\subfigure[]{\includegraphics[width=0.49\linewidth]{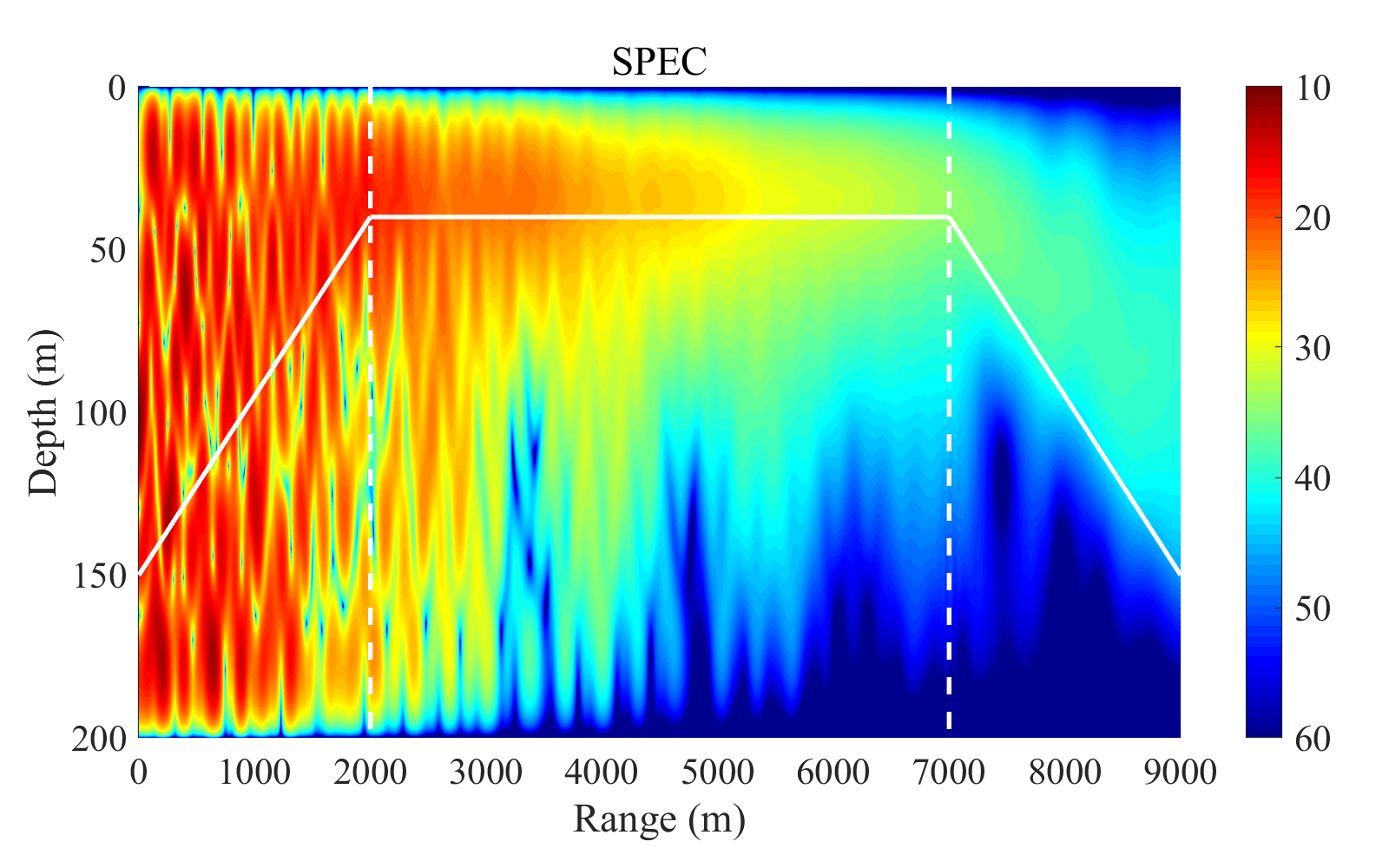}}\\
	\subfigure[]{\includegraphics[width=\linewidth]{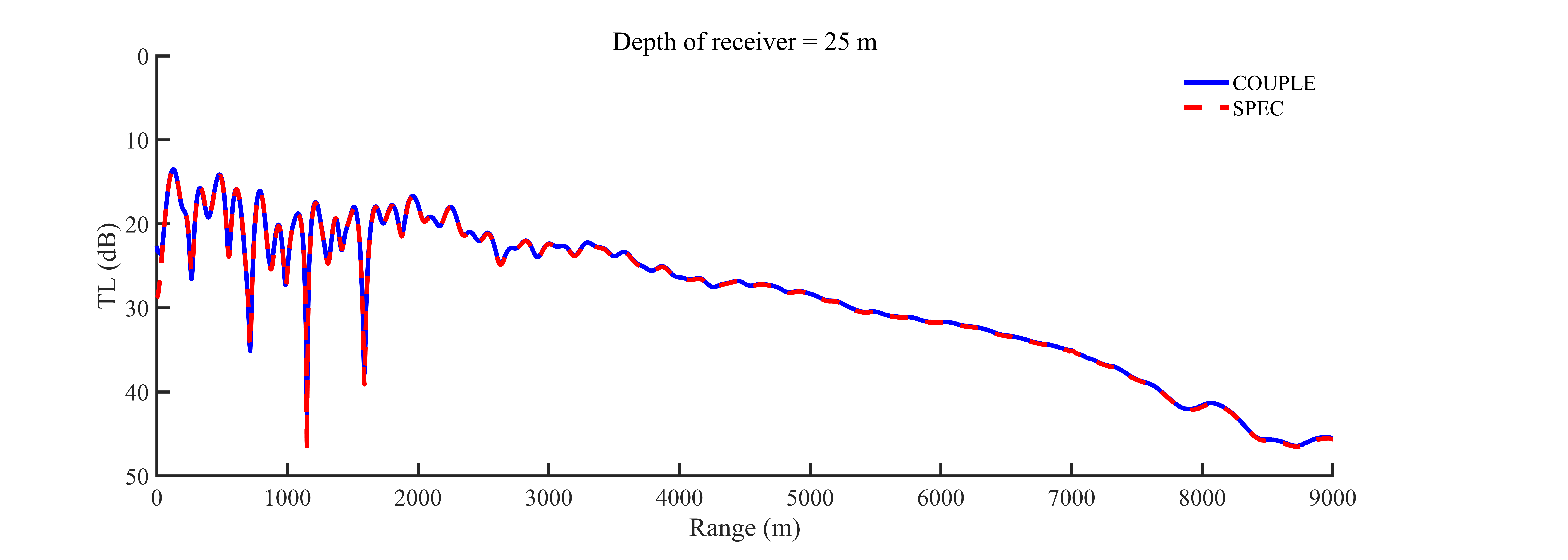}}
	\caption{Sound fields for example 10 as calculated by COUPLE (a) and SPEC (b) and the TL curves at a depth of 36 m (c); the line source is located at $x_\mathrm{s}=0$ m and $z_\mathrm{s}=100$ m).}
	\label{Figure17}
\end{figure}

\begin{figure}[htbp]
	\centering
	\subfigure[]{\includegraphics[width=0.49\linewidth]{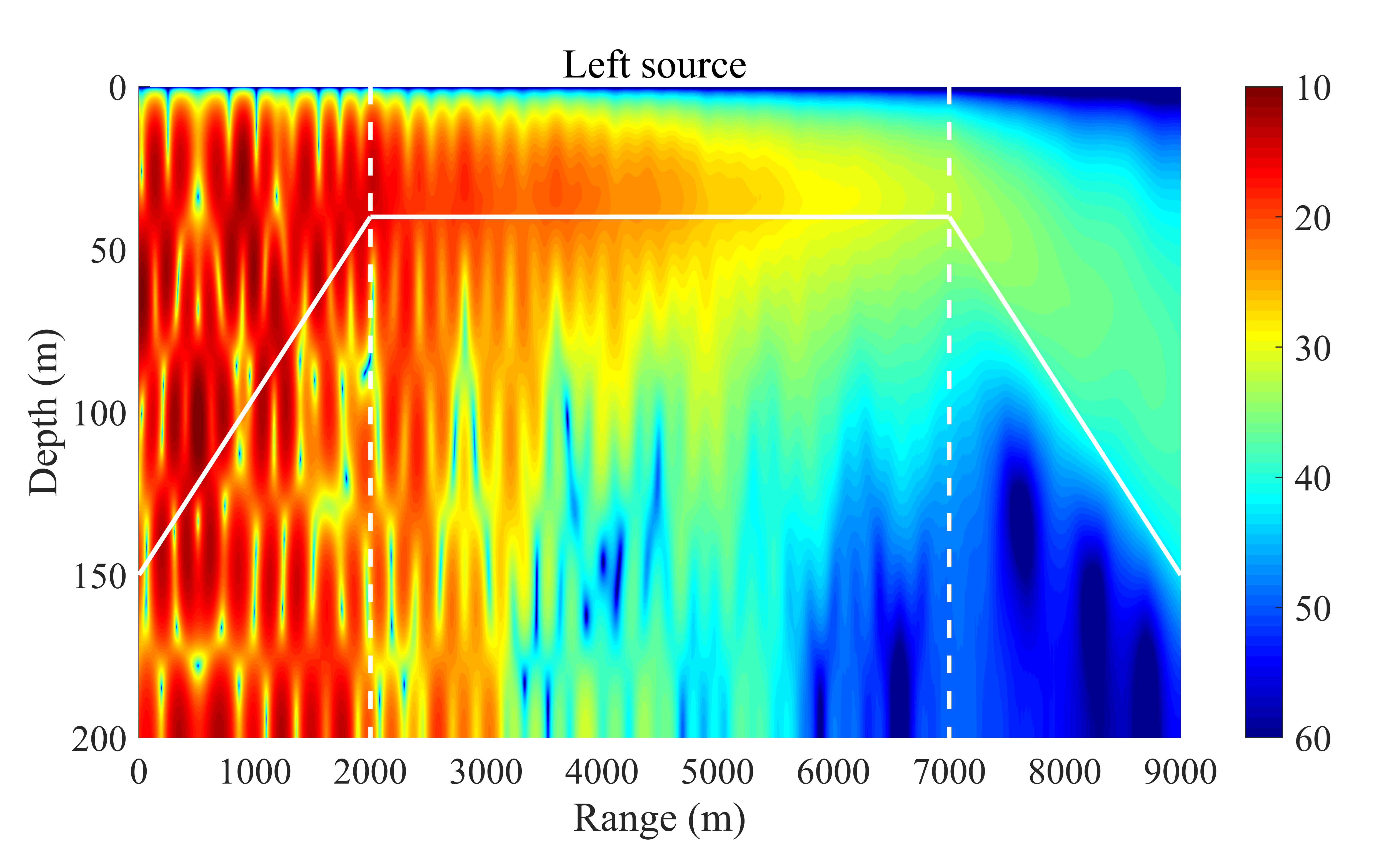}}
	\subfigure[]{\includegraphics[width=0.49\linewidth]{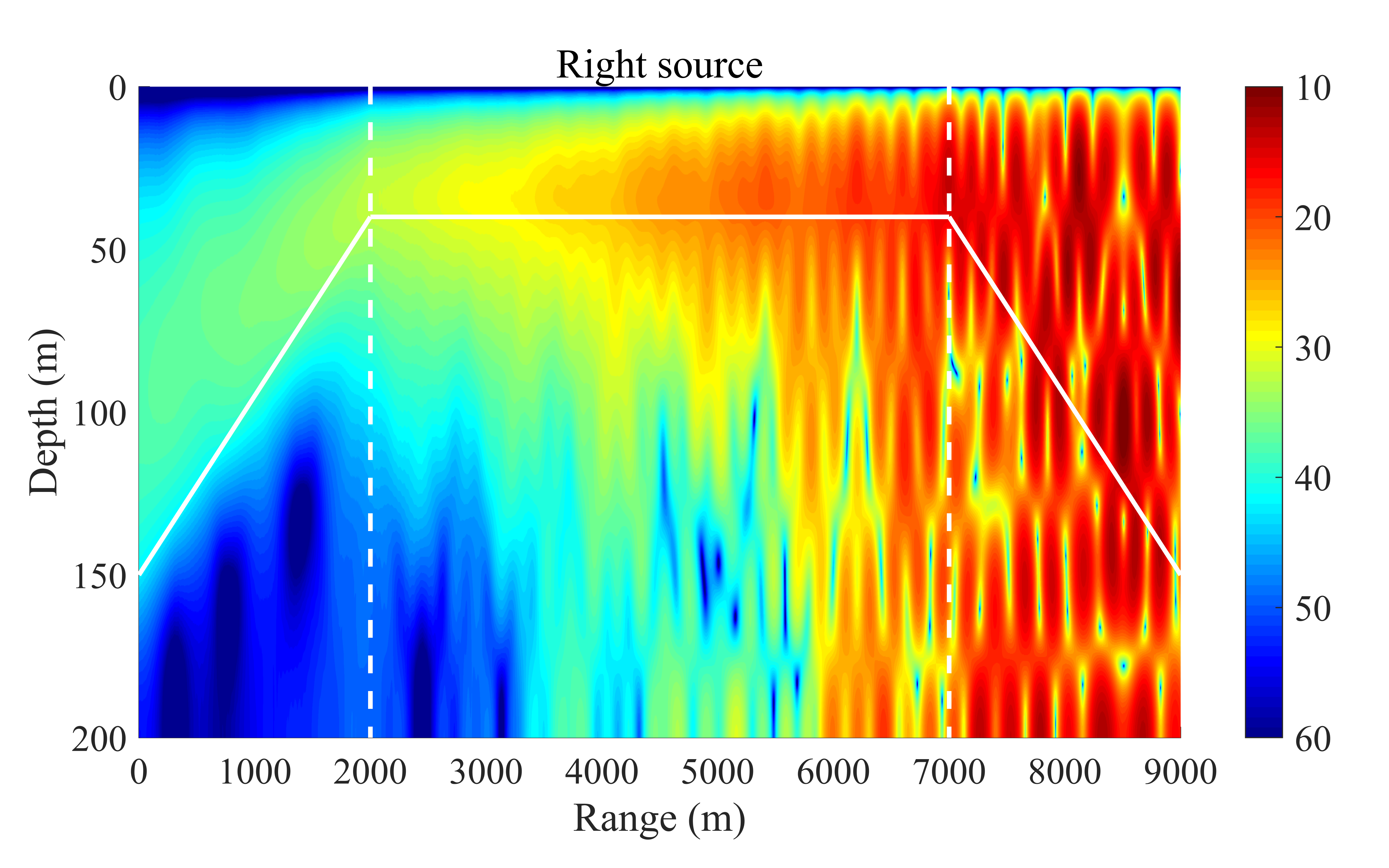}}\\
	\subfigure[]{\includegraphics[width=\linewidth]{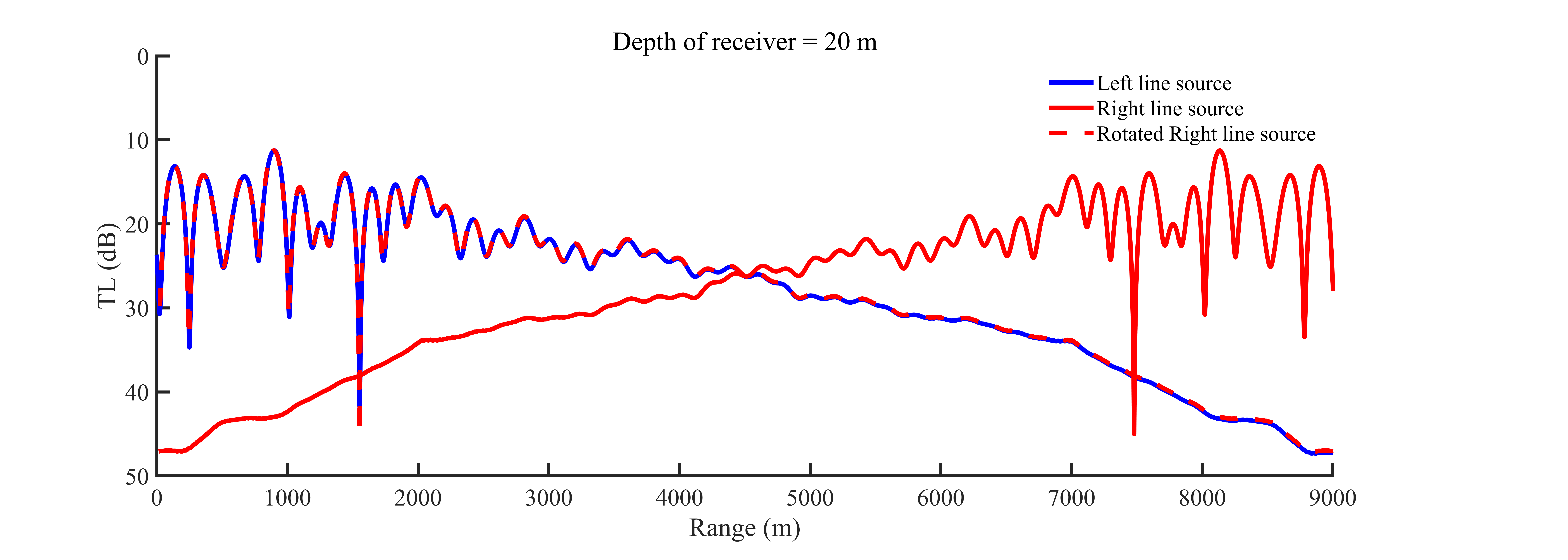}}
	\caption{Sound fields excited by sound sources at different ranges in example 10; the sound source in (a) is located at $x_\mathrm{s}=500$ m and $z_\mathrm{s}=100$ m, and the sound source in (b) is located at $x_\mathrm{s}=8500$ m and $z_\mathrm{s}=100$ m.}
	\label{Figure18}
\end{figure}

Fig.~\ref{Figure12}(d) presents a flat-topped seamount problem. First, we consider the situation in which the sound source is located at $x_\mathrm{s}$=0 and $z_\mathrm{s}$=100 m and the lower boundary is a pressure release boundary. Fig.~\ref{Figure17} shows that the results of SPEC and COUPLE are highly consistent when the same numbers of segments and coupled modes are used.
In addition, we consider line sources located at different positions above the slope. Because the range dependence in this example is symmetrical, the positions along the line source are also symmetrical. Accordingly, as seen from the SPEC results shown in Fig.~\ref{Figure18}, the sound field excited by a symmetrical line source in a symmetrical range-dependent waveguide is symmetrical as well. This outcome further verifies the correctness of the proposed algorithm and program.

It should be noted that when $x_\mathrm{s}=0$ m in COUPLE, $p_0$ is taken to be $1/\left(2\sqrt{2\pi k_\mathrm{s}}\right)$. Thus, in the examples involving comparisons with COUPLE, $p_0$ in SPEC also takes the above form instead of $p_0=\mathrm{i}\mathcal{H}_0^{(1)}(k_\mathrm{s})/4$.

\subsection{Analysis and parallelization}
\subsubsection{Analysis}
\label{sec5.1}
To compare the running time of the algorithm proposed in this paper with that of COUPLE, the test results for several of the above examples are listed in Table \ref{tab1}. The tests were performed on a Dell XPS 8930 personal computer with 16 GB of memory, and each program was executed ten times. The running times listed in the table are the average values. The compiler used was gfortran 7.5.0; all programs considered for comparison were also compiled with this compiler. For the same experiments, when using identical numbers of segments and test functions, SPEC has much shorter running times than COUPLE; these findings directly demonstrate the efficiency of the devised algorithm.

\begin{table}[htbp]
	\centering
	\caption{\label{tab1} Comparison between the running times of SPEC and COUPLE (unit: seconds).}
	\begin{tabular}{ccc}
		\hline
		No. & COUPLE &SPEC   \\
		\hline
		7  & 22.573  &1.830  \\
		9  & 13.150   &1.074 \\
		10 & 28.338   &5.118 \\
		\hline
	\end{tabular}
\end{table}

From a computational cost perspective, the computational load of the algorithm proposed in this article is concentrated in the third and fifth steps. In the former, it is necessary to solve $J$ dense matrix eigenvalue problems with a scale of $(N_w+N_b+2)$ (see Eq.~\eqref{eq:23}), and the matrix size doubles with the introduction of an acoustic half-space boundary (see Eq.~\eqref{eq:27}). In the latter, it is necessary to solve a system of linear equations of the band matrix, with a scale of $(2J-1)\times M$ (see Eq.~\eqref{eq:44}). Table \ref{tab2} lists the running times of these two steps for comparison with the running time of the entire program. These two steps obviously constitute the main computational load of the algorithm, and step 3 incurs a larger computational load than step 5.
\begin{table}[htbp]
	\centering
	\caption{\label{tab2} Comparison of the running times of the most computationally intensive steps in SPEC (unit: seconds).}
	\begin{tabular}{cccc}
		\hline
		No. & Step 3 &Step 5 & Total \\
		\hline
		7 & 0.983  &0.174 &1.790 \\
		8 & 7.033  &0.950 &9.638 \\
		9 & 0.489  &0.413 &1.125 \\
		10& 3.432  &1.244 &5.019 \\
		\hline
	\end{tabular}			
\end{table}

However, this is exactly why SPEC is more efficient than COUPLE. COUPLE uses the Galerkin method to solve range-independent modal equations. Using the Galerkin method requires first constructing a set of basis functions that satisfy the boundary conditions. Therefore, COUPLE needs to solve a nonsingular Sturm--Liouville problem first and then use the obtained solution as the basis functions. In SPEC, the Chebyshev--Tau spectral method directly uses the Chebyshev polynomials as the basis functions, without any additional basis function construction. Moreover, the Chebyshev--Tau spectral method needs only a simple spectral transformation to form the matrix eigenvalue problem, while the Galerkin method needs many numerical integrations to form the generalized matrix eigenvalue problem. Furthermore, to solve for the coupling coefficients, COUPLE performs a large number of small matrix multiplications and solves a system of linear equations, whereas the global matrix formed by SPEC is banded and thus can be efficiently solved by many mature algorithms.

\subsubsection{Parallelization}
The main computational load of the proposed algorithm can naturally be distributed in parallel. We have therefore attempted to optimize and accelerate SPEC using OpenMP multithreaded parallelization technology. Table \ref{tab3} lists the running times and speedup ratios of SPEC with different numbers of threads. In terms of the speedup, multithreaded acceleration achieves incredible results. With four threads, it is possible to reach a speedup of almost 3, which greatly reduces the running time. This remarkable effect is of extraordinary significance, as it means that SPEC can easily run in parallel and achieve good accuracy even on an inexpensive personal computer, thereby overcoming the challenges of coupled mode analysis to a certain extent. The running time on four threads further demonstrates that SPEC is far more efficient than COUPLE in practical applications.

\begin{table}[htbp]
	\centering
	\caption{\label{tab3} Acceleration effect for SPEC using OpenMP multithreaded parallelization technology (unit: seconds; the number in parentheses is the speedup ratio based on the running time on a single thread).}
	\begin{tabular}{ccccc}
		\hline
		\multirow{2}{*}{No.}&\multirow{2}{*}{Serial}&\multicolumn{3}{c}{Number of Threads}\\
		\cline{3-5}
		& &1 &2 &4 \\
		\hline
		7 &1.830  &1.802(1) &0.983 (1.83) &0.599 (3.00) \\
		8 &9.638  &9.507(1) &4.812 (1.98) &2.501 (3.80) \\
		9 &1.017  &1.081(1) &0.581 (1.86) &0.351(3.08)  \\
		10 &5.118 &5.815(1) &3.374 (1.72) &2.204 (2.64)\\
		\hline
	\end{tabular}
\end{table}

\section{Remarks and Conclusion}
\subsection{Remarks}
From the above analysis, we can intuitively summarize the following features of the algorithm and program proposed in this article:
\begin{enumerate}
	\item
	The proposed algorithm can be used to solve acoustic propagation problems for both classical line sources and generalized line sources in a planar coordinate system. At present, no open-source numerical software can solve for the sound field excited by a generalized line source at an arbitrary position.
	
	\item
	The proposed algorithm improves the method of normalizing the range solution and is unconditionally stable. This approach uses a global matrix to solve for all coupling coefficients at once; therefore, the associated numerical program (SPEC) is stable and robust.
	
	\item
	The global coupling matrices formed by the two types of line sources are band-shaped and sparse, so they can be solved efficiently.
	
	\item
	The normal mode solver based on the Chebyshev--Tau spectral method can accurately solve problems with lower boundary conditions of perfectly free, rigid boundaries as well as acoustic half-spaces.
	
	\item
	The proposed algorithm is naturally parallel, so SPEC can be easily run in parallel on a personal computer.

\end{enumerate}

\subsection{Conclusion}
In this article, we propose a numerical algorithm to solve for the sound field produced by a line source in a plane. This algorithm considers a range-dependent ocean environment and an infinite line source that can be located anywhere in the waveguide. The proposed algorithm is generally based on the coupled modes of a stepwise approximation of the environment; that is, a number of narrow range-independent segments are used to approximate the complete waveguide. The proposed algorithm uses a global matrix to solve for the coupling coefficients of all segments at once. These coupling coefficients include two components related to forward and backward propagation, so this algorithm provides a solution with full two-way coupling. To solve for local normal modes, the proposed algorithm employs the Chebyshev--Tau spectral method, which is more efficient, accurate, capable and robust than classical algorithms. This method can flexibly handle various complex acoustic profiles and seabed topographies.

It is worth mentioning that even though the algorithm proposed in this article offers greatly improved computational efficiency over traditional coupled mode algorithms, from the perspective of the model itself, the coupled modes may still be slower than methods based on rays or the parabolic approximation. Therefore, the proposed algorithm is suitable mainly for low-frequency and shallow-sea calculations or to provide accurate benchmark solutions for other models. For high-frequency, deep-sea and long-range sound fields, as analyzed in Section \ref{sec5.1}, the number of local eigensolutions that need to be found increases sharply, and the scale of the global matrix also grows rapidly; hence, the proposed algorithm needs to be further optimized and expanded to accommodate such problems.

\section*{Acknowledgments}
The authors would like to thank Roberto Sabatini of Embry-Riddle Aeronautical University for his insights, many of which have inspired the way in which the authors present this article.

This work was supported by the National Natural Science Foundation of China [grant number 61972406] and the National Key Research and Development Program of China [grant number 2016YFC1401800].

\bibliographystyle{model1-num-names}

\end{document}